\documentclass{aa}  

\usepackage{graphicx}
\usepackage{txfonts}
\usepackage{hyperref}
\usepackage[dvipsnames]{xcolor}
\hypersetup{
    colorlinks,
    linkcolor={red!50!black},
    citecolor={red!50!black},
    urlcolor={red!50!black}
}

\usepackage{lscape}
\usepackage{natbib}
\bibpunct{(}{)}{;}{a}{}{,} 

\graphicspath{{FIGURES}}

\DeclareRobustCommand{\ion}[2]{%
\relax\ifmmode
\ifx\testbx\f@series
{\mathbf{#1\,\mathsc{#2}}}\else
{\mathrm{#1\,\mathsc{#2}}}\fi
\else\textup{#1\,{\mdseries\textsc{#2}}}%
\fi}

\newcommand{\JHU}{Department of Physics and Astronomy, The Johns Hopkins University, Baltimore, MD 21218.}
\newcommand{\STScI}{Space Telescope Science Institute, Baltimore, MD 21218}
\newcommand{\ARCO}{Astrophysics Research Center of the Open University (ARCO), The Open University of Israel, Ra’anana 4353701, Israel}
\newcommand{\OUI}{Department of Natural Sciences, The Open University of Israel, Ra’anana 4353701, Israel}

\newcommand{\Harvard}{Harvard-Smithsonian Center for Astrophysics, 60 Garden Street, Cambridge, MA 02138, USA}
\newcommand{\IfA}{Institute for Astronomy, University of Hawaii, 2680 Woodlawn Drive, Honolulu, HI 96822, USA}

\newcommand{\UCSC}{Department of Astronomy and Astrophysics, University of California, Santa Cruz, CA 95064, USA}
\newcommand{\QUB}{Astrophysics Research Centre, School of Mathematics and Physics, Queen's University Belfast, Belfast BT7 1NN, UK}

\newcommand{\Northwestern}{Center for Interdisciplinary Exploration and Research in Astrophysics (CIERA) and Department of Physics and Astronomy, Northwestern University, Evanston, IL 60208, USA}
\newcommand{\DARK}{DARK, Niels Bohr Institute, University of Copenhagen, Jagtvej 155A, 2200 Copenhagen, Denmark}
\newcommand{\Illinois}{Department of Astronomy, University of Illinois at Urbana-Champaign, 1002 W. Green St., IL 61801, USA}

\newcommand{\Carnegie}{Observatories of the Carnegie Institute for Science, 813 Santa Barbara St., Pasadena, CA 91101, USA}

\newcommand{\Ozgrav}{OzGrav, School of Physics, The University of Melbourne, VIC 3010, Australia}
\newcommand{\NSFAI}{The NSF AI Institute for Artificial Intelligence and Fundamental Interactions}
\newcommand{\MIT}{Department of Physics, Massachusetts Institute of Technology, Cambridge, MA 02139, USA}
\newcommand{\OKC}{Oskar Klein Centre, Department of Astronomy, Stockholm University, AlbaNova, SE-106 91 Stockholm, Sweden}
\newcommand{\IFA}{Institute for Astronomy, University of Hawaii, 640 N.`Aohoku Pl., Hilo, HI 96720, USA}
\newcommand{\supercomputing}{Center for Astrophysical Surveys, National Center for Supercomputing Applications, Urbana, IL, 61801, USA}
\newcommand{\skai}{NSF-Simons SkAI Institute, 875 N. Michigan Ave., Chicago, IL 60611, USA}
\newcommand{\lancaster}{Department of Physics, Lancaster University, Lancaster, LA1 4YB, UK}
\newcommand{\inaf}{INAF, Osservatorio Astronomico di Capodimonte, salita Moiariello 16, I-80131, Naples, Italy}
\newcommand{\hubble}{Hubble Fellow, Department of Astronomy and Astrophysics, California Institute of Technology, Pasadena, CA 91125, USA}
\newcommand{\thai}{National Astronomical Research Institute of Thailand, 260 Moo 4, Donkaew, Maerim, Chiang Mai, 50180, Thailand}
\newcommand{\fisk}{Department of Physics, Fisk University, 1000 17th Avenue N. Nashville, TN 37208, USA}

\newcommand{\ha}{H$\alpha$}

\newcommand{\hei}{\ion{He}{i}}
\newcommand{\heii}{\ion{He}{ii}}
\newcommand{\Modot}{M$_{\odot}$}
\newcommand{\ciii}{\ion{C}{iii}}

\newcommand{\mosfit}{{\tt MOSFiT}}

\definecolor{lime}{HTML}{A6CE39}
\DeclareRobustCommand{\orcidicon}{
	\begin{tikzpicture}
	\draw[lime, fill=lime] (0,0) 
	circle [radius=0.16] 
	node[white] {{\fontfamily{qag}\selectfont \tiny iD}};
	\draw[white, fill=white] (-0.0625,0.095) 
	circle [radius=0.007];
	\end{tikzpicture}
	\hspace{-2mm}
}

\newcommand{\orcid}[1]{\href{https://orcid.org/#1}{\textcolor[HTML]{A6CE39}{\orcidicon}}}

\begin{document} 

   \title{Characterization of type Ibn SNe}

   \author{D.~Farias\inst{1}\thanks{\email{diego.farias@nbi.ku.dk}}, 
          C.~Gall\inst{1}, 
          V.~A.~Villar\inst{2} \and
          K.~Auchettl\inst{3,4} \and
          K.~M.~de~Soto\inst{2} \and
          A.~Gagliano\inst{2,5,6} \and
          W.~B.~Hoogendam\inst{7} \and
          G.~Narayan\inst{8,9,10} \and
          A.~Sedgewick\inst{1} \and
          S.~K.~Yadavalli\inst{2} \and
          Y.~Zenati \inst{11,12,13} \and
          C.~R.~Angus\inst{14} \and
          K.~W.~Davis\inst{3} \and
          J.~Hjorth\inst{1} \and
          W.~V.~Jacobson-Gal\'an~\inst{15} \and
          D.~O.~Jones\inst{16} \and
          C.~D.~Kilpatrick\inst{17} \and
          M.~J.~Bustamante~Rosell\inst{18} 
          D.~A.~Coulter\inst{11,23} \and
          G.~Dimitriadis\inst{19} \and
          R.~J.~Foley\inst{3} \and
          A.~Gangopadhyay\inst{20} \and
          H.~Gao\inst{7} \and
          M.~E.~Huber~\inst{7}\and
          L.~Izzo\inst{1,21}\and
          J. L. Johnson\inst{3} \and
          A.~L.~Piro\inst{22} \and
          A.~Rest\inst{11,23}
          C.~Rojas-Bravo\inst{3} \and
          M.~R.~Siebert\inst{23} \and
          K.~Taggart\inst{3} \and
          S.~Tinyanont\inst{24}
          }
   \institute{\DARK \and
              \Harvard \and
              \UCSC \and 
              \Ozgrav \and 
              \NSFAI \and 
              \MIT \and 
              \IfA \and
              \Illinois \and
              \supercomputing \and
              \skai \and
              \JHU \and
              \ARCO \and
              \OUI \and
              \QUB \and
              \hubble \and
              \IFA \and
              \Northwestern \and
              \fisk \and
              \lancaster\and 
              \OKC \and
              \inaf \and
              \Carnegie \and
              \STScI \and
              \thai
             }
   \date{Received XXXX; accepted XXXX}

 
  \abstract
   {
   Type Ibn supernovae (SNe) are characterized by narrow helium (\ion{He}{i}) lines from photons produced by the unshocked circumstellar material (CSM). 
   About 80 SNe Ibn have been discovered to date, and only a handful have extensive observational records. Thus, many open questions regarding the progenitor system and the origin of the CSM remain.} 
   {
    Here we investigate potential correlations between the spectral features of the prominent \ion{He}{i}~$\lambda 5876$~\AA{} line and the optical and X-ray light curve properties of SNe~Ibn.
    }
    {
    We compile the largest sample of 61 SNe~Ibn to date, of which 24 SNe have photometric and spectroscopic data from the Young Supernova Experiment and 37 SNe have archival data sets. We fit 24 SNe Ibn 
    with sufficient photometric coverage ($B$ to $z$ bands) using semi-analytical models from \mosfit{}.%
    }
   {We demonstrate that the light curves of SNe~Ibn are more diverse than previous analyses suggest, with absolute $r$-band peak magnitudes ($r_{max}$) of $-19.4\pm 0.6$~mag and rise (from $-10$ days to peak, $\gamma_{-10}$) and decay-rates (from peak to +10 days; $\gamma_{+10}$) of $-0.08 \pm 0.06$ and  $0.08\pm 0.03$ mag/day, respectively.
   We find that the majority of SNe~Ibn in the sub-sample are consistent with a low-energy explosion ($<10^{51}$ erg) of a star with a compact envelope surrounded by $\sim$0.1~\Modot{} of helium-rich CSM. The inferred ejecta masses are small ($M_{\mathrm{ej}}\sim 1$~\Modot) and expand with a velocity of $\sim$5000~km/s.
   Our spectroscopic analysis shows that the mean velocity of the narrow component of the \hei{} lines, associated to the CSM, peaks at $\sim 1100$ km/s. 
   }
   {
     The mean CSM and ejecta masses inferred for a sub-sample of SNe~Ibn indicate that their progenitors are not massive ($\sim10$~\Modot), single stars at the moment of explosion, but are likely binary systems. This agrees with the detection of potential companion stars of SNe~Ibn progenitors, and the inferred CSM properties from stellar evolution models.}


   \maketitle
%

\section{Introduction}~\label{sec:intro}

The majority of massive stars ($M_\ast \gtrsim 8$\Modot) end their lives as core-collapse supernovae (CCSNe). The photometric and spectroscopic characteristics of these explosions reflect differences in the physical properties of their progenitors, power sources, and local environments. To study these properties, CCSNe are classified into various types and subtypes. The two primary classes are distinguished by the absence (type I) or presence (type II) of hydrogen in their spectra~\citep[][and references therein]{filippenko_sne}.
Further division into subtypes (e.g., Ib, Ic, IIn) is based on photometric and spectroscopic features indicating the presence or absence of specific elements (e.g., helium), as well as evidence of circumstellar material (CSM) and its composition or interaction with the ejecta. The CSM is typically produced by the progenitor system prior to explosion, often through complex mass-loss processes~\citep[][and references therein]{Smith_rev_massloss}.
SNe interacting with a surrounding CSM gives rise to narrow ($\lesssim 3000$ km/s) emission and absorption lines of elements present in the slow-moving material~\citep[][and references therein]{Dessart_rev}. For example, type Ibn SNe~\citep{Pastorello_2007_2006jc, Foley_2006jc} are embedded in a He-rich CSM and exhibit strong, narrow \ion{He}{i} lines in the spectra.  

In this work, we focus on SNe~Ibn. These SNe are rare, and their volumetric rate indicates that they only make up about $1-2\%$ of all CCSNe~\citep{Maeda_Moriya2022,Warwick_2023tsz}. Typically, SNe~Ibn are characterized by rapidly evolving light curves, with rise times of about 10 days. Past peak brightness, their optical magnitude declines by about $0.1$~mag/day~\citep{Hosseinzadeh_2017,Ho_2021}. 
Outbursts preceding the SN explosion have been observed in some SNe~Ibn, such as 2006jc~\citep{Pastorello_2007_2006jc}, 2019uo~\citep{Gangopadhyay_2020,Strotjohann_2021} and 2023fyq~\citep{Brennan_23fyq,Dong_23fyq}.
Around peak brightness, the spectra of SNe~Ibn exhibit relatively narrow ($v \sim 3000$ km/s) {\hei} lines on top of a blue ($T \gtrsim 10000$ K) continuum~\citep[][]{Pastorello_2014av}. Continuous interaction between the SN shock, ejecta and the CSM produces intermediate velocity components ($v \sim 4000$ km/s) of the \hei{} lines~\citep{Pastorello_2015_LSQ}. Broad ($v\sim 10000$ km/s) spectral features, if present, are likely associated with emission coming from the SN ejecta~\citep{Fraser_2020rev}. About one month from peak brightness, the spectra of SNe~Ibn are characterized by emission lines of calcium, magnesium, and a blue pseudo-continuum at about $5500$~\AA{} due to the emission of a forest of iron lines~\citep{Dessart_2022}.
Several SNe~Ibn such as SN~2019uo~\citep{Gangopadhyay_2020}, SN~2019wep~\citep{Gango_2022} and SN~2023emq~\citep{Pursiainen_2023} show strong emission of highly ionized, narrow carbon lines at early phases. Such lines are characteristic of type Icn SNe~\citep{GalYam_2022,Fraser_2021csp,Davis2022ann}.
The presence of carbon lines in the early spectra of a growing number of SNe Ibn may indicate a continuum between the Ibn and Icn types~\citep{Pursiainen_2023}.

The nature of the progenitor system of SNe~Ibn remains elusive. Several suggestions of progenitor systems that can account for the observed characteristics have been put forward. For the prototypical example of the Ibn class, SN~2006jc, deep pre-explosion imaging revealed a precursor outburst, detected two years before the SN explosion, strongly supporting a scenario with a Wolf-Rayet (WR) progenitor star undergoing episodic mass loss~\citep{Pastorello_2008_2006jc}. This scenario is also supported by the observed photometric homogeneity in a sample of SNe~Ibn~\citep{Hosseinzadeh_2017} and the ejecta masses ($M_\mathrm{ej}\simeq 10$~\Modot) estimated by light curve modeling~\citep{Karamehmetoglu_2017,Gangopadhyay_2020,Kool_2021_2020bqj,ASASSN14ms}.
Alternatively, binary interaction between two lower-mass stars, such as an accreting helium star ($M_\ast\lesssim 5$~\Modot) and a compact object (e.g., a neutron star) have also been proposed for SNe~Ibn 2006jc and 2019uo~\citep{Tsuna_06jc_outburst}. A similar scenario invoking an unstable mass transfer and merging of the helium and neutron star was proposed for SN~2023fyq based on the brightness and timescale of the observed precursor emission ~\citep{Dong_23fyq,2024_Tsuna_merger_23fyq}. Furthermore, late-time photometric observations at the location of SN~2006jc revealed the presence of a potential surviving companion star~\citep{Maund_2016,Sun_2020}. Theoretical modeling suggests that such systems can create massive CSM shells through multiple mass transfer episodes~\citep[][]{Wu_Fuller}. In stellar evolution models, it is easier to obtain He-rich SN ejecta masses of $\sim 4$~{\Modot} from binary systems than from single stars ~\citep{Dessart_2022,Takei_ibn}.
To date, only two sample studies of SNe~Ibn exist,~\citet[P16;][]{Pastorello_2014av} and~\citet[H17;][]{Hosseinzadeh_2017}, which primarily focus on the observational properties. More than 40 SNe Ibn have been discovered since the most recent study (H17), some of which are studied in detail ~\citep[e.g., ][]{Karamehmetoglu_2017,Ho_2021,2018bcc_Karamehmetoglu,Kool_2021_2020bqj,Pellegrino_2022_19deh,BenAmi_2022}. In this work, we present the photometric and spectroscopic observations of 24 SNe Ibn enabled by the Young Supernova Experiment~\citep[YSE;][]{Jones_2021,Aleo_DR1YSE}. 

\begin{figure}[t]
\includegraphics[width=\columnwidth]{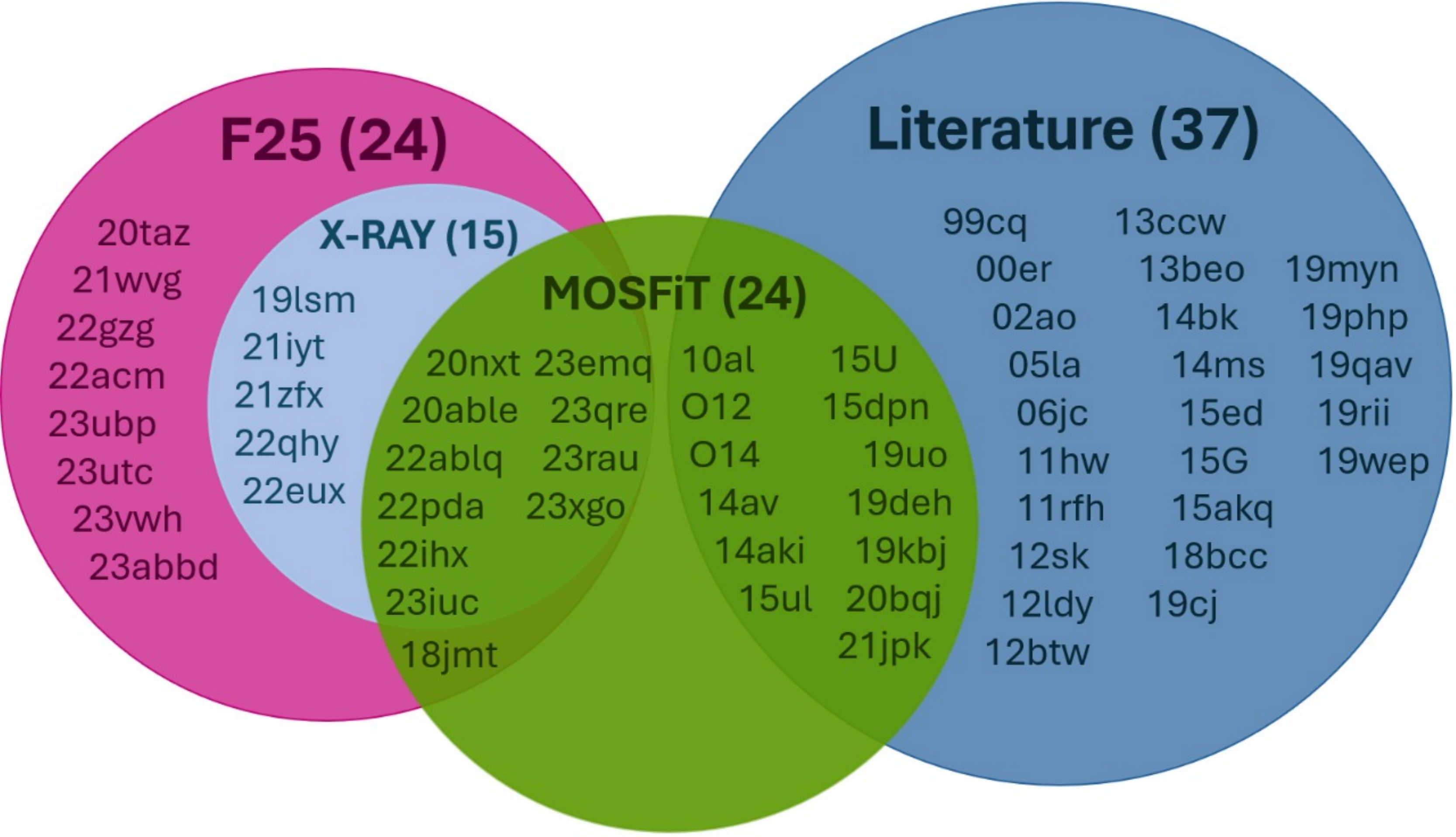}{
\caption{\label{fig:test_sample}
Visualization of the four different SN Ibn samples (F25, X-RAY, \mosfit{}, and Literature) analyzed in this work, with the individual member SN labeled. For details, see Tables~\ref{tab:photo_props} and~\ref{tab:spectroscopic}).  
}
}
\end{figure}

YSE is an ongoing transient survey that uses a 7\% time allocation of the Panoramic Survey Telescopes and Rapid Response System~\citep[Pan-STARRS;][]{Pan-STARRS_Chambers} (PS1 \& PS2) to scan $\sim750$~deg$^{2}$ of the sky with an approximate $3$-day cadence to a depth of $gri\approx 21.5$~mag and $z\approx20.5$~mag~\citep{Aleo_DR1YSE}. Hereafter, these 24 SNe Ibn will be referred to as the `F25' sample (Fig.~\ref{fig:test_sample}). In addition, 15 out of the 24 SNe in the F25 sample have X-ray observations (`X-RAY' sample), which are analyzed in this work. 

We complement the F25 sample with archival photometric and spectroscopic data of 37 SNe~Ibn, referred to as `Literature' sample (see Fig.~\ref{fig:test_sample}, Tables~\ref{tab:photo_props} and~\ref{tab:spectroscopic}). In total, we study the photometric and spectroscopic properties of 61 SNe~Ibn (F25 + Literature), the largest sample to date assembled for this class. 
From the combined sample (F25 + literature), we select 24 Type Ibn SNe with well-sampled light curves (hereafter referred to as the `\mosfit{}' sample; see Fig.~\ref{fig:test_sample}). These SNe are modeled using \mosfit{} with a self-consistent fitting approach. Such an approach is essential for exploring the physical parameters associated with the CSM and the explosion properties of SNe Ibn at the population level, as light curve modeling is sensitive to the chosen model implementation, the set of free parameters, and the prior distributions used in the fits.
We also present the analysis of prominent \hei{} emission lines in the spectra of 59 out of the 61 SNe~Ibn. We measure the velocities of these lines for 43 SNe with spectra taken from the Weizmann Interactive Supernova Data Repository (WISeREP)\footnote{\url{https://www.wiserep.org}}, and this work (see Table~\ref{tab:spectroscopic}). Additionally, we include measurements of \hei{} emission line velocities of 16 SNe~Ibn presented in P16, H17,~\citet{Gango_2022} and~\citet{Wang_2024_Ibn}. 

The paper is organized as follows: Section \ref{sec:obs} describes the acquisition and reduction of photometric and spectroscopic data of the 24 YSE-observed SNe~Ibn (F25 sample) which are first presented in this work. Section \ref{sec:photana} outlines the methods used to estimate photometric properties, such as peak magnitudes and color evolution. The light curve modeling approach of the \mosfit{} sample is detailed in Section \ref{sec:mod}. Section \ref{sec:xray} discusses the X-ray analysis of 15 YSE-observed SNe~Ibn in the X-RAY sample. Section \ref{sec:specana} presents the analysis method and results of the decomposition of the \hei~$\lambda 5876$ \AA{} profiles of 59 SNe~Ibn from our total (F25+Literature) sample. In Section \ref{sec:discuss}, we discuss the results and conclude in Section \ref{sec:concl}.


\section{The F25 sample}
\label{sec:obs}

The F25 sample (see Fig.~\ref{fig:test_sample}) of SNe~Ibn consists of objects that are either observed by the YSE collaboration between the years 2020 and 2023, or have `well-sampled' light curves unpublished at the beginning of 2024. Here, well-sampled light curves refer to photometric data obtained at a $1-4$ day cadence between  $-3$ days to $+50$ days relative to peak brightness and in at least three different filter bands. In total, 24 SNe~Ibn fulfill these criteria.

There are 15 SNe in the F25 sample with photometric and/or spectroscopic follow-up observations obtained with resources by the YSE collaboration. Two SNe in this group, SN~2021iyt and SN~2021zfx, are YSE discoveries. Seven of the F25 SNe have extensive photometric and spectroscopic data coverage. These seven YSE Ibn SNe are SN~2020nxt~\citep{Qinan_2020nxt}, SN~2020able,~SN~2021iyt, SN~2021zfx, SN~2022gzg, SN~2022qhy and SN~2023iuc.
The eight remaining SNe, SN~2022eux, SN~2022ihx, SN~2022pda, SN~2023emq~\citep{Pursiainen_2023},~SN~2023qre, SN~2023rau, SN~2023xgo~\citep{Gango_2023xgo} and SN~2023abbd have sporadic data coverage obtained by YSE. 
Nine SNe of the F25 sample are not YSE-related. For five of these SNe~Ibn (SN~2020taz, SN~2021wvg, SN~2023ubp, SN~2023utc and SN~2023vwh), we obtained the photometric data from the Pan-STARRS Survey for Transients~\citep[PSST;][]{Pan-STARRS_Chambers}. Data for two more SNe Ibn, SN~2018jmt and SN~2022ablq are from ~\citet{Wang_2024_Ibn} and ~\citet{Pellegrino_2022ablq}, respectively. 
Additionally, we include two SNe~Ibn from public surveys (SN~2019lsm and SN~2022acm) with yet unpublished data that fulfill our selection criterion. For this work, we re-processed all publicly available photometric data. Details of the photometric data of the F25 sample and their re-processing are given in Sec.~\ref{subsec:photdata}. All data used in our analysis will be published alongside the paper. 

\subsection{Photometric data}
\label{subsec:photdata}

We describe the data acquisition and reduction of photometric data of the SNe in the F25 sample obtained by eight different facilities below. The photometric data are reported in Table~\ref{tab:short_phot}.

\subsubsection{Optical and UV}

For twelve SNe~Ibn, optical photometric data were obtained with the Pan-STARRS telescopes (PS1 and PS2) located at Haleakala Observatory, Hawaii, USA~\citep{Pan-STARRS_Chambers}. Data was acquired in the same survey mode as employed by YSE~\citep{Jones_2021,Aleo_DR1YSE} and accessed through \texttt{YSE-PZ} \citep{Coulter_YSEPZ} and the Pan-STARRS Survey for Transients~\citep[PSST;][]{PSST}. 
For SNe~2020nxt~\citep{Qinan_2020nxt}, 2020able~\citep{Aleo_DR1YSE}, 2021iyt, 2021zfx, 2022gzg, 2022qhy and 2023iuc, YSE obtained optical photometry in $griz$ bands. PSST SNe~2021wvg, 2023ubp, 2023utc, 2023vwh and 2023abbd were observed in $wgrizy$ bands.
 All images from Pan-STARRS were reduced following the procedure detailed in~\citet{Aleo_DR1YSE}. Basic data processing was performed by the PS1 Image Processing pipeline~\citep[][]{Magnier2020}. Template images were taken from the PS1 $3\pi$ sky survey~\citep{Pan-STARRS_Chambers}. 
 Template convolution and differencing photometry were performed using the \texttt{Photpipe}~\citep{Rest_2005,Rest_2014} pipeline. 
 The difference fluxes are calibrated using nightly zero-points calculated from photometric standard stars from the PS1 $3\pi$ sky survey. 

Fourteen SNe~Ibn also have ZTF $g$- and $r$-band photometry included in this work: SNe~2019lsm, 2020taz, 2021wvg, 2022acm, 2022gzg, 2022ihx, 2022pda, 2023emq, 2023iuc, 2023rau, 2023ubp, 2023utc, 2023vwh and 2023xgo. We retrieved all available public photometric data of these SNe from the Lasair broker~\citep{Lasair} and no further processing was performed.

SN~2023rau was observed in $griz$ bands with the 1-m Lulin Optical Telescope located at Lulin Observatory, Taiwan.  All images were processed using bias and flat-field calibration frames following standard methods.  We then performed photometry and pixel-level calibration in {\tt Photpipe} following methods similar to those described above for the Pan-STARRS photometry. 

Fifteen SNe were observed with the Ultraviolet and Optical Telescope \citep[UVOT;][]{Roming_2005} on the {\it Neil Gehrels Swift Observatory}, in $V,B,U,UVW1, UVM2, UVW2$ bands. SNe~2020nxt, 2022eux, 2022ihx, 2022qhy, 
2022ablq, 2023emq, 2023qre and 2023xgo were observed in all six UVOT bands. SNe 2020able, 2021iyt, and 2023iuc lack only $V$ band and SN~2022pda lacks $V$ and $B$ band observations.
We used the {\tt uvotsource} tool from the {\tt HEASoft v6.26} package to perform aperture photometry within a 3\arcsec\ aperture centered on all UVOT SNe. We subtracted the background of our science images using observations of the site of the SNe once these have faded. We report the measured magnitudes in each UVOT band for the observed SNe~Ibn at a $>3\sigma$ level.

SN~2023xgo was observed in $g^{\prime}$, $r^{\prime}$, $i^{\prime}$, and $z^{\prime}$ bands with the Thacher 0.7-m telescope in Ojai, California~\citep{Thacher_Swift}. Through {\tt Photpipe}, we performed pixel-level calibration using bias and flat-field frames obtained in the same night and with the instrumental configuration as our science frames. We then performed PSF photometry using {\tt DoPhot} and calibrated the resulting  $griz$ magnitudes from the Pan-STARRS1 catalog~\citep{Flewelling16}. 
 
SNe~2022qhy, 2023emq, 2023qre and 2023rau were observed in $uBVgri$ bands by the Direct 4k$\times$4k camera on the 1-m Henrietta Swope telescope located at Las Campanas Observatory, Chile.  For details on the complete reduction process, see the description in \citet{Kilpatrick_2018}.

Fifteen SNe~Ibn were observed by ATLAS~\citep{ATLAS_Tonry} in either orange ($o$) or cyan ($c$) bands of which SNe~2020nxt, 2020able, 2021iyt, 2022gzg, 2023emq and 2023xgo were observed in both bands. SNe~2021zfx, 
2022eux, 2022ihx, 2022pda, 2023rau, 2023ubp, 2023utc, 2023vwh and 2023abbd were only observed in $o$ band. Following the procedure described in~\citet{ATclean_sofia} and \citet{Farias_2024}, we extracted photometry at eight different circular regions surrounding each of these SNe at each epoch and in each band. If the flux measurements in these regions are not consistent with zero, then the epoch is flagged and it is not considered as a detection. The final product is a 1-day binned light curve calculated as a 3$\sigma$-cut weighted mean in each band.

For five SNe Ibn we obtained photometry with the Sinistro cameras on the Las Cumbres Observatory Global Telescope (LCOCT) 1-m telescope network~\citep{Brown_2013}. The LCOGT data for SN~2020nxt are published in~\citet{Qinan_2020nxt} and adopted in this study. 
LCOGT unpublished photometric observations of
SNe 2022ihx ($gp,rp,ip$), 2022pda ($U,V,gp,rp,ip$) and 2023iuc ($gp,rp,ip$) are presented in this work. Additionally, we also reprocessed the LCO $B,V,gp,rp,ip,z$ images for SN~2018jmt~\citep{Wang_2024_Ibn}.
We used the LCO {\tt BANZAI} pipeline \citep{McCully_2018} to pre-process the imaging data (bias and flat-fielding). Image subtraction is done using {\tt Photpipe} pipeline \citep{Rest_2005, Jones_2021}. Thereafter, we perform PSF photometry using {\tt DoPhot}~\citep{Schechter_1993}. To obtain the final magnitudes we calibrate the data using $u$-band SDSS~\citep{Alam_2015} together with $griz$ Pan-STARRS1 photometric standards observed in the vicinity of the SNe.

\subsubsection{X-ray}~\label{sec:data:xray}

The X-RAY sample consists of 15 SNe~Ibn (SNe~2019lsm, 2020nxt, 2020able, 2021iyt, 2021zfx, 2022eux, 2022ihx, 2022pda, 2022qhy, 2022ablq, 2023emq, 
2023iuc, 2023qre, 2023rau, and 2023xgo) which were observed with the {\it Swift} X-ray Telescope (XRT, \citealt{burrows05}) in photon counting mode~(see Fig.\ref{fig:test_sample}). To place constraints on the presence of any X-ray emission, we reprocessed all XRT observations from level one data using the package \texttt{xrtpipeline} version 0.13.7 and applied the standard filter and screening criteria\footnote{\url{http://swift.gsfc.nasa.gov/analysis/xrt_swguide_v1_2.pdf}} and the most recent calibration files. For the vast majority of SNe in the X-RAY sample, we use a source region with a radius of 30\arcsec centered on the position of the SN and a source-free background region to constrain the X-ray emission. For those SNe which are located close to complex regions (i.e., near the center of their host or a bright X-ray region), we used a source region with a radius of 20\arcsec (SN~2020able), 15\arcsec (SN~2020nxt, 2022ihx, 2022pda, 2023xgo) or 10\arcsec (SN~2022qhy, 2022ablq). To increase the signal-to-noise ratio (S/N), we stack these individual observations together using \texttt{xselect} version 2.5b to either place the strongest constraint on the X-ray emission arising from the SN or in multiple time bins to obtain an X-ray light curve. 

With the exception of SN~2020nxt,  SN~2022qhy, SN~2022ablq, and SN~2023emq, we find no significant X-ray emission associated with individual nor the stacked observations of the SN. However, the emission towards  SN~2020nxt and SN~2022qhy are very close to the center of their host galaxies, so it is likely that the detected emission is not entirely from the SNe. For those observations for which we did not detect any X-ray emission, we derived $3\sigma$ upper limits to the count rate in the 0.3-10.0 keV energy range. All count rates were corrected for the size of the aperture. To convert the 0.3-10.0 keV count rate into an unabsorbed X-ray flux (and luminosity) in the same energy band, we assumed an absorbed thermal bremsstrahlung model with a temperature of 2.3 keV, similar to that used by \citet{Pellegrino_2022ablq}. With the exception of SN~2022ablq, for which we took the column density from \citet{Pellegrino_2022ablq}, all SNe had their column densities derived using the \ion{H}{i} maps published in \citet{2016A&A...594A.116H}.

\subsection{Optical spectroscopic data}
\label{subsec:specdata}

Seventy-four spectra of the SNe~Ibn in F25 sample were obtained from multiple observing facilities through programs within the YSE collaboration. Additionally, 23 archival spectra are taken from the Weizmann Interactive Supernova Data Repository (WISeREP). The archival spectra from WISeREP are classification spectra from different collaborations such as Spectroscopic Classification of Astronomical Transients collaboration (SCAT; \citealp{Tucker_2022_SCAT}), and were taken using the SuperNova Integral-Field Spectrograph (SNIFS; \citealp{Lantz_2004_SNIFS}); and the Global Supernova Project collaboration using the FLOYDS-N/S and Goodman spectrograph.
A log of the spectroscopic observations is presented as Table~\ref{tab:spectroscopic_table}.
SNe~2023iuc, 2023qre and 2023rau were observed with the Wide Field Spectrograph (WiFeS), mounted at the Australian National University (ANU) 2.3-m telescope located at the Siding Spring Observatory~\citep{Dopita_2007}. WiFeS is an integral-field spectrograph with a field-of-view of $38 \times 25$ arcsec. Each spectrum was taken in ‘Nod \& Shuffle’ mode using the R = 3000 grating which covers  3200-9800 \AA range. Each observation was reduced using {\tt PyWiFeS}~\citep{PyWiFeS}, with sky subtraction done using a 2D sky spectrum that was taken during the observations \citep{2024PASA...41...68C}. 

SNe~2020nxt, 2020able, 2021iyt, 2021zfx, 2022ihx, 2022pda,  2023emq, 2023qre, 2023rau and 2023xgo were observed with the Kast dual-beam spectrograph~\citep{Miller_Kast} on the Shane 3-m telescope at the Lick Observatory. To reduce Kast spectra, we used the {\tt UCSC Spectral Pipeline}\footnote{\url{https://github.com/msiebert1/UCSC\_spectral\_pipeline}} \citep{Siebert2019}, a custom data-reduction pipeline based on procedures outlined by \citet{Foley03}, \citet{silverman12}, and references therein.  The two-dimensional (2D) spectra were bias-corrected, flat-field corrected, adjusted for varying gains across different chips and amplifiers, and trimmed. Cosmic-ray rejection was applied using the {\tt pzapspec} algorithm to individual frames. Multiple frames were then combined with appropriate masking. One-dimensional (1D) spectra were extracted using the optimal algorithm \citep{Horne86}. The spectra were wavelength-calibrated using internal comparison-lamp spectra with linear shifts applied by cross-correlating the observed night-sky lines in each spectrum to a master night-sky spectrum. Flux calibration was performed using standard stars at similar airmass to the science exposures, with ``blue'' (hot subdwarfs; i.e., sdO) and ``red'' (low-metallicity G/F) standard stars. We correct for atmospheric extinction. By fitting the continuum of the flux-calibrated standard stars, we determine the telluric absorption in those stars and apply a correction, adopting the relative airmass between the standard star and the science image to determine the relative strength of the absorption. We allow for slight shifts in the telluric A and B bands, which we determine through cross-correlation. For dual-beam spectrographs, we combine the sides by scaling one spectrum to match the flux of the other in the overlap region and use their error spectra to correctly weight the spectra when combined. More details of this process are discussed elsewhere \citep{Foley03, silverman12, Siebert2019}.

SNe~2021iyt, 2023emq, 2023qre and 2023rau were observed with the Goodman spectrograph~\citep{Clemens_Goodman} on the NOIRLab 4.1-m Southern Astrophysical Research (SOAR) telescope at Cerro Pachón, Chile. We used the {\tt UCSC Spectral Pipeline} to reduce the Goodman spectra as described above for Kast.

SNe 2021iyt, 2022pda, 2022ihx, 2023emq, 2023iuc, 2023rau and 2023xgo were observed with the Alhambra Faint Object Spectrograph and Camera (ALFOSC) on the 2.56-m Nordic Optical Telescope (NOT). All ALFOSC spectra were obtained using a 1.0\arcsec\ slit with grisms \#4. We used standard routines within {\tt IRAF}\footnote{\url{https://iraf-community.github.io}} and {\tt PYRAF}\footnote{\url{https://github.com/iraf-community/pyraf}}  to reduce, extract, wavelength calibrate and flux calibrate the spectra. 

SNe 2020nxt, 2020able, 2021iyt and 2022gzg were observed with the Low Resolution Imaging Spectrometer~\citep[LRIS;][]{Oke_1995,LRIS} mounted on the Keck I telescope, Hawaii. We used the {\tt UCSC Spectral Pipeline} to reduce the LRIS spectra as described above for Kast and Goodman.
SNe 2021zfx, 2022gzg and 2023abbd were observed with 
Gemini Multi-Object Spectrograph (GMOS) at Gemini South Observatory through Large and Long Program LP-204 \citep[PI: W. Jacobson-Galán][]{LLPwjg}. For all of these spectroscopic observations, standard CCD processing, spectrum extraction and flux calibration were accomplished with the {\tt DRAGONS} pipeline \citep{dragons}.

\section{Photometric Analysis}
\label{sec:photana}

We present a consistent approach to analyze the light curves of all SNe~Ibn in the F25 + Literature sample (see Sect.~\ref{sec:obs}, 
Fig.~\ref{fig:test_sample} and Table~\ref{tab:photo_props}). Our analysis aims to identify correlations between the photometric properties within the type Ibn class.

\subsection{Data Pre-processing}
\label{subsec:preprocess}

The majority of the 61 SNe in the F25 + Literature sample have light curves in multiple bands at a high cadence ($\sim 3$ days). However, there are periods of missing data in different bands, usually as a result of unfavorable weather. To consistently analyze the light curves of the SNe in each band, we construct uniformly sampled light curves (1-day cadence) from the photometric data using a Gaussian process~\citep[GP; ][]{GPs} with a Matern32 kernel\footnote{\url{https://george.readthedocs.io/en/latest/user/kernels/}} implemented in {\tt george}~\citep{george}. The GP interpolation is performed on the flux in each band versus time space, rather than in magnitude space, to incorporate low S/N observations, which nevertheless constrain the shape of the GP at all phases. We optimize the length scale of the kernel of the GP by maximizing the log-likelihood of the modeled light curve and the data using {\tt scipy optimize}~\citep{scipy}. Thereafter, we mask the interpolated light curves to ensure the S/N of each data point exceeds 3.

\subsection{Peak magnitude}
\label{subsec:peakmag}

We chose the interpolated $r$-bands (e.g., ZTF-$r$, ATLAS-$o$, PS1-$r$, depending on availability) to estimate the time of the peak ($t_{\rm max}$) and the peak $r$-band magnitude in a given $r$-band ($r_{\rm max}$) of the 61 SNe~Ibn.
We did not utilize the maximum of the interpolated light curve for the peak magnitude given that simple stationary kernels are prone to cause the interpolated posterior mean of the light curve to being undersmooth or oversmooth, while also underestimating the error on the time of peak~\citep{stevancelee}. 
Instead, we use Monte Carlo Markov Chain (MCMC) as implemented in the {\tt emcee}\footnote{\url{https://emcee.readthedocs.io/en/stable/}}~\citep{emcee} python package to fit a second order polynomial to the interpolated $r$-band light curves between $\pm 10$ days of peak. We used 100 walkers, 10000 iterations and a burn-in period of 1000 steps (see~Table~\ref{tab:photo_props}).
Thirteen of the 61 SNe lack of photometric data around the peak and thus, the peak time and magnitude cannot be estimated with this method. These 13 SNe are SNe~2000er, 2002ao, 2005la, 2006jc, 2011hw, LSQ12btw, 2014bk, ASASSN-14ms, 2015G, 2019wep, 2020bqj, 2022qhy and 2023abbd.
For SNe~2000er, 2002ao, 2005la, 2006jc, 2011hw, LSQ12btw, 2014bk, ASASSN-14ms and, 2015G, we retrieve the values reported in Table 4 from H17. The peak dates and magnitudes of SNe~2019wep and~2020bqj were obtained from~\citet{Gango_2022} and~\citet{Kool_2021_2020bqj}, respectively. For SNe 2022qhy and 2023abbd in the F25 sample, we assumed the peak time and peak magnitude from the first detection in $V$ and $o$ bands, respectively, as these are closest available photometric bands to $r$ for these objects.

\subsection{Dust Extinction}
\label{subsec:ext}

For all SNe in the F25 sample, the Galactic reddening $E(B-V)_{\rm MW}$ is the mean color excess along the line-of-sight towards each SN which is taken from the dust maps of~\citet{SF_2011}. The values of $E(B-V)_{\rm MW}$ are reported in Table~\ref{tab:summary}. The Galactic extinction in different bands is obtained using the dust extinction law of~\citet{fitzpatrick_extinction} through the {\tt dust\_extinction}~\citep{dust_extinction_paper} package assuming $R_{V}=3.1$. We apply the extinction correction after pre-processing the light curves (Sect.~\ref{subsec:preprocess}.)

For all SNe in F25 that do not have reported host extinction in previous studies, we follow a similar approach to~\citet{Shivvers_2015U} to estimate $E(B-V)_{\rm host}$. We fit a blackbody (BB) function modified to account for dust extinction to the spectral energy distribution (SED) at the time of $r$-band maximum. 
From the interpolated light curves at $t_{\rm max}$, we construct the SED of each SN by requiring at least three filters with effective wavelengths between $4500$~\AA{} and $9000$~\AA{} (i.e., $B$ to $z$ bands). We find 16 SNe that fulfill our ``three filter" criterion. We explore the posterior distributions of three free parameters $T_{BB}$, $R_{BB}$ and $E(B-V)_{\rm host}$ in our SED BB fitting. To do so, we use the nested sampler {\tt dynesty}~\citep{Speagle_2020}, assuming uniform prior distributions on the three parameters. The results of this analysis for SNe~2020able, 2022ihx, 2023iuc and 2023xgo are shown in Fig.~\ref{fig:extinction}.
For SNe~2018jmt and 2023emq, the absence of \ion{Na}{i}D lines suggests that $E(B-V)_{\rm host}$ is negligible~\citep{Wang_2024_Ibn,Pursiainen_2023}. For these SNe, we obtain $E(B-V)_{\rm host} \lesssim 0.1$ mag. In contrast, SN~2022ablq appears significantly extinguished by dust along the line of sight towards the SN in the host galaxy. \citet{Pellegrino_2022ablq} estimate $E(B-V)_{\rm host} = 0.18\pm 0.1$ mag, whereas we obtain $E(B-V)_{\rm host} = 0.35$ mag, with an uncertainty of $\approx 0.1$ mag, within $1\sigma$ from the value reported in the previous work. 
For eight SNe, for which no independent extinction estimates are available and for which the SEDs at $t_{\rm max}$ cannot be constructed due to lack of data, we assume zero extinction associated to the host galaxies. These SNe are 2020nxt, 2021zfx, 2022qhy, 2023qre, 2023ubp, 2023utc, 2023vwh and 2023abbd. For SN~2023qre, this assumption is further confirmed by the inferred host extinction through light curve modeling (Sect.~\ref{sec:mod}). For all SNe in the Literature sample, we take the total (MW + host) extinction estimates from their respective references as provided in Table~\ref{tab:photo_props}.

\subsection{Color evolution}
\label{subsec:color}

\begin{figure}[t]
\includegraphics[width=\columnwidth]{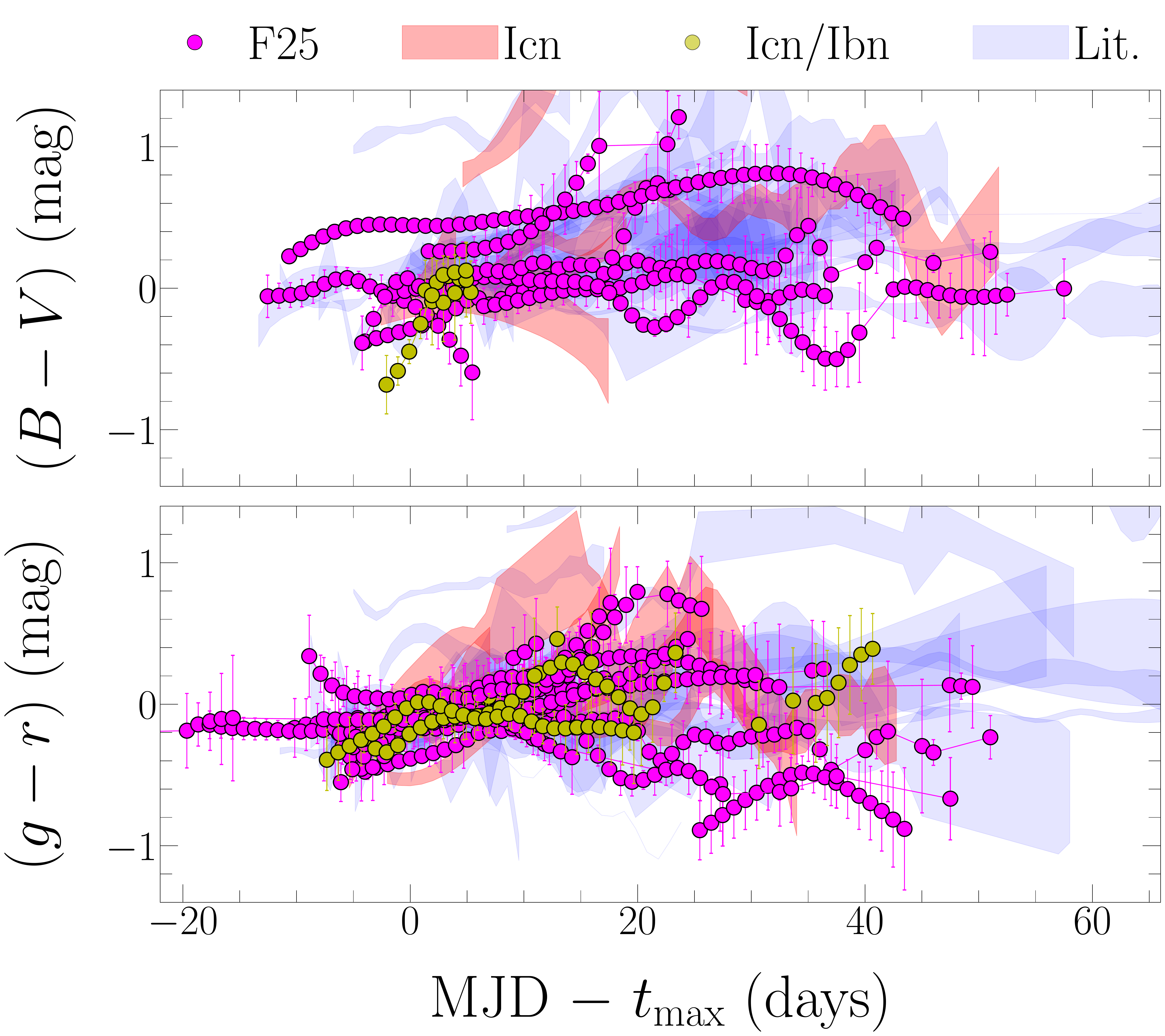}
\caption{Interpolated $B-V$ (upper panel) and $g-r$ (lower panel) curves for the 24 SNe~Ibn from F25 (magenta filled circles), the Icn/Ibn (2023emq-like; yellow filled circles) and the color curves ($\pm1\sigma$) of the Literature (Lit.) sample (blue regions). The color curves ($\pm1\sigma$) of five SNe Icn are included for comparison (red regions). The photometric data of the SNe Icn are retrieved from~\citet{GalYam_2022,Perley_2021csp,Fraser_2021csp,Pellegrino_Icn,Davis2022ann}.} 
\label{fig:color}
\end{figure}

Figure~\ref{fig:color} shows the evolution of the interpolated $B-V$ and $g-r$ light curves without any corrections for host or Galactic extinction. It is evident that neither SNe~Ibn nor SNe Icn form a homogeneous group, as there is a spread of approximately one magnitude in the color curves throughout the SN evolution.
We note that the ``Icn/Ibn" group refers to SNe whose classification spectra have been reported to resemble that of SN~2023emq (see Table~\ref{tab:summary}). This similarity is based on the presence of a spectral feature at $5696$~\AA{}, which is associated with the \ion{C}{iii} emission line. The SNe classified as Icn/Ibn SNe are SN~2023emq, 2023qre, 2023rau, and 2023xgo (this work).
This sample of SNe Icn/Ibn do not form a homogeneous group either. 

Figure \ref{fig:color} illustrates that the evolution of the $B-V$ and $g-r$ color curves is too diverse to be a reliable indicator of extinction. Therefore, it is challenging to create color templates for SNe Ibn, similar to those created for stripped-envelope SNe~\citep{Stritzinger_host_templates}.

\subsection{Absolute $R/r$-band light curve}
\label{subsec:abs}

\begin{figure}[t]
\includegraphics[width=1.0\columnwidth]{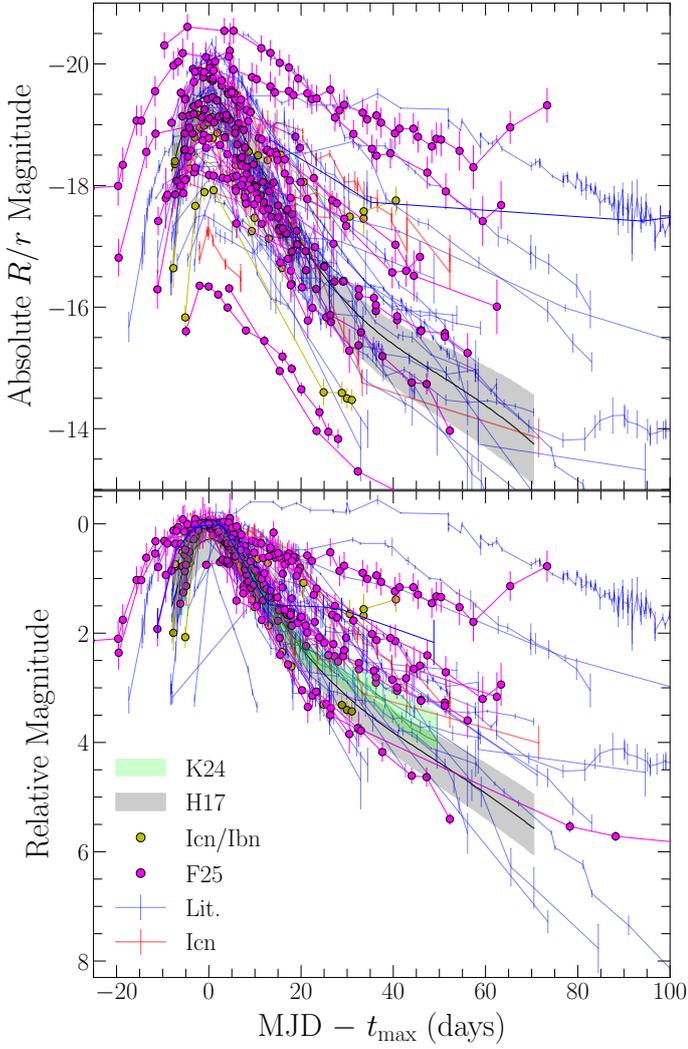}
\caption{{\it Upper panel}: Absolute $R/r$-band-like light curves of all SNe~Ibn from F25 sample that are first reported in this work (magenta dots) and Literature sample (blue lines). 
Gray (H17) regions correspond to the average light curve and $1.96\sigma$ error bars of 18 type Ibn SNe from~\citet{Hosseinzadeh_2017}.
Yellow dots correspond to the likely Icn/Ibn SNe~2023emq~\citep{Pursiainen_2023}, 2023qre, 2023rau and 2023xgo~(this work). Red lines correspond to the five Icn SNe discovered up to date.
{\it Lower panel}: Same as {\it upper panel}, with the subtraction of the peak magnitude of each SN. Green (K24) regions correspond to the median light curve and the 25\% and 75\% percentiles from~\citet{Somayeh_2024}.
}
\label{fig:ABS}
\end{figure}

We compute the extinction-corrected, absolute magnitudes of the $R/r$-band light curves for all 61 SNe~Ibn in the F25 + Literature sample shown in Fig.~\ref{fig:ABS}. To do so, we derive the distance modulus for all SNe assuming $\Lambda$CDM cosmology with $H_0$ = 67.8 km s$^{-1}$ Mpc$^{-1}$ and $\Omega_m$ = 0.307~\citep{Planck_13}. The redshifts and the extinction measurements (see Sect.~\ref{subsec:ext}) used are listed in Table~\ref{tab:summary}.  
For consistency, we chose not to propagate distance uncertainties (directly from redshift uncertainties or independent-redshift distance measurements) into the absolute magnitude LCs since we do not have such information for all SNe in the sample. 
The upper panel of Fig.~\ref{fig:ABS} shows that the absolute peak $R/r-$band magnitude ranges from about $-16.5$ to $-20.5$ mag. The weighted mean peak absolute magnitude in $R/r$-bands, and its associated standard deviation is $-19.39 \pm 0.62$ mag. Since both the F25 and Literature sample are magnitude-limited samples, the mean peak absolute magnitude should not be considered as the characteristic brightness of SNe Ibn.
Figure~\ref{fig:ABS} lower panel shows that the range of time from first detection to peak is $\sim -20$ to $\sim -5$ days. This range is in agreement with the rise times of SNe~Ibn reported by H17.
From Fig.~\ref{fig:ABS}, there is more dispersion in the brightness ($\pm 1$ mag) at late epochs ($+50$ days) in comparison to the Ibn templates~\citep[H17,][]{Somayeh_2024}, $\sim 0.4$ mag. Templates from H17 and~\citet[][]{Somayeh_2024} were created using 18 and 13 SNe from the total (F25 + Ibn) sample, respectively.

To quantify the light curve evolution at different epochs, we calculate the slope of the interpolated $R/r$-band light curves with respect to $t_{\rm max}$ by fitting a linear function to the observed light curve data in discrete time intervals. We define four light curve slope parameters, one for each of the four intervals $-10$ to $+0$, $+0$ to $+10$, $+10$ to $+20$ and $+20$ to $+30$ days. We term the parameters $\gamma_{-10}$, $\gamma_{+10}$, $\gamma_{+20}$, $\gamma_{+30}$, respectively. 
We estimate the slopes by fitting the linear model using a simple Gaussian likelihood with flat priors on the parameters. The sampling is performed with the \texttt{emcee} python package, using the same configuration (number of walkers and steps) as used to estimate the peak magnitude in Sec.~\ref{subsec:peakmag}.
The weighted means of the $\gamma_{-10}$, $\gamma_{+10}$, $\gamma_{+20}$, and $\gamma_{+30}$ are $-0.08\pm 0.06$, $0.08 \pm 0.03$, $0.06 \pm 0.04$, and $0.05 \pm 0.04$ mag/day, respectively.
To find out if a correlation exists between different light curve parameters, we calculate the Spearman's rank-order correlation coefficient~\citep{Spearman}, with the associated $p$-value, using {\tt pymcspearman}~\citep{Curran_mcspearman,pymcspearman}.
As a rough guide~\citep[see][and references therein]{corder2014nonparametric},  $|\rho_s| \sim 0$, $|\rho_s| \sim 0.1$, $|\rho_s| \sim 0.3$, $|\rho_s| \sim 0.5$ and $|\rho_s| \sim 1.0$ correspond to no, weak, moderate, strong and perfect correlation between the variables. The $p$-value associated with $\rho_s$ indicates the statistical significance of $\rho_s$.

\begin{figure}[t]
\includegraphics[width=\columnwidth]{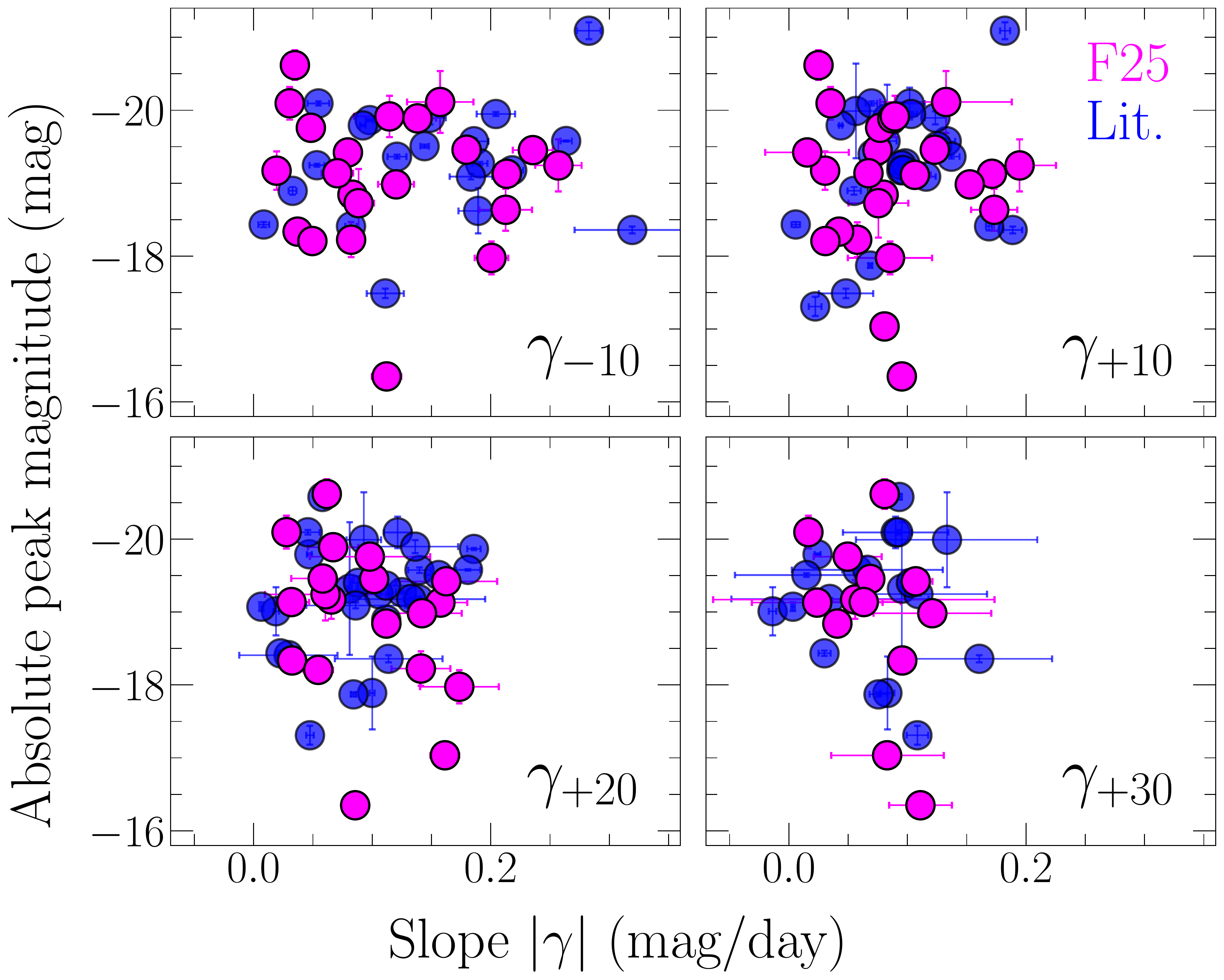}
\caption{Absolute ($R/r$-band) peak magnitude versus the absolute value of the slope parameters of the ($R/r$-band) light curves ($\gamma$). The SNe~Ibn from the F25 and Literature sample are presented as magenta and blue circles, respectively. Panels {\it upper left, upper right, lower left} and {\it lower right} show the light curve slopes $\gamma_{-10}$  ($-10$ to $t_{\rm max}$),  $\gamma_{+10}$ ($t_{\rm max}$ to day $+10$), $\gamma_{+20}$ ($+10$ to $+20$ days), and $\gamma_{+30}$ ($+20$ to $+30$ days), respectively.
} 
\label{fig:mabs_slope}
\end{figure}
\begin{figure}[t]
\includegraphics[width=\columnwidth]{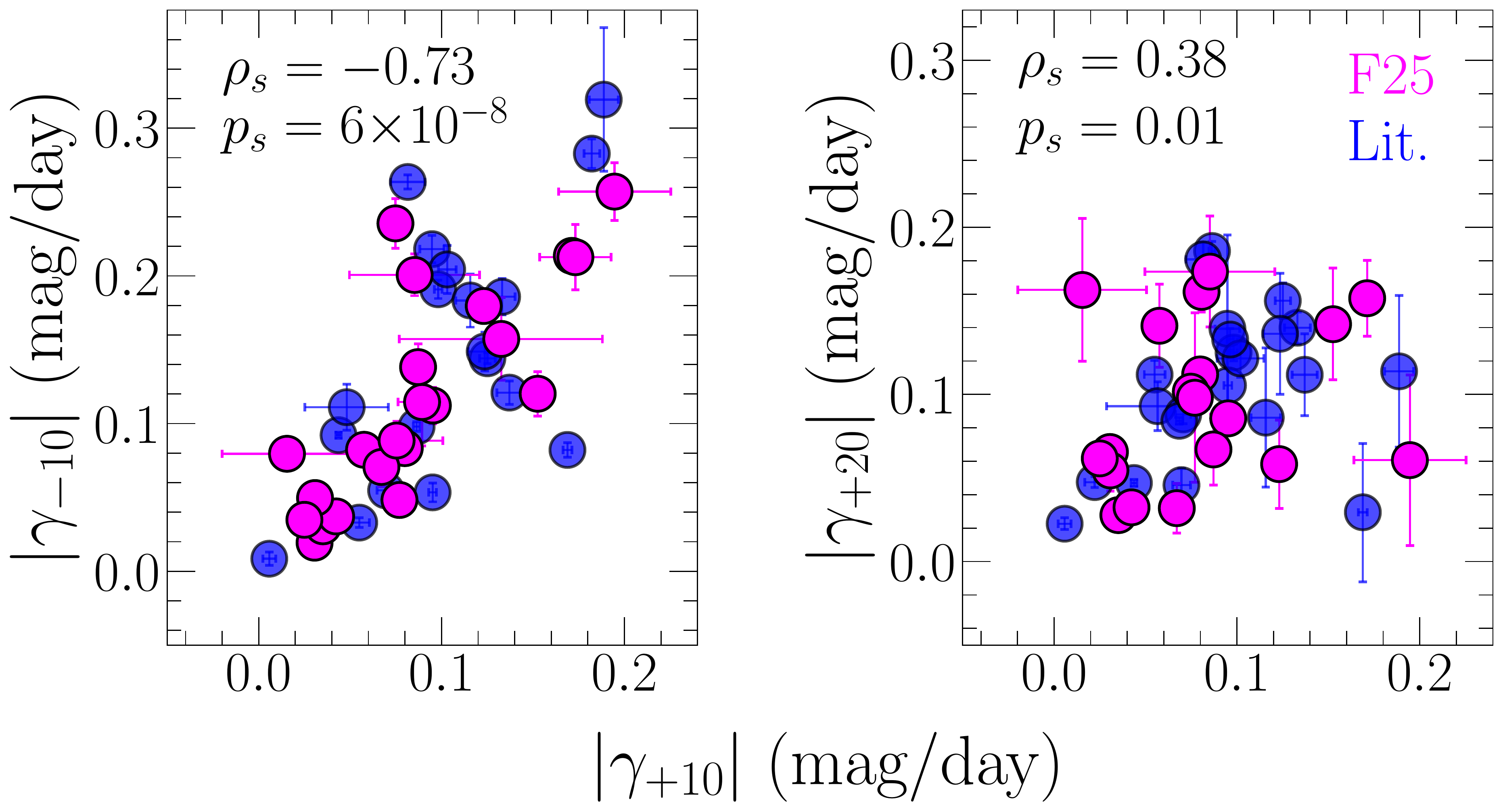}
\caption{Absolute values of the $R/r$-band slopes $\gamma_{-10}$ ({\it left panel}) and $\gamma_{+20}$ ({\it right panel}) versus $\gamma_{+10}$. The numbers within each plot are the Spearman's coefficients and $p$-values for $\gamma_{-10}$ versus $\gamma_{+10}$ ({\it left panel}) and for $\gamma_{+20}$ versus $\gamma_{+10}$ ({\it right panel}). The SNe~Ibn from the F25 and Literature sample are presented as magenta and blue circles, respectively.
} 
\label{fig:slope_slope}
\end{figure}

Figure~\ref{fig:mabs_slope} shows that there is no statistically-significant correlation between the absolute $R/r$-band peak magnitude and any of the slope parameters from $-10$ $(\gamma_{-10})$ to $+30$ $(\gamma_{+30})$ days. 
The calculated Spearman's rank coefficients and associated $p$-values in parenthesis, $\rho_{s,-10} = 0.08$ (0.45), $\rho_{s,+10} = -0.14$ (0.29), $\rho_{s,+20} = -0.02$ (0.47), $\rho_{s,+30} = 0.14$ (0.37), result in low values, supporting that there is no correlation between $R/r$-band peak magnitudes and light curve slopes.
Additionally, we investigate if there is a correlation between slope parameters $\gamma_{+10}$, $\gamma_{-10}$ and $\gamma_{+20}$~(Fig.~\ref{fig:slope_slope}). The corresponding Spearman's rank coefficient and $p$-values are $\rho_{s,-10,+10}$ = $-0.73$ $(6\times 10^{-8})$ and $\rho_{s,+10,+20}$ = $0.38$ $(0.01)$. These results reveal a correlation between rise and decay slopes at $\pm 10$ days of peak magnitude, with steeper (faster) rises associated with steeper (fast) declines. A moderate, positive correlation is found for $\gamma_{+10}$ and $\gamma_{+20}$. The correlation found between the rise and decline slopes suggests that the powering mechanism behind the rapid-evolving SNe Ibn might be short-lived.

\section{Light Curve Modeling}
\label{sec:mod}

In order to investigate the properties of the SN~Ibn progenitor systems and surrounding CSM, we model the multiband light curves of SNe using \mosfit\footnote{\url{https://mosfit.readthedocs.io/en/latest/}}\citep{mosfit}. This work represents the largest consistently modeled sample of type Ibn SNe in the literature. {\mosfit} estimates the posterior distributions of the parameters for a specific model given the photometric observations and a set of user-defined prior distributions for such parameters. 

For this purpose, we select 24 SNe from the F25 + Literature sample that have photometric data covering the $B$, $V$, $g$, $c$, $r$, $o$, $R$, $I$ and $z$ bands, each with a cadence of $\lesssim 5$ days and are well-sampled prior to maximum. The selected SNe~Ibn are SN~2010al~\citep{Pastorello_2015_2011hw}, 
OGLE-2012-SN-006~\citep{Pastorello_OGLE12},
SN~2014av~\citep{Pastorello_2014av}, 
OGLE-2014-SN-131~\citep{Karamehmetoglu_2017},
iPTF14aki and iPTF15ul~\citep{Hosseinzadeh_2017},
SN~2015U~\citep{Pasto_PSN,Shivvers_2015U}, 
PS15dpn~\citep{Smartt_15dpn},
SN~2018jmt~\citep{Wang_2024_Ibn},
SN~2019uo~\citep{Gangopadhyay_2020},
SNe~2019deh and~2021jpk~\citep{Pellegrino_2022_19deh},
SN~2019kbj~\citep{BenAmi_2022},
SN~2020bqj~\citep{Kool_2021_2020bqj},
SN~2020nxt~\citep{Qinan_2020nxt},
SN~2022ablq~\citep{Pellegrino_2022ablq}, 
SN~2023emq~\citep{Pursiainen_2023},
SNe~2020able, 2022ihx, 2022pda, 2023iuc, 2023qre, 2023rau and 2023xgo (from F25).
We term this sample  `\texttt{MOSFiT} sample' (see Fig.~\ref{fig:test_sample}). 

We adopt the pure CSM-SN ejecta interaction model of {\mosfit} (CSI) described in detail in \citet{Chatzopoulos_2012,Chatzopoulos_2013,Villar_2017}. We chose the CSI model based on the moderate nickel masses found typically in SNe~Ibn from light curve modeling~\citep[$\lesssim 0.2$~\Modot; ][]{Kool_2021_2020bqj,Gango_2022,BenAmi_2022,Farias_2024} and the absence of a nickel radioactive tail~\citep[e.g.,][]{Gorbikov_13beo,Shivvers_2015U}.
The CSI model describes the total SN luminosity with a combination of CSM parameters and SN ejecta parameters. The CSM parameters are the optical CSM opacity ($\kappa$), the CSM mass ($M_{\rm CSM}$), the inner radius of the CSM ($R_0$), the CSM density ($\rho_{\rm CSM}$) and the index of the density profile ($s$) of the CSM ($\rho_{\rm CSM} \propto r^{-s}$). The SN ejecta parameters are the total mass of the SN ejecta ($M_{\rm ej}$), the inner ($\delta$) and outer ($n$) indexes of the density profile of the ejecta ($\rho_{\rm inner,ej} \propto r^{-\delta}$ and $\rho_{\rm outer,ej} \propto r^{-n}$), and the ejecta velocity ($v_{\rm ej}$). 
The parameter ``n'' describes the exponent of the outer velocity profile of the progenitor itsef, therefore it is a measure of the compactness of the progenitor envelope.
Additional parameters of the CSI model are the hydrogen column density of the host galaxy ($n_{\rm H,host}$),    
the explosion time relative to the first detection ($t_{\rm exp}$), the minimum temperature before the photosphere starts to recede ($T_{\rm min}$), the efficiency coefficient of converting kinetic energy to radiation ($\epsilon$) and a white noise term ($\sigma$) to account for underestimation of the reported photometric uncertainties.
We fix the inner slope of the SN ejecta ($\delta=1$) and the efficiency coefficient of kinetic energy to radiation ($\epsilon=0.5$). The former has a minimal effect on the light curve, while the latter has been to shown to vary between $\sim0.3-0.7$ in Type IIn SNe (see \citealt{dessart2015numerical, Ransome2024}).
In order to test if the contribution of the radioactive nickel decay to the light curves is small for SNe~Ibn, we also model the multiband light curves of all 24 SNe using the $^{56}$Ni radioactive decay + CSM interaction (RD+CSI) from {\mosfit}. This adds two more parameters with respect to the CSI model; the $\gamma$-ray opacity ($\kappa_{\gamma}$) of the SN ejecta and the nickel fraction of the ejecta mass ($f_{\rm Ni}$), following the prescription developed by~\citet{Arnett_gammaray,Chatzopoulos_2009}.
Each free parameter has the same prior distribution in each individual SN except for the lower boundary of the uniform distribution of the explosion time, which is set based on the observations of each object.
The prior distributions of the parameters in the {\mosfit} modeling are either uniform ($\mathcal{U}$) or log-uniform ($\log\mathcal{U}$), with the exception of the distribution of the ejecta velocity. For the latter, we assume a Gaussian ($\mathcal{G}$) distribution centered at $6000$ km/s with a standard deviation of $2000$ km/s and lower and upper limits of $3000$ and $20000$ km/s, respectively. This is justified because velocity components larger than about $8000$ km/s are rare in the line profiles of SNe~Ibn~\citep[e.g., ][see also Sect.~\ref{sec:specana}]{Pastorello_2014av}.

For each SN in the \mosfit{} sample, we sample the posterior distributions of the parameters for both the CSI and RD+CSI models using the {\tt dynesty} sampler implemented in \mosfit. Tables~\ref{tab:mosfit} and~\ref{tab:nimosfit} present the median values and corresponding 1$\sigma$ confidence intervals for each parameter (e.g., $\rho_{\rm CSM}$) derived from the posterior distributions for each SN in the \mosfit{} sample, for the CSI and RD+CSI models, respectively. 
Tables~\ref{tab:mosfit} and~\ref{tab:nimosfit} show that the median values of 10 \mosfit{} parameters for the CSI and RD+CSI encompass wide ranges, most notably $R_{0}$ and $\rho_{\rm  CSM}$.
For 20 out of the 24 SNe in the \mosfit{} sample, the CSM and ejecta masses inferred by the CSI and RD+CSI models are below $3$~\Modot.
For the CSI case, we find that $n<9$ for 22 SNe in the \mosfit{} sample. These values are smaller than the characteristic density profile of a red supergiant star~\citep[$n=12$;][]{1999Matzner_CCSNe}, and more akin to a progenitor with a compact envelope~($n \lesssim 10$) such as a Luminous Blue Variable or Wolf-Rayet star~\citep{1969Colgate_SNe}. For the RD+CSI, values of $n$ are larger than for the CSI case ($ n > 9$), although consistent within the typical uncertainties of the parameter in both models.
Moreover, for the CSI model, $s > 1$ for 22 out of 24 SNe. These values imply that the CSM geometry is more wind-like, i.e, formed by steady mass loss ($s=2$) rather than an eruptive episode (shell-like, $s=0$). However, the values of $s$ for RD+CSI do not show any preference.  
For the nickel masses in the RD+CSI case, we obtain $M_{\rm Ni} \lesssim 0.01$~\Modot{} for 20 out of 24 SNe in the \mosfit{} sample. This is consistent with the lower limit of nickel masses obtained for SNe Ibc~\citep[0.03 -- 0.015~\Modot;][]{Meza_SESNe,TailNi_Ibc}.
Assuming that the energy of the SN explosion is proportional to $M_{\rm ej}v_{\rm ej}^{2}$, the explosion energies of the SNe Ibn in the \mosfit{} sample are below $10^{51}$ erg. 

\begin{figure*}[hbt!]
\includegraphics[width=\textwidth]{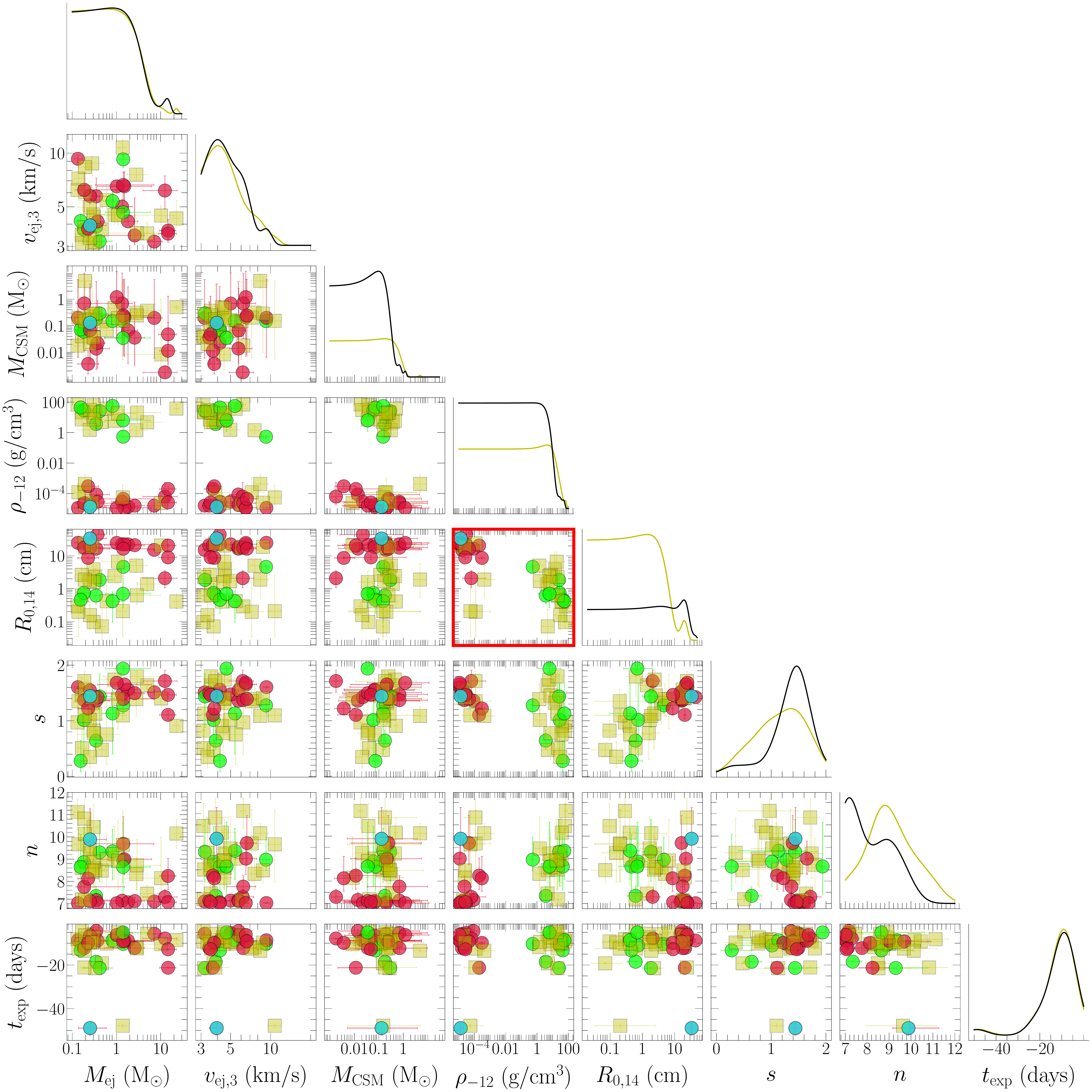}
\caption{Median values $\pm 1\sigma$ uncertainties of {\mosfit} parameters for all 24 SNe in the \mosfit{} sample. The parameters shown here are $M_{\rm ej}$, $v_{\rm ej}$, 
$\rho_{\rm CSM} \times 10^{-12}$
(labeled in the figure as $\rho_{-12}$), 
$R_{0}\times 10^{14}$ (labeled in the figure as $R_{0,14}$),
$s$, $n$ and $t_{\rm exp}$ for the CSI (circle) and RD+CSI (square) models. 
The blue circle in each plot highlights the type Ibn OGLE-2014-SN-131.
For the CSI sample, the lime and crimson circles correspond to the SNe with $\rho_{\rm CSM}$ above or below $10^{-14}$ g/cm$^{3}$, respectively.
Probability density functions (PDFs) produced with a Gaussian kernel density estimator are shown along the diagonal.
Black and yellow curves correspond to the PDF of a specific parameter for the CSI and RD+CSI models, respectively.
The subplot with red axes highlights the correlation between $\rho_{\rm CSM}$ and $R_0$, further explored in Fig.~\ref{fig:R0_vs_rho}.
} 
\label{fig:mosfit_phot_csm}
\end{figure*}

Figure~\ref{fig:mosfit_phot_csm} shows the median values of each parameter for each SN in the \mosfit{} sample, including the probability density functions (PDFs) in the diagonal.  
The PDFs are obtained using the (Gaussian) kernel density estimation (KDE) package from {\tt scipy.stats}\footnote{\url{https://docs.scipy.org/doc/scipy/reference/generated/scipy.stats.gaussian_kde.html}}.
Evidently, the PDF of $n$ for the CSI model is skewed towards the lower boundary of their prior distributions ($n\sim 7$). The explosion dates (relative to the peak magnitude) cluster close to $\sim -10$ day. The clear outlier at $t_{\rm exp} = -48.9^{+2.2}_{-3.2}$ days, OGLE-2014-SN-131, has the longest rise-time of all SNe~Ibn~\citep{Karamehmetoglu_2017}.
The most intriguing findings of the light curve modeling are the bimodal distributions for $\rho_{\rm CSM}$ and $R_0$, and their potential correlation. 
The modeling results show that the densest CSM (i.e., $\rho_{\rm CSM} \gtrsim 10^{-14}$ g/cm$^{3}$), is located closest to the progenitor star (i.e., $R_0 \lesssim 2\times 10^{14}$ cm). For the SNe~Ibn best described by these values, the median values of the ejecta parameters, $M_{\rm ej}$ and $v_{\rm ej}$, are also tightly constrained, with $M_{\rm ej} \lesssim 1$~\Modot{} and $v_{\rm ej} \lesssim 5000$ km/s. 
For SNe~Ibn with less dense CSM (i.e., $\rho_{\rm CSM} < 10^{-14}$ g/cm$^{3}$), we find that $R_0 > 2\times 10^{14}$ cm, $n\lesssim10$  and $s\gtrsim 1$.
For the CSI case, we note that the three SNe~Ibn in our {\tt MOSFiT} sample (OGLE-2012-SN-006, SN~2021jpk and SN~2022pda) with the largest ejecta masses ($> 10$~\Modot) have low CSM densities.
Irrespective of the model and density, the PDF of the CSM mass of the SNe of the \mosfit{} sample peaks at $\sim 0.1$~\Modot{}, and of the ejecta masses at $\sim 1.0$~\Modot{}.

As can be seen from Fig.~\ref{fig:mosfit_phot_csm}, the PDFs of $v_{\rm ej}$, $M_{\rm ej}$ and $t_{\rm exp}$ for the RD+CSI model are similar to those of the CSI model, suggesting that these parameters are insensitive to the addition of the radioactive powering source. The largest differences between the shapes of the PDFs of the two models are noticeable for the index of the density profile of the CSM ($s$) and the outer SN ejecta ($n$).
The bimodality in the PDFs of $\rho_{\rm CSM}$ and $R_{0}$ for the CSI case is not as prominent in the RD+CSI scheme. Potential explanations behind the difference in the bimodality between both models are discussed in Sect.~\ref{sec:discuss}.
Overall, the values of the host extinction ($A_{V}$) for the CSI model is about a factor of 100 less than the one resulting from the RD+CSI model. The result of the CSI model suggests that there is negligible host galaxy extinction. Since about half of the SNe in the \mosfit{} sample have accurate host galaxy extinction from previous studies, we expect that $A_{V}\approx 0$ for the \mosfit{} sample.

Since 13 SNe out of the {\mosfit} sample have been extensively studied elsewhere (see Table~\ref{tab:photo_props} for references), we can broadly compare these results with our CSI and RD+CSI modeling results.
Nickel masses have been estimated for SNe~2014av~\citep{Pastorello_2014av},~2015U~\citep{Shivvers_2015U}, 2019uo, 2019deh, 2021jpk, iPTF15ul~\citep{Pellegrino_2022_19deh}, PS15dpn~\citep{Wang_15dpn},~2019kbj~\citep{BenAmi_2022},~2023emq~\citep{Pursiainen_2023} and~2020bqj~\citep{Kool_2021_2020bqj}. These studies obtained nickel masses within $10^{-2} - 10^{-1}$~\Modot{}, about one dex higher than our results. 
Ejecta masses of SNe~2018jmt~\citep{Wang_2024_Ibn}, 2019uo, 2021jpk, iPTF15ul, 2019kbj~\citep{BenAmi_2022}, and 2023emq have been estimated to be $\lesssim 2$~\Modot{}. These values agree with our median values for the majority of the SNe in the \mosfit{} sample, for any configuration. However,~\citet{Kool_2021_2020bqj},~\citet{Karamehmetoglu_2017} and~\citet{Wang_15dpn} reported ejecta masses $>10$~\Modot{} for SNe~Ibn~2020bqj, OGLE-2014-SN-131 and PS15dpn, respectively. 
We only obtain such massive ejecta for SN~2020bqj from the RD+CSI model ($\sim 20$~\Modot) and SNe~2021jpk, 2022pda and OGLE-2012-SN-006 for the CSI model ($\sim 10$~\Modot).  
Discrepancies in the nickel and the ejecta masses are related to the underlying models used to estimate these parameters. For example,~\citet{Pastorello_OGLE12} did not use a CSI model to estimate the ejecta and nickel masses inferred for OGLE-2012-SN-006. Instead, they used a pure radioactive decay model, obtaining an ejecta mass of $\sim 2$~\Modot{}, with a nickel mass of $\sim 1$~\Modot{}. This nickel mass is larger than the maximum value observed stripped-envelope SNe~\citep[$\sim 0.6$~\Modot;][]{TailNi_Ibc}.
The CSM masses estimated in previous studies (e.g.,~\citet{Pursiainen_2023,Pellegrino_2022_19deh} are typically of the order of $\sim 0.1$~\Modot{}, consistent with our findings.

\subsection{Modeling Caveats}

\mosfit{} and similar codes~\citep[e.g., ][]{Chatzopoulos_2013,redback} aim at modeling the light curves of different types of SNe. Hence, these models explore a large parameter space and the results are affected by the prior distributions of the parameters. 
Here, we test whether a change of the prior distributions of $v_{\rm ej}$, a Gaussian distribution centered at $\mu=8000$ km/s with $\sigma=2000$ km/s, and of $\rho_{\rm CSM}$, a log-Uniform distribution between $10^{-18}$ and $10^{-8}$ g/cm$^{3}$, $\mathcal{U}(10^{-18}, 10^{-8})$, affect the results obtained for the CSI and RD+CSI modeling. 
We refer to the original set and the set of test prior distributions as {\tt PRIOR A} and {\tt PRIOR B}, respectively.

Independent of the choice of prior ({\tt A} or {\tt B}), the main relations between different parameters shown in Fig.~\ref{fig:mosfit_phot_csm} are preserved. However, for some individual SNe, there are large discrepancies between the median values of several parameters, such as $\rho_{\rm CSM}$ and $R_0$. 
These discrepancies involve changes from either low to high values or vice versa. These changes could imply that the posterior distributions of the \mosfit{} parameters are multi-modal, and our sampling routine may not be adequate to capture these complex distributions.  

\begin{figure}[t]
\includegraphics[width=\columnwidth]{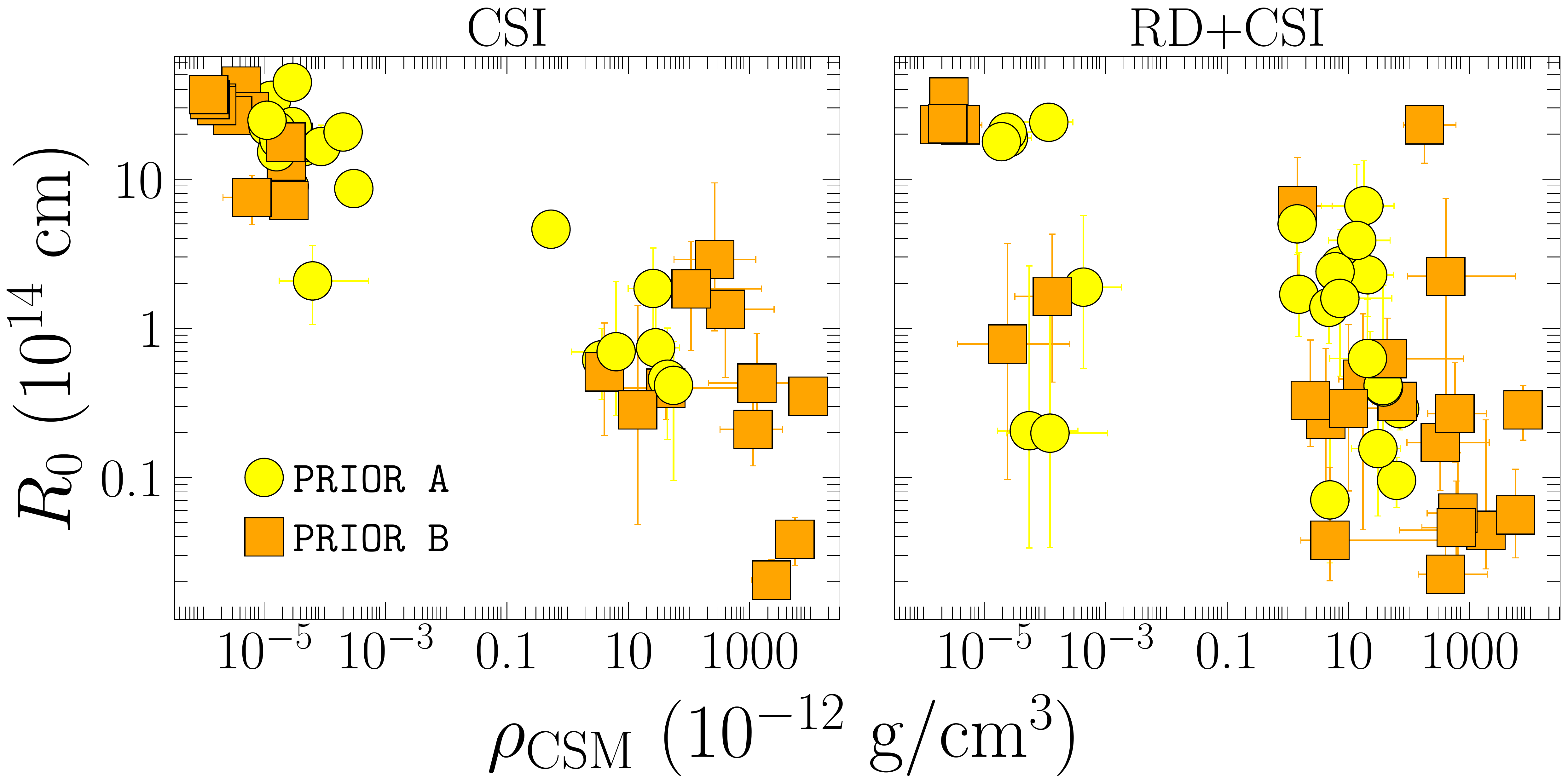}
\caption{{\it Left panel:} Median values of the posterior distributions of $R_{0}$ versus $\rho_{\rm CSM}$ for each SN in the \mosfit{} sample under the CSI model, using prior distributions {\tt A} (yellow circles) and {\tt B} (orange squares).
{\it Right panel:} Same as the {\it left panel}, for the RD+CSI case. 
} 
\label{fig:R0_vs_rho}
\end{figure}

Figure~\ref{fig:R0_vs_rho} shows the relation between between $R_{0}$ and $\rho_{\rm CSM}$ for the CSI and RD+CSI model, given the {\tt PRIOR A} or {\tt PRIOR B}. 
The width of the prior distribution of $\rho_{\rm CSM}$ has a direct impact on the median value of the posterior distribution of $\rho_{\rm CSM}$, with a preference for large densities ($\sim10^{-8}$ g/cm$^{3}$) for {\tt PRIOR B}.
The Spearman's rank coefficients for the CSI model are both $\rho_{s} \sim -0.9$, with a $p$-value $p_{\rm val} < 10^{-6}$, suggesting a strong anti-correlation. This anti-correlation is not as prominent for the RD+CSI model, perhaps due to the larger number of free parameters involved in the latter. 
However, we emphasize that this anti-correlation might not be real. Instead, the evident relation between $R_0$ and $\rho_{\rm CSM}$ might be a degeneracy of the model.

In summary, irrespective of the choice of the prior distribution and the model, the correlation between $\rho_{\rm CSM}$ and $R_0$ remains, suggesting that there is a degeneracy inherent to the modeling of the circumstellar interaction.
Nevertheless, both both CSI and RD+CSI models share some results such as CSM masses below 1~\Modot{},
ejecta masses below 10~\Modot{}, ejecta velocities below 10000 km/s, and $s\gtrsim 0.5$.

\section{X-ray analysis}
\label{sec:xray}

Prior to this study, only SNe~Ibn~2006jc~\citep{Immler_2006jc} and 2022ablq~\citep{Pellegrino_2022ablq} had reported X-ray detections. In this work, we analyze the XRT observations of 15 SNe (see the X-RAY sample in Fig.~\ref{fig:test_sample} and Sect.~\ref{sec:data:xray}) of the F25 sample (Table~\ref{tab:xray}). Except for SN~2022ablq, SN~2022qhy, SN~2020nxt and SN~2023emq, we do not detect X-ray emission at the positions of the SNe. Furthermore, the detected X-ray emission of SNe~2020nxt and 2022qhy is likely from the host galaxy. 
The X-ray emission associated with SN~2023emq on day $+11$ is only a 2$\sigma$ detection. Since we only retrieve upper limits from XRT observations of SN~2023emq after this epoch, the X-ray emission at day $+11$ might be a spurious detection.

The powering source of the soft (0.1 - 10 keV) and hard ($>$ 10 keV) X-ray light curve of Type Ibn SNe is primarily free-free emission from electrons at the forward (i.e. circumstellar) shock region~\citep{Inoue_xray_ibn}. There are only small contributions~\citep{Margalit} from inverse Compton scattering of photons by the shocked electrons at the interaction region of the forward shock~\citep{Chevalier_Xray_review} and possible a co-existing reverse shock~\citep[e.g.,][]{Fransson_2010jl}. The X-ray light curves ($L_{X}$) of SNe~2006jc and 2022ablq share similar shapes, which can be parametrized by two power laws with respect to the X-ray peak luminosity~\citep{Pellegrino_2022ablq}. The pre-maximum X-ray light curves of SNe~2006jc and 2022ablq follow a power law defined as $L_{X}\propto t$. The post-maximum X-ray light curve of SN~2006jc follows a power law $L_{X}\propto t^{-3.1}$, while $L_{X}\propto t^{-1.8}$ for SN~2022ablq.

\begin{figure}[t]
\includegraphics[width=\columnwidth]{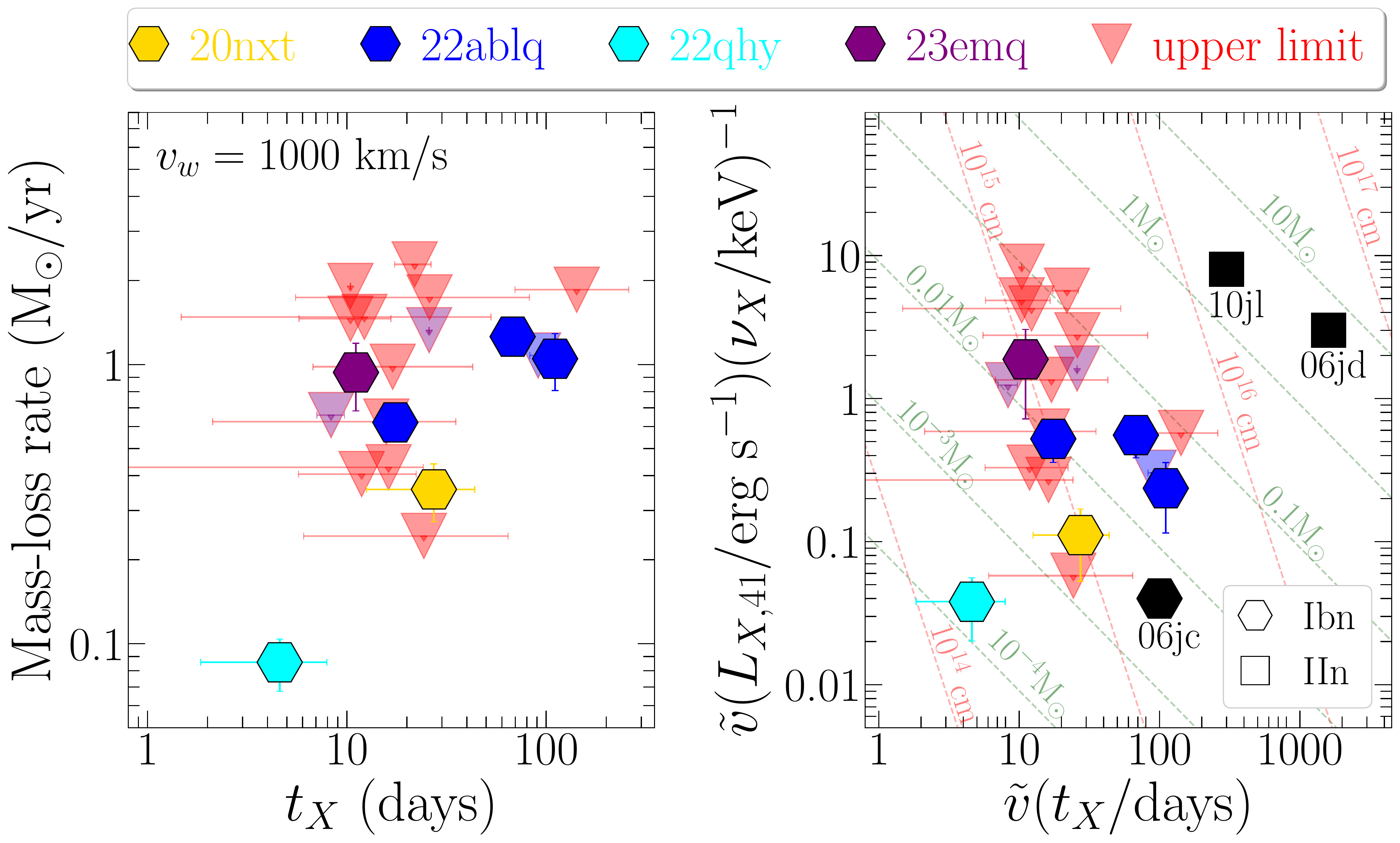}
\caption{{\it Left panel}: Mass-loss rates versus the duration of the X-ray event of 15 Ibn SNe of the F25 sample observed with {\it Swift} XRT. The mass-loss rates were estimated using Eq.~\ref{eq:xray}. {\it Right panel}: Parameter space of~\citet{Margalit}, $\tilde{v}L_{X}/\nu_{X}$ versus $\tilde{v}t_X$. Here we included the X-ray estimates from the type Ibn SN~2006jc~\citep{Immler_2006jc}, and the type IIn SNe 2006jd~\citep{Chandra_2006jd} and 2010jl~\citep{Chandra_2010jl}.
Green and red lines correspond to the slopes of constant CSM radii and CSM masses using Eqs.~\ref{sample:eq:mcsm} and~\ref{sample:eq:RCSM}, respectively.
The origin of the X-ray emission associated to SN~2020nxt and SN~2022qhy is likely the host galaxy.} 
\label{fig:x-ray}
\end{figure}

 The evolution of the post-maximum X-ray light curves of both SNe imply that the CSM is not necessarily a result of a steady mass-loss process~\cite[index of the density profile, $s=2$, e.g.,][]{Dwarkadas_Xray}, with a typical  decline $L_{X} \propto t^{-1}$ to $t^{-2}$~(\citealt{Fransson_1996_Xray}, and see e.g., Figure 13 of \citealt{2024PASA...41...59P} and references therewithin). 
Nevertheless, for simplicity and given the single detections/upper limits of several SNe~Ibn in the X-RAY sample, we set $s=2$ for our X-ray analysis. 
Under the assumption of a negligible contribution of the reverse shock to the total X-ray luminosity,~\citet{Dwarkadas_2016} approximated the luminosity at 1 keV ($L_{\rm CS,1 keV}$) as:
\begin{align}~\label{eq:xray}
L_{\rm CS,1\,keV} &= 1.4\times 10^{38} \xi T_{8}^{-0.236}\frac{e^{-0.116/T_{8}}}{(3-s)} \left [ \frac{\dot{M}_{-5}}{v_{w}} \right ]^2\\
&\times v_{\rm ej,\, 4}^{3-2s} \left [ \frac{t_X}{11.57} \right ]^{3-2s} \quad \text{ erg/s/keV}\nonumber,
\end{align}
where $T_8$ is the shock temperature in units of $10^{8}$ K, $\dot{M}_{-5}$ is the mass-loss rate scaled to $10^{-5}$~M$_{\odot}$/yr, $v_w$ is the wind velocity in terms of 10 km/s, $v_{\rm ej,\,4}$ is the maximum ejecta velocity in units of $10^4$ km/s, $\xi\sim 0.85$ and $t_{X}$ is the duration of the X-ray event.
We can determine $\dot{M}$ from Eq.~\ref{eq:xray} assuming $T$, $v_{w}$, $v_{\rm ej}$, $L_{\rm CS,\,1\,keV}$ and $t_{X}$. From the mean values of the narrow and broad velocity components of \hei~$\lambda 5876$~\AA{} in Sect.~\ref{sec:specana}, we estimate $v_{w}\approx 1000$ km/s and $v_{\rm ej}\approx 5000$ km/s. 
Furthermore, we adopt a conservative value of the shock temperature~\citep[$T_{8}\sim1;$][]{T8_xray}. The quantities $L_{\rm CS,\, 1\, keV}$ and $t_X$ are estimated directly from the \textit{Swift} XRT observations~(Table~\ref{tab:xray}). 
The duration of the X-ray event is defined as the average date of the combined XRT observations relative to the discovery date of the supernova. The total X-ray luminosity at 1 keV ($L_{\rm CS,\, 1\, keV}$) is estimated as the XRT luminosity scaled by the passband width of XRT, from 10 keV to 0.3 keV.
The left panel of Fig.~\ref{fig:x-ray} shows the mass-loss rates inferred for each SN based on the X-ray data. Most of these measurements represent upper limits. By construction, higher wind velocities results in correspondingly higher mass-loss rate estimates. 
For the vast major of SN in our sample, we observe that the upper limits of the mass-loss rates from the X-ray analysis are largely consistent with the results obtain using \mosfit{} (Tables~\ref{tab:mosfit} and~\ref{tab:nimosfit}). 
We find that the mass-loss rates of SN~2022ablq is about $1$~\Modot/yr from our X-ray analysis, about twice the values reported in~\citet[$\lesssim 0.5$~\Modot/yr,][]{Pellegrino_2022ablq}.

\citet{Margalit} studied the X-ray luminosity coming from the shock due to free-free emission and Compton ionization, for different shock regimes (radiative or adiabatic). They constructed the parameter space $\tilde{v}\mathit{L}_{X}/\nu_{X}$ versus $\tilde{v}t_X$, where $\tilde{v}=v_{\rm sh}/v_{\rm rad}$ is the ratio between the shock velocity ($v_{\rm sh}$) and the limiting velocity ($v_{\rm rad}$), and determines if a shock is either adiabatic or radiative. Using these variables, they obtain analytical expressions of the mass and radius of the CSM,

 \begin{align}
 M_{\rm CSM} &\approx 0.11 \left ( \frac{\tilde{v} L_{X}/\nu_{\rm keV}}{10^{41} {\rm erg/s}} \right ) \left ( \frac{\tilde{v}t_{X}}{100\,{\rm days}} \right ) \, {\rm M}_{\odot}~\label{sample:eq:mcsm}\\
 R_{\rm CSM} &\approx 4.5\times 10^{15}   \left ( \frac{\tilde{v} L_{X}/\nu_{\rm keV}}{10^{41} {\rm erg/s}} \right )^{1/4} \left ( \frac{\tilde{v}t_{X}}{100\,{\rm days}} \right )^{3/4} {\rm cm}~\label{sample:eq:RCSM}.
 \end{align}

Given the lack of a rich spectroscopic follow-up of several SNe in the X-RAY sample to trace the velocity of the shock and evaluate $\tilde{v}$ as done by~\citet{Pellegrino_2022_19deh}, we assume $\tilde{v}=1$. Similarly to~\citet{Margalit}, we assume a characteristic frequency $\nu_{X} = 1$ keV. Using Eqs.~\ref{sample:eq:mcsm} and~\ref{sample:eq:RCSM}, we constrained the values of $M_{\rm CSM}$ and $R_{\rm CSM}$ for the SNe in the X-RAY sample, independently from the light curve analysis in Table~\ref{tab:mosfit}. The right panel of Fig.~\ref{fig:x-ray} shows that the CSM in SNe~Ibn, including SN~2006jc~\citep{Immler_2006jc}, is much less massive ($\sim0.1$~\Modot) and more compact ($R_{\rm CSM} < 10^{16}$ cm) than the most extreme type IIn SNe~2010jl~\citep{Chandra_2010jl} and 2006jd~\citep{Chandra_2006jd}.

\section{Analysis of helium lines}
\label{sec:specana}

For SNe in the F25 sample as described in Sect.~\ref{subsec:specdata}, we model the prominent \hei~$\lambda 5876$~{\AA} emission line in order to characterize the progenitor wind and ejecta velocities over time. All SNe of the F25 sample have at least one spectrum. The SN with the largest (unpublished) spectral series is SN 2020able, with spectra taken from $-14$ days to +20 days. The \hei~$\lambda 5876$ emission line is visible in all spectra. 

We designed a {\tt python} program using the Markov Chain Monte Carlo (MCMC) sampler {\tt emcee} to obtain the posterior distributions of the set of parameters of the different components of the spectral lines. For this purpose, we subtract the local continuum from the \hei{}~$\lambda 5876$~\AA{} by fitting a linear function to the background regions of the line. The two background regions are defined centered on $ \pm 200$~\AA{} of the $\lambda 5876$~\AA{} \hei{} line complex, with a width of $\sim50$~\AA{}.
We added an extra scatter parameter to the fitting to account for the white noise of the spectrum. The purpose of including this term is to estimate the noise in spectra published without any reported uncertainties, mainly from WISeREP.
Thereafter, we fit the continuum-subtracted \hei{} lines using a linear combination of Gaussian profiles. One Gaussian profile is fit to each distinct component of the \hei{} line complex. The \hei{} line complex is composed of a mixture of narrow and broad components. Some of the individual components also exhibit entire P-Cygni profiles, consisting of both emission and absorption components. Thus, the number of profiles to be fit for the \hei{} line complex depends on the number of components, which we determined by visual inspection of each spectrum. 
We categorize any component of the \hei{}~$\lambda 5876$~\AA{} line complex with a measured velocity as the full width at half maximum (FWHM) lower than $3000$ km/s as `narrow'. Similarly, a `broad' component refers to any emission component of the \hei{}~$\lambda 5876$~\AA{} line with a FWHM larger than $3000$ km/s. Typically, the narrow emission component represents the velocity of the CSM, $v_{\rm CSM}$ (see Sect.~\ref{sec:intro}). 
In the case of a P-Cygni narrow profile, we determine the velocity from either the minimum of the narrow absorption component or the FWHM of the narrow emission component. 
We assume that the velocity of the broad component corresponds to the lower limit of the ejecta velocity.
To fit the continuum and the Gaussian profiles of the \hei{}~$\lambda 5876$~\AA{} lines, we used 100 walkers for 10,000 steps with a MCMC fitter to sample the parameter space and obtain the posterior distributions of each parameter. 

For each SN Ibn in the F25 sample that has a time-series of spectra, we choose the largest velocity of all \hei{}~$\lambda 5876$~\AA{} broad components as the lower limit of the ejecta velocity for that SN. Similarly, we choose the lowest velocity of the \hei{}~$\lambda 5876$~\AA{} narrow components as the CSM velocity of that SN. However, we emphasize that this velocity is not the true velocity of the CSM.
Instead, it is the lowest velocity (an upper limit) measurable from the available spectra without accounting for the spectral resolution of the instruments.
For 21 out of 37 SNe in the Literature sample, we retrieve the spectra from WISeREP or private communication (e.g., SNe~2019wep, 2022ablq, and 2023emq) and analyze the \hei{}~$\lambda 5876$~\AA{} emission lines following the same procedure as for the SNe in F25 (outlined above).
For 14 SNe of the Literature sample~(see Table~\ref{tab:spectroscopic_table}), not all the spectroscopic data from their analysis were published~\citep[e.g., P16, ][]{Smartt_15dpn}. In these cases, we take the velocity measurements of either the narrow or broad components that were reported. For the specific case of the P16 sample, we retrieve the minimum velocity of the narrow component as the CSM velocity, and the maximum velocity of the broad component as the potential ejecta velocity of a SN (Table 7 in P16).
For two SNe of the Literature sample, PTF11rfh and SN~2019qav, we could not obtain any estimates of the narrow or broad velocity components of the \hei{}~$\lambda 5876$. In total, we collect the velocity estimates of 59 SNe for the F25 (24) + Ibn (35) samples. All velocity estimates of the 59 SNe are listed in Table~\ref{tab:spectroscopic}. 

\begin{figure}[t]
\centering
\includegraphics[scale=0.4]{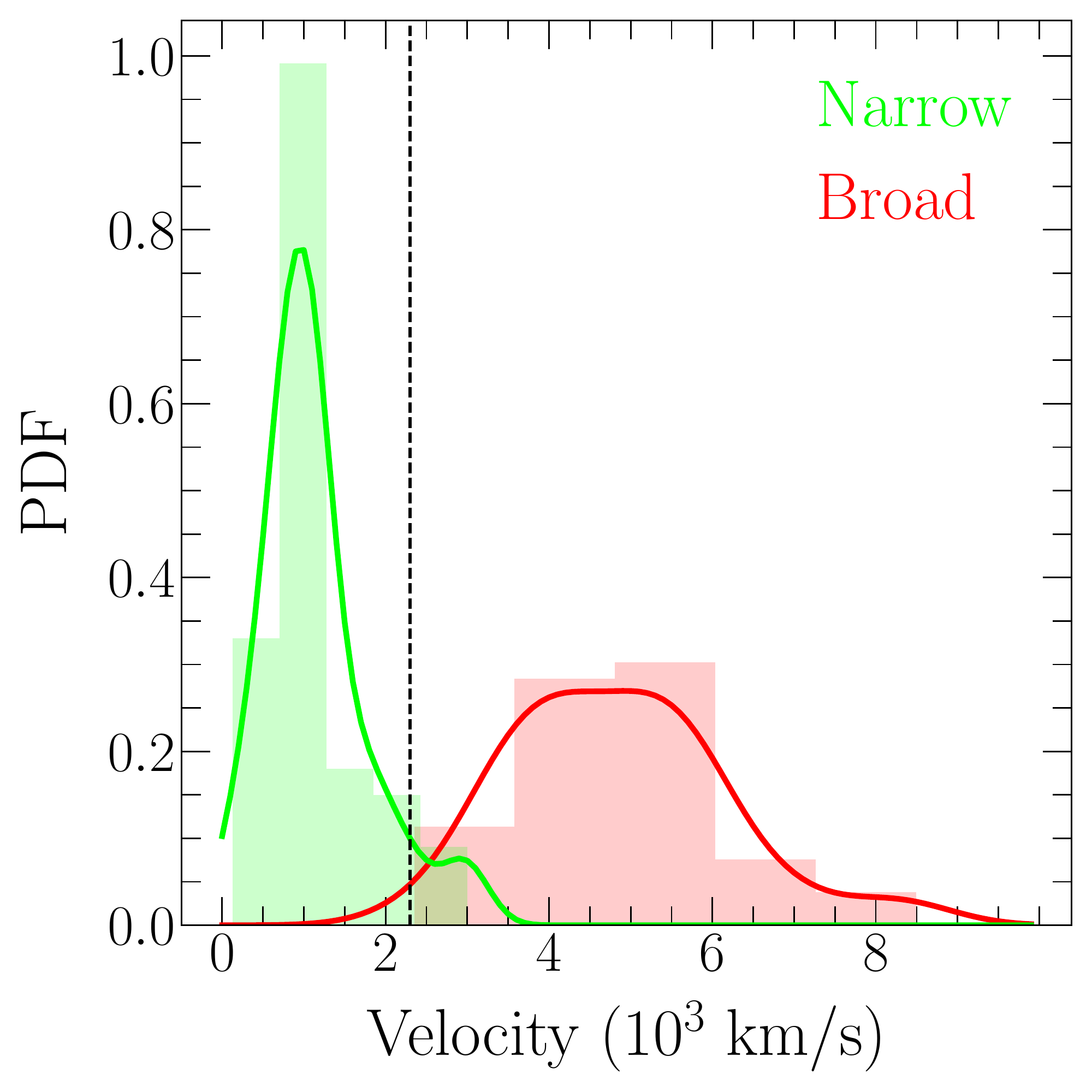}
\caption{Probability density function (PDF) of the narrow (lime) and broad (red) velocity components of \hei~$\lambda5876$ lines of 59 SNe~Ibn derived using KDE (solid lines). Lime and red histograms correspond to the empirical velocity measurements used to estimate the PDFs. 
The black dashed line marks $v=2300$ km/s.} 
\label{fig:histograms_vel}
\end{figure}

Figure~\ref{fig:histograms_vel} shows the distribution and PDFs of the final CSM and ejecta velocities derived from the narrow and broad components of the complex \hei~$\lambda5876$ emission line profile for the 59 SNe~Ibn. The PDFs were obtained using the method described in Sect.~\ref{sec:mod}.
Furthermore, we drew $10^{6}$ samples from their PDFs to estimate the mean and standard deviation of the PDFs.
The PDF of the narrow components peaks at a velocity of $1144\pm 487$ km/s and the PDF of the broad components peaks at a velocity of $4847 \pm 2038$ km/s. The latter is consistent with previous works (e.g., P16, H17) and supports our assumption of the mean value of the prior distribution of the ejecta velocity for the \mosfit{} light curve modeling  (Sect.~\ref{sec:mod}).

By integrating the PDF of the narrow velocity over the velocity range, we note that over 90\% of the our estimates are at $v\lesssim 2300$ km/s. Therefore, we adopt $v=2300$ km/s as a tentative threshold to classify type Ibn SNe. However, this threshold is purely observational and dependent on the time at which the spectra were obtained, the resolution or S/N of the available spectra and how rich/poor the spectroscopic follow up was for a particular SN. Nevertheless, it is evident that unless the observation was performed with a low-resolution instrument ($R\lesssim 120$ at \hei{}~$\lambda 5876$~\AA{}), or at late phases in the evolution of the SN, the threshold represents a practical benchmark for observers to characterize the Ibn class.

To visualize the variety of possible origins of the different components comprising the spectral lines, 
Fig.~\ref{fig:He_postmax} shows a compilation of the continuum-subtracted \hei~$\lambda5876$~{\AA} line profiles of different SNe~Ibn and at different epochs between one to three weeks after maximum light. 
About 50\% of the SNe in the F25 sample exhibit narrow P-Cygni profiles of \hei{}~$\lambda 5876$~\AA{} throughout their evolution. For the remaining SNe~Ibn at these epochs, the \hei{}~$\lambda 5876$~{\AA} is a pure emission line. 
Highlighted in Fig.~\ref{fig:He_postmax} is also the presence of {\ha} next to \hei~$\lambda6678$~\AA{} which becomes stronger at later epochs in some SN Ibn spectra. 
This is not uncommon for SNe~Ibn~\citep[see e.g.,][]{Pastorello_2008_2005la} since the He-rich CSM might be mixed with H-rich material which was not completely expelled from the progenitor star/system before the SN explosion~\citep{Farias_2024}. 
In higher resolution spectra of some SNe~Ibn, e.g., SNe~2021iyt,~2022gzg,~2022ihx,~2023iuc and~2023qre, [\ion{N}{ii}]~$\lambda\lambda 6583,6548$~\AA{} emission lines are identified next to {\ha}, suggesting a host galaxy origin for the hydrogen lines for some fraction of these objects. 

At several days prior to peak (about $-7$ days), the emission lines are likely `flash-ionization features'~\citep{Khazov}, such as observed in SNe~Ibn 2010al, 2015U, 2019cj~\citep{Wang_2024_Ibn}, 2019uo, 2019wep~\citep{Gango_2022} and 2023emq.
Appendix Fig.~\ref{fig:flashers} shows the early spectra of all Ibn-like SNe known to date which have flash-ionization features.
In the F25 sample, we can ascribe a flash-ionization origin
~of the blended complex of {\heii}~$\lambda 4686$, \ion{N}{iii}~$\lambda\lambda 4634, 4641$ and {\ciii}~$\lambda\lambda 4648, 4650$  emission lines for nine SNe (2019lsm, 2020able, 2021iyt, 2022eux, 2022ihx, 2023iuc, 2023qre, 2023rau and 2023xgo), typically observed in early spectra of Type II SN~\citep{wynn_flash}.
Though blended, the spectral feature at $\sim 4600${\AA} shows wide wings, characteristic of the broadening of narrow lines due to the photons scattering off of the electrons in the ionized CSM.
Furthermore, for eight SNe~Ibn in F25 we identify the {\ciii} emission line at $\approx 5696$~{\AA} (SNe~2019lsm, 2020able, 2021iyt, 2022ihx, 2023iuc, 2023qre, 2023rau, and 2023xgo) The presence of this line indicates a high ionization of the CSM and has so far only been observed in one SN Ibn, SN~2023emq~\citep{Pursiainen_2023}, although it is more common in spectra of type II SNe, such as SN 1998S~\citep{Fassia_98S}. 

SNe that have identifiable He lines, and thus belong to a class of type Ib SNe, may not exhibit a narrow component at the time of classification. This can be of physical origin, e.g., the ejecta has already swept up the entire CSM and thus no signatures of it remain in the spectra. This could be the case at either late epochs or early on, when only a marginal amount of CSM close to the progenitor is present. Alternatively, CSM velocities much less than about 1000 km/s can be missed in low(er) resolution spectra and with poor S/N.
Some SNe~Ibn have also been found with no broad He velocity components in their spectra. One such example is SN~2011hw, which does not exhibit any velocity component larger than $2350$ km/s since its discovery up to $+60$ days past maximum light~\citep{Pastorello_2014av}. On the other hand, it shows resolved narrow components of \hei{}~$\lambda 5876$~\AA{} ($v_{\rm narrow}\approx 200$ km/s), likely originating from the unshocked CSM. A SN can have either an unclear or missing broad velocity component of an emission line if the SN ejecta has been slowed to velocities hovering around $\sim 2300$ km/s through the interaction with a dense CSM. A broad velocity component could also be hidden by the photosphere created ahead of the ejecta in the CSM. Independent of the origin or mechanism leading to an absent broad velocity component, it would only affect the classification of a SN Ibn if, additionally, no narrow component is present or can be identified.

With our derived estimates of the CSM velocity ($v_{\rm CSM}$), we can estimate the average mass-loss rate for each SN in the \mosfit{} sample, following Eq. 4 from~\citet{Ransome2024}:
\begin{equation}~\label{eq:mdot}
\langle \dot{M} \rangle \approx \frac{4\pi \rho_{\rm CSM} R_{0} v_{\rm CSM}}{3-s} R_{\rm CSM,outer}^{2-s},
\end{equation}
with
\begin{equation}~\label{eq:rcsm}
R_{\rm CSM,outer} = \left ( \frac{3 M_{\rm CSM}}{4\pi \rho_{\rm CSM} R_{0}^{s}}  + R_{0}^{3-s}\right )^{\frac{1}{3-s}},
\end{equation}
using samples from the full joint posterior distribution of $\rho_{\rm CSM}, R_{0}, M_{\rm CSM}$ and $s$ for each SN, thereby preserving correlations among parameters (Sect.~\ref{sec:mod}, Tables~\ref{tab:mosfit} and~\ref{tab:nimosfit}).

Using Eq.~\ref{eq:mdot}, the mass-loss rates of the 17 SNe of \mosfit{} sample for the CSI case are $\dot{M} \lesssim 0.1$~\Modot/yr~(see Table~\ref{tab:mosfit}). This value ($\gtrsim 10^{-3}$~\Modot/yr) is consistent with the typical mass-loss rates found in the late stages of the mass-transfer episodes for binary systems of helium star + compact object~\citep[e.g., ][]{Wu_Fuller}. For the RD+CSI case, the mass-loss rate values are larger than for the CSI case, with 11 SNe having mass-loss rates over $1$~\Modot{}/yr.  
The values of the outer radius of the CSM ($R_{\rm outer,CSM}$) for 18 SNe in the \mosfit{} sample (RD+CSI) are $\lesssim 10^{16}$ cm, consistent with the constraints found from our X-ray analysis. Larger values of $R_{\rm outer,CSM}$ are found for the CSI case. For the latter, 50\% of the SNe in the \mosfit{} sample have $R_{\rm outer,CSM}>10^{16}$ cm.

Mass-loss rates have been estimated for several SNe~Ibn from light curve modeling~\citep{Wang_15dpn,Gango_2022,Shivvers_2015U,Pellegrino_2022_19deh} and X-ray analysis~\citep{Pellegrino_2022ablq}. Except for PS15dpn~($\sim 8$~\Modot{}/yr), the mass-loss rates estimated in the previous studies are $\sim 1$~\Modot/yr, similar to the values we estimate for the CSI and RD+CSI models (Tables~\ref{tab:mosfit} and~\ref{tab:nimosfit}).
However, for SN~2022ablq, we obtain a mass-loss rate of $8.6^{+2.0}_{-1.5}$ M$_{\odot}$/yr from \mosfit{} modeling (from either CSI or RD+CSI models), as opposed to $\approx 1$~M$_{\odot}$/yr from our X-ray analysis. The discrepancy between the mass-loss rate values of SN~2022ablq are due to differences of the assumed wind velocity, and the several assumptions we have had to make in order to evaluate Eq.~\ref{eq:xray}. For instance, by using $v_{w}=1000$ km/s in Eq.~\ref{eq:mdot}, the \mosfit{} mass-loss rate for SN~2022ablq decreases to $\approx 3$~\Modot/yr. 

\section{Discussion}
\label{sec:discuss}

We have presented a comprehensive analysis of a rich set of photometric and spectroscopic data of 24 SNe~Ibn (F25 sample) observed by the Young Supernova Experiment between 2019 and 2023. 
We augment the F25 sample with a photometric and spectroscopic dataset of 37 SNe Ibn (Literature sample) from the literature (see Fig.~\ref{fig:test_sample}, Sect.~\ref{sec:photana}, Table~\ref{tab:photo_props}) Our detailed light curve analysis shows that there exists a high degree of heterogeneity in brightness, color, light curve shape and evolution, which is not apparent in previous studies (e.g., P16, H17) of small samples of SNe~Ibn. In particular, the normalized $R/r$-band light curves of the 61 SNe~Ibn in our F25 + Literature sample (Fig.~\ref{fig:ABS}, lower panel) show a large spread ($> 1.5$ mag) with respect to known SNe~Ibn templates at late times ($\sim +50$ days)~\citep[H17;][]{Somayeh_2024}. Furthermore, there is no noticeable common evolutionary path  for either $B-V$ or $g-r$ color curves for any SN in our F25 + Literature sample (Fig.~\ref{fig:color}). Based on our measurements, we do not find any correlation between the peak magnitudes and decay slope for our SNe~Ibn (Sect.~\ref{sec:photana}). 
However, there is a clear correlation between the rise and decay slopes of our SNe~Ibn, with faster evolving SNe having steeper slopes. 
The study of the origin of the rapid-evolving nature of the LCs of SNe Ibn is beyond the scope of this paper. However, its origin might be related to, for example, short photon diffusion times~\citep[e.g.,][]{Chevalier_Irwin}, short-lived powering mechanisms such as SN ejecta-CSM interaction~\citep[e.g.,][]{Wang_15dpn}, or asymmetries in the SN ejecta/CSM ~\citep{circumbinary_disk_suzuki}, among others.

Typically, the main mechanism that powers the light curves of stripped-envelope SNe is the release of the shock energy deposited in the SN ejecta~\citep{Villar_2017}. Other potential heating sources include the chain of radioactive decay of $^{56}$Ni, and interaction between the SN ejecta and the CSM. In the absence of the latter two, the light curve of a stripped-envelope SN is mostly determined by the kinetic energy of the SN explosion (or the ejecta velocity), and the ejecta mass. 
This means that for a fixed kinetic energy, the larger the ejecta mass, the longer the duration of the light curve. For a fixed ejecta mass, the more energetic the SN explosion, the higher the SN luminosity. Since slow-evolving SNe~Ibn are rare~(Fig.~\ref{fig:ABS}), the ejecta masses of most of the SNe Ibn are not expected to be large ($\sim$ 10~\Modot{}). 
Furthermore, the absence of a radioactive tail in the light curves of SNe Ibn suggests that the nickel mass must also be small~\citep[$\lesssim 0.1$~\Modot;][]{Moriya_2016}.  
Thus, for the majority of our sample, a direct, rapid interaction between the SN ejecta and a nearby, compact CSM is the most plausible scenario. 
However, as suggested by the diversity seen in the photometry and spectroscopy, the mechanism powering the slow light curve evolution of some SNe~Ibn in our sample may be more complex. It is possible that the SN ejecta-CSM interaction might cause photometric evolution that mimics the effect of large ejecta masses.

To elucidate possible mechanisms driving the photometric heterogeneity of SNe~Ibn found in this study, we modeled the well-sampled light curves of 24 SNe~Ibn (\mosfit{} sample, see Sect.~\ref{sec:mod}), using the {\tt csm} (CSI) and {\tt csmni} (RD+CSI) models implemented in {\mosfit}. 
We find that median rise times ($\sim 10$ days), CSM ($\approx 0.1$~\Modot) and ejecta masses ($\sim 1.0$~\Modot), and the index of the density profile of the SN ejecta ($n \lesssim 10$) inferred from most of the SNe in the \mosfit{} sample are consistent between both models~(Table~\ref{tab:mosfit}, Fig.~\ref{fig:mosfit_phot_csm}). 
Rise times of $\sim 10$ days are also estimated from the photometric analysis in H17. A rapid photometric evolution suggests that radioactive $^{56}$Ni decay does not play a major role as an additional powering source, specifically at late times. Our RD+CSI modeling leads us to conclude that the light curves in our \mosfit{} sample are powered by the radioactive decay of small amounts ($< 0.01$~\Modot{}) of $^{56}$Ni. 
These nickel masses are smaller than the median masses found for stripped-envelope SNe~\citep[0.08~\Modot;][]{TailNi_Ibc} and consistent with the little-to-no nickel masses derived from the bolometric light curves of several SNe~Ibn~\citep{Maeda_Moriya2022}. This again supports the interpretation of the power source being primarily SN ejecta interacting with a nearby CSM. 
The low CSM and ejecta masses inferred from light curve modeling suggest that the compact progenitor of most SNe in our sample must not necessarily be a massive Wolf-Rayet star at the moment of explosion~\citep[$\sim$ 10~\Modot;][]{WR_masses} as hypothesized in early studies~\citep[e.g., ][]{Foley_2006jc}.
Recently,~\citet{Ercolino_theoreticalModel} found that a binary system composed of a low-mass ($\sim 3$~\Modot) helium-rich star and a main sequence star can produce an asymmetric CSM with $M_{\rm CSM}\lesssim 0.8$~{\Modot}. 
Such scenario is suggested for the progenitor system of SN~2020nxt~\citep{Qinan_2020nxt}.
In~\citet{Ercolino_theoreticalModel}, the CSM formation happens through several mass transfer episodes via Roche-lobe overflow or ``effective ROLF'' $\rm(1\pm e)x_{RLOF}$~\citep[see section 2.3 in][]{Hamers_Dosopoulou2019}. The CSM can then be understood as a circumbinary disk~\citep{circumbinary_disk}. The helium star, if it explodes as a CCSN, would eject $\lesssim 2~$\Modot. 
Similar results were found for a low-mass helium star with a neutron star as a companion~\citep{Wu_Fuller}.
Such binary scenarios were also proposed to explain the precursor emission observed in SNe~2006jc~\citep{Foley_2006jc},~2019uo~\citep{Strotjohann_2021} and 2023fyq~\citep{Dong_23fyq}. 

Although consistent between different models and prior distributions of the parameters, our \mosfit{} modeling results are obtained assuming a spherically symmetric CSM geometry. An asymmetric CSM, such as produced by the binary progenitor system proposed for SNe~Ibn~\citep{Ercolino_theoreticalModel} that is
observed from specific viewing angles, could be responsible for the appearance (or absence) of P-Cygni profiles in SNe~Ibn. This hypothesis has been proposed by H17 in order to account for the roughly equal ratio between SNe Ibn with P-Cygni versus emission lines.
Alternatively, optical depth and/or density effects are suggested for the origin of P-Cygni versus emission lines~\citep[][]{2018bcc_Karamehmetoglu}.
About 50\% of the SNe in the F25 sample show P-Cygni profiles in the {\hei}~$\lambda 5876$~{\AA}, a ratio similar to that found by H17. Therefore, the assumption of a spherical symmetric CSM would break down. This is a limitation of the available tools to model SNe observations, and more detailed models accounting for an asymmetric CSM geometry, and detailed spectropolarimetric observations of SNe~Ibn are needed to more completely understand the geometry of the CSM and the nature of the progenitor systems.

\begin{figure}[t]
\includegraphics[width=\columnwidth]{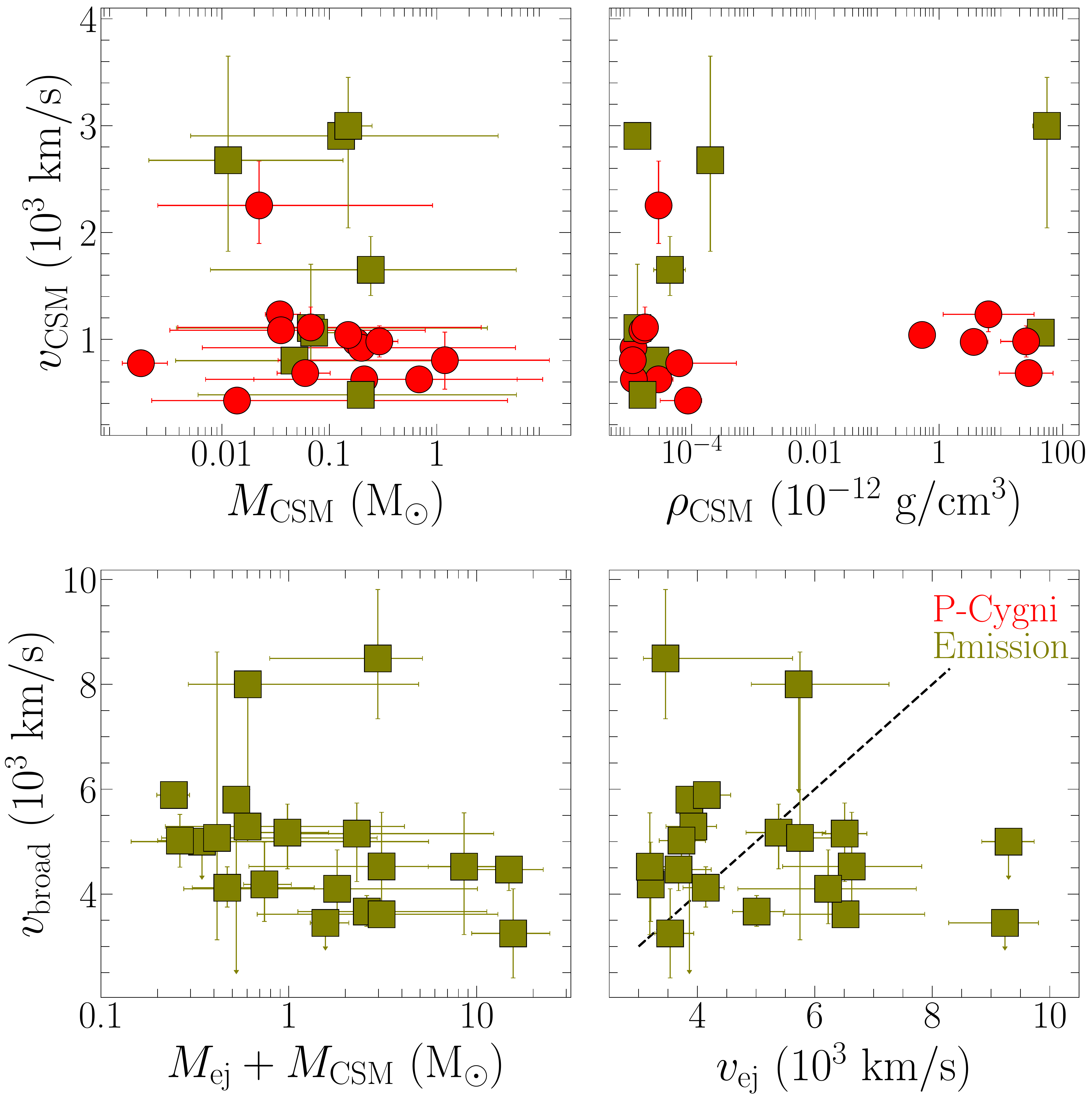}
\caption{{\it Upper left:} Velocity of the CSM versus the CSM mass retrieved from the CSI modeling of 24 SNe~Ibn of the \mosfit{} sample.
 The velocity of the CSM was estimated either from the minimum of the absorption component of the P-Cygni profile (red circles) or the FWHM of the emission component (olive squares) of {\hei}~$\lambda5876$~{\AA} line.
{\it Upper right}: Similar to {\it upper left}, velocity of the CSM versus the CSM density. {\it Bottom left:} the broadest velocity component of {\hei}~$\lambda5876$~{\AA} line through the entire spectroscopic evolution versus the sum of the CSM and ejecta masses from {\mosfit} CSI modeling. 
{\it Bottom right}: Similar to {\it bottom left}, the broadest velocity component of {\hei}~$\lambda5876$~{\AA} line versus the ejecta velocity from {\mosfit} CSI modeling.
}
\label{fig:phot_vs_spec}
\end{figure}

Our spectroscopic analysis of $59$ SNe~Ibn shows that throughout the evolution, the median velocity of the broad component of the \hei~$\lambda5876$~\AA{} line ($v_{\rm broad}$) is $\approx 5000$ km/s. 
This velocity might be considered a lower limit of the ejecta velocity ($v_{\rm ej}$) in SNe~Ibn. The extent to which this differs from the true value is difficult to assess.
Assuming that the  {\mosfit} CSI modeled ejecta velocity is accurate, then Fig.~\ref{fig:phot_vs_spec} (bottom right panel) shows that $v_{\rm broad}$ does not follow a 1:1 relation with {\tt MOSFiT} $v_{\rm ej}$.
In principle, $v_{\rm ej} < v_{\rm broad}$ is plausible if the SN ejecta has decelerated during the interaction with the CSM. In this case, prominent emission lines will not have a broad velocity component.  
However, several SNe~Ibn show $v_{\rm broad} > v_{\rm ej}$, with some extreme cases such as SNe 2019kbj and 2022ihx ($> 8000$ km/s). In the latter case, an asymmetric CSM configuration may be hiding a portion of the SN ejecta that is interacting with the CSM, making a broad velocity component visible.
Furthermore, despite the large uncertainties, we do not observe any correlation between $M_{\rm CSM}$ and $v_{\rm CSM}$~(Fig.~\ref{fig:phot_vs_spec}, upper left panel). 
Similarly, there is no apparent correlation between either the CSM velocity with the CSM density or the broad velocity and the sum of the SN ejecta and CSM masses (Fig.~\ref{fig:phot_vs_spec}, upper right and bottom left panels, respectively). The lack  of any apparent correlation between the spectroscopic and the (modeled) photometric features shows that it is not straightforward to reconcile spectroscopic and photometric-derived information of SNe~Ibn.
These results reinforce that the CSM configuration is more complex than currently captured by light curve models such as \mosfit{}. 
Moreover,~\citet{Wang_2024_Ibn}
found that the light curves of the SNe Ibn~2018jmt and 2019cj may be well-fitted by more than one CSM region, also indicating an asymmetric CSM.
Further development of such models accounting for either multiple CSM shells or asymmetric geometries is critical to understanding the explosion physics of SNe~Ibn. Moreover, consideration in future tools may also be given to  additional information derived from the spectroscopic evolution of SNe Ibn, such as the different line velocities of characteristic elements.

\section{Conclusions}
\label{sec:concl}
In this work, we analyzed the photometric and spectroscopic data of 24 SNe Ibn observed through the Young Supernova Experiment from 2020 to 2023~(see Tables~\ref{tab:summary} and \ref{tab:xray}). In addition, we collected 37 SNe Ibn from previous studies, creating the largest sample of SNe~Ibn to date~(see Fig.~\ref{fig:test_sample} and Tables~\ref{tab:photo_props} and \ref{tab:spectroscopic}).
From the photometric analysis, we conclude that SNe~Ibn are more heterogeneous than reported in previous studies, with a large range of peak magnitudes, photometric slopes and color evolution. 
We did not find any statistically significant correlation between the $R/r$-band peak magnitude and the slopes of the light curve from -10 to +0 ($\gamma_{-10}$), +0 to +10 ($\gamma_{+10}$), +10 to +20 ($\gamma_{+20}$), and +20 to +30 ($\gamma_{+30}$) days with respect to the time of the maximum light. A strong correlation was found between rise ($\gamma_{-10}$) and decay slope at +10 days ($\gamma_{+10}$).
The spectroscopic analysis of the {\hei} lines confirms that the majority of these SNe do not show velocity components of about $10000$ km/s, as expected for stripped-envelope Ib SNe~\citep{Liu_2016}. The average CSM velocity, as derived from our He emission line analysis of 59 SNe~Ibn, is about $1100$ km/s and is much larger than the average CSM velocity measured for SNe IIn~\citep[$\approx 400$ km/s;][]{Ransome2024}.
Furthermore, we identify the early emission of \ciii{} at $\approx 5696$~{\AA} previously only observed in SN~2023emq~\citep{Pursiainen_2023} as a flash-ionization feature commonly observed in 7 other SNe in our F25 sample.

Light curve modeling of well-sampled multi-band light curves of 24 SNe~Ibn using an SN ejecta/CSM interaction (CSI) model as the primary powering source, suggests that the progenitor star of most of these SNe does not have an extended ejecta ($n < 10$). It must be surrounded by $\approx 0.1$~\Modot{} of CSM material and ejecting $\sim 1$~\Modot{} of material in a low-energy explosion ($<10^{51}$ erg) in comparison to normal stripped-envelope SNe~\citep[$\gtrsim 10^{51}$ erg;][]{Taddia_2018_Ib}. However, these results are obtained assuming a spherically symmetric CSM, which does not capture the full complexity of the CSM as expected for a binary system progenitor.

Our X-ray analysis of 15 SNe~Ibn of the F25 sample supports the CSM properties derived from the optical light curve modeling.
These results are consistent with a binary progenitor system composed of an (exploding) helium star of $\approx 3$~{\Modot} and a compact object such as a neutron star~\citep{Ercolino_theoreticalModel}.
Nevertheless, the class of SNe~Ibn is diverse and heterogeneous, and very high ejecta masses of $\sim 10$~\Modot{} are inferred from the light curve modeling of some SNe (e.g., SNe~2020bqj and 2022pda). Therefore, massive He-rich WR stars may be the progenitors of some SNe~Ibn, as initially suggested for SN~2006jc ~\citep{Foley_2006jc}, indicating that this class of supernovae may have multiple progenitor channels. More detailed observations might yield more ways to disentangle SNe~Ibn from these channels.
We also explored different combinations of models and prior distributions of their associated parameters. The combination of CSI model and the radioactive decay of $^{56}$Ni, RD model, does not significantly alter our results. Furthermore, the nickel masses inferred in the RD+CSI model are $< 0.01$~\Modot{}, which are smaller than the values typically found for stripped-envelope SNe~\citep{Meza_SESNe}. 
Irrespective of the model, minor alterations of the prior distributions of the \mosfit{} model parameters allowed us to pinpoint a potential degeneracy between the inner radius of the CSM and the CSM density. This means that a dense ($\gtrsim 10^{-14}$ g/cm$^{3}$) CSM must be close ($\lesssim 10^{14}$ cm) to the exploding star.    
Our work represents the first attempt to derive the physical properties of the progenitor system of a large sample of SNe~Ibn using a consistent methodology. However, without independent constraints on the properties of the CSM we caution against the validity of modeling individual objects under limited assumptions.
For instance, the index of the CSM density profile ($s$) plays a crucial role in shaping the optical and X-ray light curves.~\citet{Maeda_Moriya2022} found that the decline phase of the bolometric light curves of different SNe~Ibn (e.g., SNe 2006jc, 2011hw, and 2014av) can be well fitted with a steeper CSM density profile where \(s \gtrsim 2\). 
Similarly,~\citet{Wang_2024_Ibn} found reasonable fits to the bolometric light curves of SNe 2018jmt and 2019cj using a broken power-law density profile, characterized by an inner, flat region (\(s \sim 0\)) and an outer, steeper component (\(s \sim 3\)). 
From the X-ray perspective,~\citet{Pellegrino_2022ablq} argued that the X-ray light curves of SNe 2006jc and 2022ablq suggest a complex CSM density profile, which also follows a broken power-law. 
Recent studies by~\citet{Farias_2024,Gango_2024} have confirmed that the CSM surrounding interacting SNe is complex, and may be composed of different shells expelled at various times during the evolution of the progenitor star.
Therefore, future theoretical and computational studies may consider the results of our work to expand current modeling tools to infer the physical properties of type Ibn SN explosions. 
A self-consistent progenitor system must be able to reconcile both the photometric and spectroscopic analysis in this work.

\begin{acknowledgements}
We thank Miika Pursiainen and Craig Pellegrino for supplying the spectra of SN~2023emq and SN~2022ablq, respectively.\\
This work is supported by a VILLUM FONDEN Young Investigator Grant (project number 25501) a Villum Experiment grant (VIL69896) and by research grants (VIL16599, VIL54489) from VILLUM FONDEN. 
Parts of this research were supported by the Australian Research Council Centre of Excellence for Gravitational Wave Discovery (OzGrav), through project number CE230100016.
The Villar Astro Time Lab acknowledges support through the David and Lucile Packard Foundation, National Science Foundation under AST-2433718, AST-2407922 and AST-2406110, as well as an Aramont Fellowship for Emerging Science Research.
KdS thanks the LSST-DA Data Science Fellowship Program, which is funded by LSST-DA, the Brinson Foundation, the WoodNext Foundation, and the Research Corporation for Science Advancement Foundation; her participation in the program has benefited this work.
This work is supported by the National Science Foundation under Cooperative Agreement PHY-2019786 (The NSF AI Institute for Artificial Intelligence and Fundamental Interactions, \url{http://iaifi.org/}).
This material is based upon work supported by the National Science Foundation Graduate Research Fellowship Program under Grant Nos.\ 1842402 and 2236415. Any opinions, findings, conclusions, or recommendations expressed in this material are those of the authors and do not necessarily reflect the views of the National Science Foundation.
D.O.J. acknowledges support from NSF grants AST-2407632 and AST-2429450, NASA grant 80NSSC24M0023, and HST/JWST grants HST-GO-17128.028, HST-GO-16269.012, and JWST-GO-05324.031, awarded by the Space Telescope Science Institute (STScI), which is operated by the Association of Universities for Research in Astronomy, Inc., for NASA, under contract NAS5-26555.
G.N acknowledges NSF support from AST-2206195 for his involvement in this work. G.N is additionally funded by NSF CAREER grant AST-2239364, supported in-part by funding from Charles Simonyi, OAC-2311355, AST-2432428, as well as AST-2421845 and funding from the Simons Foundation for the NSF-Simons SkAI Institute. G.N is also supported by the DOE through the Department of Physics at the University of Illinois, Urbana-Champaign (\# 13771275), and the HST Guest Observer Program through HST-GO-16764. and HST-GO-17128 (PI: R. Foley).
W.J.-G. is supported by NASA through Hubble Fellowship grant HSTHF2-51558.001-A awarded by the Space Telescope Science Institute, which is operated for NASA by the Association of Universities for Research in Astronomy, Inc., under contract NAS5-26555.
L.I acknowledges financial support from the INAF Data Grant Program 'YES' (PI: Izzo).
C.D.K. gratefully acknowledges support from the NSF through AST-2432037, the HST Guest Observer Program through HST-SNAP-17070 and HST-GO-17706, and from JWST Archival Research through JWST-AR-6241 and JWST-AR-5441.
We thank the Swope observers for their valuable contributions: Abdo Campillay, Yilin Kong Riveros and Jorge Anaís.
M.R.S is supported by the STScI Postdoctoral Fellowship.
Y.Z acknowledges visitor support from the Kavli Institute for Cosmology, Cambridge, where part of this work was completed.\\
The data presented here were obtained in part with ALFOSC, which is provided by the Instituto de Astrofísica de Andalucía (IAA) under a joint agreement with the University of Copenhagen and NOT.
Research at Lick Observatory is partially supported by a generous gift from Google. Based in part on observations obtained at the SOAR telescope, which is a joint project of the Ministério da Ciência, Tecnologia e Inovações (MCTI/LNA) do Brasil, the US National Science Foundation`s NOIRLab, the University of North Carolina at Chapel Hill (UNC), and Michigan State University (MSU).
 We acknowledge the use of public data from the Swift data archive. This work makes use of data taken with the Las Cumbres Observatory global telescope network. The LCO group is funded by NSF grants AST-1911151 and AST-1911225. This work has made use of data from the ATLAS project. ATLAS is primarily funded to search for near-Earth asteroids through NASA grants NN12AR55G, 80NSSC18K0284, and 80NSSC18K1575; byproducts of the NEO search include images and catalogs from the survey area. The ATLAS science products have been made possible through the contributions of the University of Hawaii Institute for Astronomy, the Queen’s University Belfast, and the Space Telescope Science Institute.
 YSE-PZ was developed by the UC Santa Cruz Transients Team. The UCSC team is supported in part by NASA grants NNG17PX03C, 80NSSC19K1386, and 80NSSC20K0953; NSF grants AST-1518052, AST-1815935, and AST- 1911206; the Gordon \& Betty Moore Foundation; the Heising-Simons Foundation; a fellowship from the David and Lucile Packard Foundation to R.J.F.; Gordon and Betty Moore Foundation postdoctoral fellowships and a NASA Einstein Fellowship, as administered through the NASA Hubble Fellowship program and grant HST-HF2-51462.001, to D.O.J.; and an NSF Graduate Research Fellowship, administered through grant DGE-1339067, to D.A.C.
\end{acknowledgements}

\bibliographystyle{aa} 
\bibliography{bibl}

\begin{appendix} 

\section{Figures}

\begin{figure}[t]
\includegraphics[width=\columnwidth]{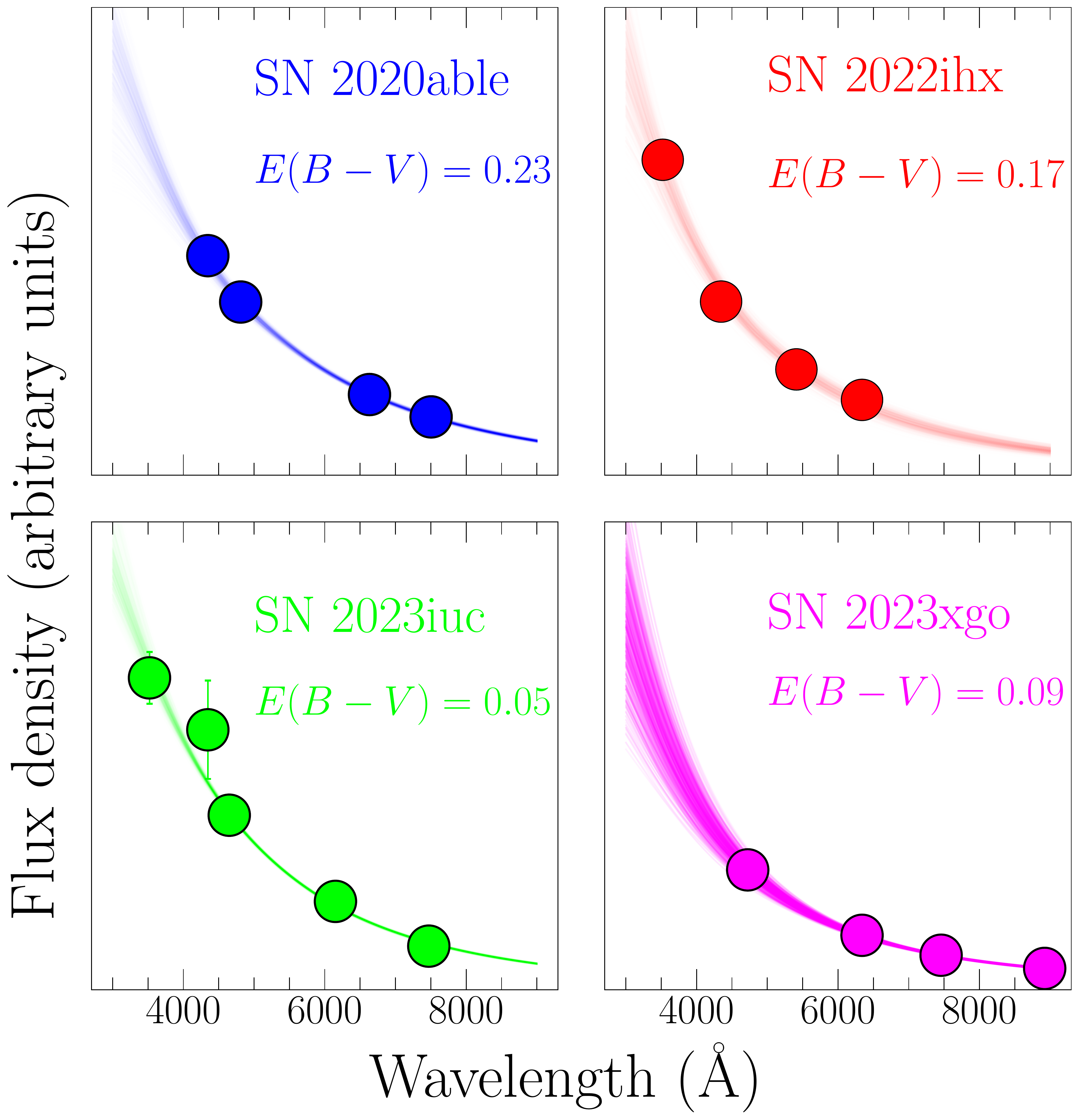}{
\caption{\label{fig:extinction} Collection of the SED ($f_{\lambda}$ versus wavelength) fitting of the interpolated light curves at maximum light of four SNe~Ibn in the F25 sample. The fitting function is a modified blackbody, as described in Sect.~\ref{subsec:ext}. The values of the color excess reported in each plot were obtained assuming $R_{V}=3.1$.
}}
\end{figure}

\begin{figure}[t]
\includegraphics[width=\columnwidth]{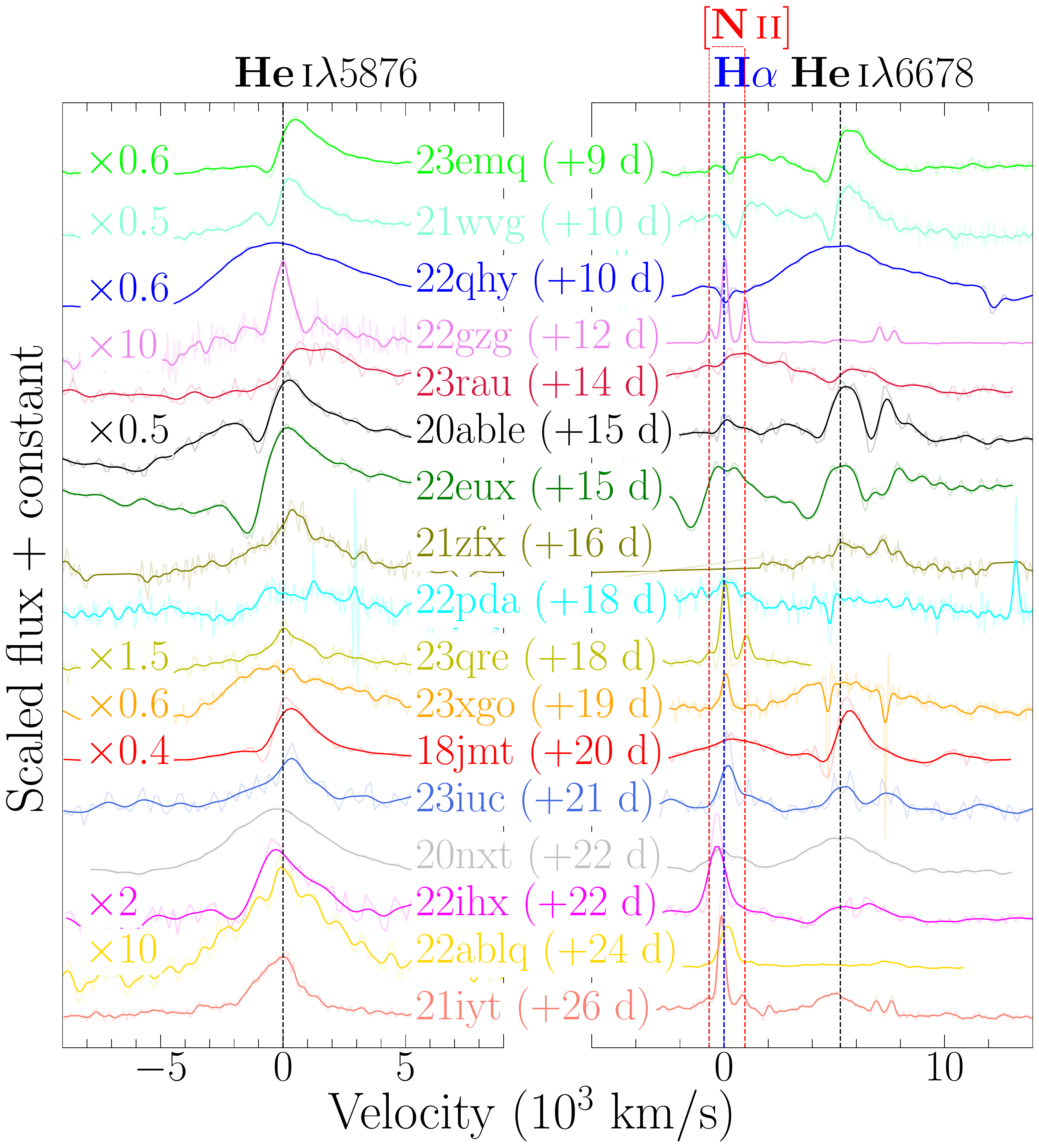}
\caption{{\it Left panel:} Collection of the continuum-subtracted spectral region surrounding \hei~$\lambda 5876$ {\AA} of SNe in the F25 sample ($F_{\lambda}$ versus wavelength). {\it Right panel:} Similar to the {\it left panel}, highlighting the {\ha} spectral region instead.} 
\label{fig:He_postmax}
\end{figure}

\begin{figure}[t]
\includegraphics[width=\columnwidth]{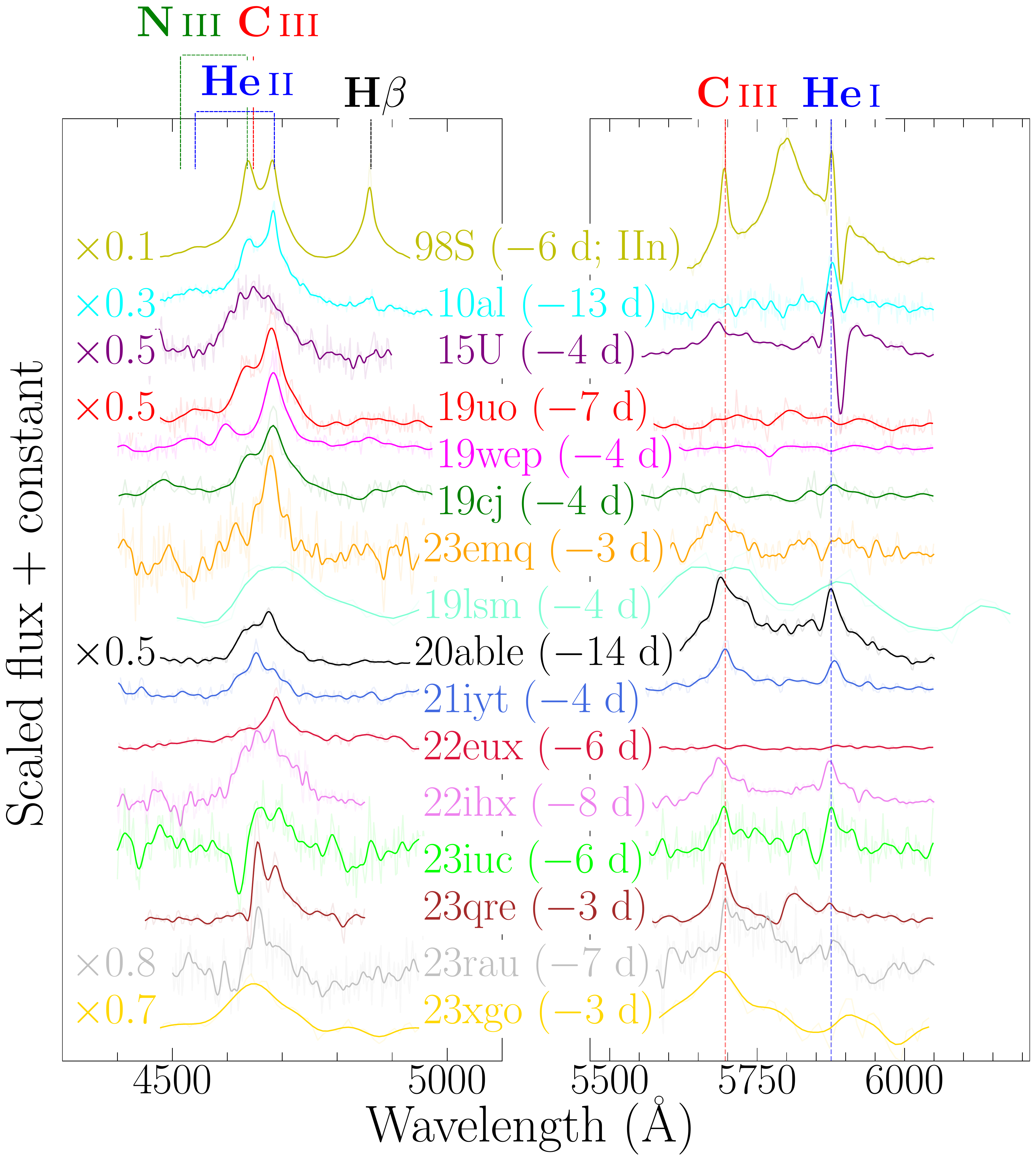}
\caption{Collection of all 15 SNe~Ibn that shows flash-ionized features ($F_{\lambda}$ versus wavelength). For comparison, we include the type IIn 1998S. Nine (from SN~2019lsm to bottom) are first reported in this work. 
} 
\label{fig:flashers}
\end{figure}

\section{Tables}

\begin{table*}
\centering
\caption{\label{tab:short_phot} Photometric data of the 24 SNe from the F25 sample. All magnitudes are reported in the AB system$^{\square}$.}
\begin{tabular}{lccccc}
\hline\hline
Object & MJD (days) & mag & $\Delta$mag & Band & Facility\\
\hline
SN~2019lsm & 58684.2062 & 20.033 & 0.184 & $g$ & ZTF\\
SN~2019lsm & 58684.2067 & 19.886 & 0.171 & $g$ & ZTF\\
SN~2019lsm & 58692.1968 & 18.139 & 0.058 & $g$ & ZTF\\
SN~2019lsm & 58692.1973 & 18.130 & 0.089 & $g$ & ZTF\\
SN~2019lsm & 58689.1754 & 18.413 & 0.091 & $r$ & ZTF\\
SN~2019lsm & 58689.1759 & 18.460 & 0.081 & $r$ & ZTF\\
SN~2019lsm & 58692.1768 & 18.362 & 0.158 & $r$ & ZTF\\
SN~2019lsm & 58692.1773 & 18.265 & 0.116 & $r$ & ZTF\\
SN~2019lsm & 58698.1775 & 18.538 & 0.105 & $r$ & ZTF\\
SN~2019lsm & 58705.1643 & 18.948 & 0.152 & $r$ & ZTF\\
SN~2019lsm & 58685.5000 & 19.256 & 0.187 & $o$ & ATLAS\\
SN~2019lsm & 58689.5000 & 18.460 & 0.146 & $o$ & ATLAS\\
SN~2019lsm & 58693.5000 & 18.307 & 0.072 & $o$ & ATLAS\\
SN~2019lsm & 58705.5000 & 19.267 & 0.324 & $o$ & ATLAS\\
SN~2019lsm & 58691.5000 & 18.221 & 0.029 & $c$ & ATLAS\\
SN~2019lsm & 58703.5000 & 18.814 & 0.088 & $c$ & ATLAS\\
SN~2019lsm & 58707.5000 & 18.830 & 0.348 & $c$ & ATLAS\\
SN~2019lsm & 58696.0772 & 18.101 & 0.090 & $U$ & \textit{Swift}\\
SN~2019lsm & 58699.4604 & 18.393 & 0.103 & $U$ & \textit{Swift}\\
SN~2019lsm & 58701.2547 & 18.798 & 0.135 & $U$ & \textit{Swift}\\
\hline
\end{tabular}
\tablefoot{
\tablefoottext{$\square$}{Full table available in electronic form.}
}
\end{table*}
\begin{table*}
\centering
\caption{\label{tab:spectroscopic_table} Log of spectroscopic observations of the SNe in the F25 sample$^{\square}$.}
\begin{tabular}{lcccccc}
\hline\hline
Object & Observation date &  Phase (days)$^{a}$ & Wavelength range (\AA) & Dispersion (\AA) & Instrument & Analysis$^{b}$\\  
\hline
SN~2018jmt & 2018/12/16 & $+3$ & $3638$-$9233$ & 5.52 & EFOSC2$^{\dag}$ & $\checkmark$\\
\hline
SN~2019lsm & 2019/07/27 & $-4$ & $3777$-$9223$ & 25.6 & SEDM$^{\dag}$ & $\checkmark$\\
SN~2019lsm & 2019/07/30 & $-1$ & $3645$-$9240$ & 5.52 & EFOSC2$^{\dag}$ & $\checkmark$\\
\hline
SN~2020taz & 2020/09/24 & $+5$ & $4002$-$9852$ & 2.0 & DIS$^{\dag}$ & $\checkmark$\\
\hline
SN~2020able & 2020/12/08 & $-14$ & $3499$-$10001$ & 2.29 & FLOYDS$^{\dag}$ & $\checkmark$\\
SN~2020able & 2020/12/11 & $-11$ & $3162$-$10148$ & 1.18 & LRIS & $\checkmark$\\
SN~2020able & 2021/01/06 & $+15$ & $3405$-$10494$ & 2.54 & Kast & $\checkmark$\\
SN~2020able & 2021/01/11 & $+20$ & $3406$-$9995$ & 2.53 & Kast & $\checkmark$\\
\hline
\end{tabular}
\tablefoot{
\tablefoottext{$\square$}{\footnotesize Full table available in electronic form.}\\
\tablefoottext{$a$}{\footnotesize Phases estimated with respect to MJD$_{\rm max}$~(Table~\ref{tab:photo_props}).}\\
\tablefoottext{$b$}{\footnotesize The symbol $\checkmark$ ($\times$) indicates that the corresponding spectrum was (not) utilized for the spectroscopic analysis in Sect.~\ref{sec:specana}.}\\
\tablefoottext{$\dag$}{\footnotesize The spectrum is retrieved directly from WISeREP. For the following SNe, the spectra from WISeREP has already been published: SN~2018jmt~\citep{Wang_2024_Ibn}, SN~2020nxt~\citep{Qinan_2020nxt}, 
SN~2020taz~\citep{wang_sample_ibn}, 
SN~2022ablq~\citep{Pellegrino_2022ablq}, SN~2023emq~\citep{Pursiainen_2023}, 
SN~2023utc~\citep{wang_sample_ibn}, 
and SN~2023xgo~\citep{Gango_2023xgo}.}
}
\end{table*}
\begin{sidewaystable*}
\caption{Summary of type Ibn SNe in the F25 sample.}
\resizebox{\textwidth}{!}{
\begin{tabular}{llllccccccl} 
\hline\hline             
Object 
& RA (h:m:s) 
& DEC ($^{\circ}:':''$) 
& Host Galaxy
& MJD$_{\rm disc}$ (days)
& $E(B-V)_{\rm MW}$\tablefootmark{\footnotesize \rm a}
& $E(B-V)_{\rm host}$\tablefootmark{\footnotesize \rm b,f}
& Redshift
& X-ray\tablefootmark{\footnotesize \rm c}
& MOSFiT\tablefootmark{\footnotesize \rm d}
& References\tablefootmark{\footnotesize \rm e}\\
\hline
2018jmt & 06:54:47.10 & $-$59:30:10.80 & PGC 370943 & 58457.29 & 0.102 & 0.0 & 0.036 & No & Yes &
[1,2,3]\\
2019lsm & 14:35:02.55 & +14:46:36.85  & WISEA J143502.18+144639.6 & 58689.28& 0.016 & 0.27$^{+0.20}_{-0.17}$ & 0.0417 & Yes & No & [4,5,6] \\
2020nxt & 22:37:36.23 & +35:00:07.68 &  WISEA J223736.70+350006.5 & 59033.54 & 0.067 & $-$ & 0.022 & Yes & Yes &
[7,8,9]\\
2020taz & 22:26:06.27 & +10:33:29.65 & LEDA 1381253 & 59103.32 & 0.088 & 0.09$^{+0.06}_{-0.06}$ & 0.05 & No & No & [10,11] \\
2020able & 09:26:02.26 & +24:31:18.07 & SDSS J092602.93+243115.1 & 59185.53 & 0.027 & 0.23$^{+0.05}_{-0.11}$ & 0.025 & Yes 
& Yes & [12,13] \\
2021iyt & 10:18:44.41 & +00:03:19.34 & 2dFGRS TGN357Z086 &  59313.36 & 0.040 & 0.15$^{+0.13}_{-0.10}$ & 0.071 & Yes & No & [14,15,16] \\
2021wvg & 17:59:12.18 & +68:53:02.71 & WISEA
J175914.81+685257.5 & 59450.27 & 0.034 & 0.23$^{+0.08}_{-0.12}$ & 0.085 & No & No & [17,18] \\
2021zfx & 21:09:23.74 & +09:45:39.20 & 2MASX J21092358+0945425 & 59479.38 & 0.079 & $-$ & 0.069 & Yes & No & [19,20]\\
2022acm & 17:35:55.07 & +50:48:23.76 & WISEA J173555.46+504823.9 & 59600.53 & 0.031 & 0.16$^{+0.06}_{-0.09}$ & 0.056 & No & No & [21,22]\\
2022eux & 05:55:31.23 & $-$15:40:34.25 & WISEA J055532.08-154043.0 & 59654.26 & 0.138 & 0.16$^{+0.03}_{-0.05}$ & 0.025 & Yes & Yes & [23,24,25]\\
2022gzg & 12:43:37.52 & +01:23:03.78 & 2dFGRS TGN391Z177 & 59677.30 & 0.015 & 0.19$^{+0.18}_{-0.13}$ & 0.088 & No & No & [26,27,28]\\
2022ihx & 19:16:38.41 & +61:41:15.48 & WISEA J191638.31+614115.1 & 59690.45 & 0.054 & 0.17$^{+0.07}_{-0.09}$ & 0.033 &  Yes & Yes & [29,30]\\ 
2022pda & 22:04:13.89 & $-$18:49:40.46 & LEDA 135305 & 59779.56 & 0.021 & 0.09$^{+0.10}_{-0.06}$ & 0.062 & Yes & Yes & [31,32,33,34]\\
2022qhy & 06:01:08.92 & $-$23:39:59.83 & 2MASS J06010909-2340043 & 59792.16 & 0.029 & $-$ & 0.006 &  Yes & No & [35,36]\\
2022ablq & 12:13:06.47 & +17:05:56.23 & Mrk 0762 &  59907.58 & 0.031 & 0.18 & 0.014 & Yes & Yes &
[37,38,39,40] \\
2023emq & 13:34:22.03 & $-$23:43:51.45 & LEDA 797708 & 60035.98 & 0.105 & 0.0 & 0.033 & Yes & Yes &
[41,42,43]\\
2023iuc & 14:38:55.64 & +02:17:14.15 & SDSS J143855.53+021711.5 & 60082.25 & 0.038 & 0.05$^{+0.06}_{-0.03}$ & 0.076 & Yes &  Yes & [44,45,46]\\
2023qre & 20:50:59.18 & +06:08:18.50 & 2MASX
J20505921+0608141 & 60181.27 & 0.079 & $-$ & 0.047 & Yes & Yes & [47,48]\\
2023rau & 00:20:55.21 & $-$01:45:28.90 & LEDA 1110852 & 60186.36 & 0.029 & 0.18$^{+0.05}_{-0.05}$ & 0.067 & Yes & Yes & [49,50]\\
2023ubp & 23:10:31.06 & +02:57:15.36 & KUG 2307+026 & 60222.14 & 0.050 & $-$ & 0.056 & No & No & [51,52]\\
2023utc & 09:11:59.15 & +53:43:02.60 & SDSS J091159.16+534304.2 & 60228.94 & 0.0145 & $-$ & 0.013 & No& No & [53,54] \\ 
2023vwh & 04:09:59.62 & $-$06:05:52.52 & WISEA J040959.54-060553.3 & 60244.32 & 0.084 & $-$ & 0.06 & No & No & [55,56,57] \\
2023xgo & 05:04:19.19 & +67:37:21.15 & WISEA J050420.81+673723.2 & 60257.28 & 0.142 & 0.09$_{-0.05}^{+0.05}$ & 0.013& Yes & Yes & [58,59,60,61]\\
2023abbd & 23:02:53.530 & +26:01:35.83 & UGC 12327 & 60299.24 & 0.096 & $-$ & 0.042 & No & No  & [62,63]\\
\hline
\end{tabular}
}
\tablefoot{
\tablefoottext{a}{\footnotesize Mean value of $E(B-V)$ from \citet{SF_2011}.}\\
\tablefoottext{b}{\footnotesize Details of the host extinction estimate are described Sect.~\ref{subsec:ext}.}\\
\tablefoottext{c}{\footnotesize Boolean column representing if the SN was observed in X-ray.}\\
\tablefoottext{d}{\footnotesize Boolean column representing if the SN has been modeled using \texttt{MOSFiT}.}\\
\tablefoottext{e}{\footnotesize [1,2,3]~\citet{TNS_2018jmt_class,Vallely_TESS,Wang_2024_Ibn},
                                [4,5,6]~\citet{TNS_2019lsm_discovery,TNS_2019lsm_class,TNS_2019lsm_flash},
                                [7,8,9]~\citet{TNS_2020nxt_discovery,TNS_2020nxt_class,Qinan_2020nxt},
                                [10,11]~\citet{TNS_2020taz_discovery,TNS_2020taz_class},
                                [12,13]~\citet{TNS_2020able_discovery,TNS_2020able_class},
                                [14,15,16]~\citet{TNS_2021iyt_discovery,TNS_2021iyt_class,TNS_2021iyt_class_v2},
                                [17,18]~\citet{TNS_discovery_2021wvg,TNS_2021wvg_class},
                                [19,20]~\citet{TNS_2021zfx_discovery,TNS_2021zfx_class},
                                [21,22]~\citet{TNS_2022acm_discovery,TNS_2022acm_class},
                                [23,24,25]~\citet{TNS_2022eux_discovery,TNS_2022eux_class_wrong,TNS_2022eux_class},
                                [26,27,28]~\citet{TNS_2022gzg_discovery,TNS_2022gzg_class,TNS_2022gzg_class_v2},
                                [29,30]~\citet{TNS_2022ihx_discovery,TNS_2022ihx_class},
                                [31,32,33,34]~\citet{TNS_2022pda_discovery,TNS_2022pda_class,Hiramatsu_2022pda};Cai in prep.,
                                [35,36]~\citet{TNS_2022qhy_discovery,TNS_2022qhy_class},
                                [37,38,39,40]~\citet{TNS_2022ablq_discovery,TNS_2022ablq_class_Other,TNS_2022ablq_class_TDE,TNS_2022ablq_class_Ibn,Pellegrino_2022ablq},
                                [41,42,43]~\citet{TNS_2023emq_discovery,TNS_2023emq_class,Pursiainen_2023},
                                [44,45,46]~\citet{TNS_2023iuc_discovery,TNS_2023iuc_class_wrong,TNS_2023iuc_class}
                                [47,48]~\citet{TNS_2023qre_discovery,TNS_2023qre_class},
                                [49,50]~\citet{TNS_2023rau_discovery,TNS_2023rau_class},
                                [51,52]~\citet{TNS_2023ubp_discovery,TNS_2023ubp_class},
                                [53,54]~\citet{TNS_2023utc_discovery,TNS_2023utc_class},
                                [55,56,57]~\citet{TNS_2023vwh_discovery,TNS_2023vwh_class_wrong,TNS_2023vwh_class},
                                [58,59,60,61]~\citet{TNS_2023xgo_discovery,TNS_2023xgo_class_Icn,TNS_2023xgo_class_Ibn,Gango_2023xgo},
                                [62,63]~\citet{TNS_2023abbd_discovery,TNS_2023abbd_class}.
                                }\\
\tablefoottext{f}{\footnotesize The extinction in $R/r$-like bands was estimated directly from the posterior samples of $E({B-V})_{\rm host}$ instead of the value tabulated here. Reported values SN~2018jmt, SN~2022ablq and SN~2023emq are retrieved from the respective references.}
}\label{tab:summary}
\end{sidewaystable*}
\begin{table*}
\centering
\caption{\label{tab:mosfit}Median and 1$\sigma$ confidence intervals of  10 (+ 2 derived) parameters of the CSI model implemented in {\tt MOSFiT} for the 24 SNe Ibn of the \mosfit{} sample.}
\resizebox{\textwidth}{!}{%
\begin{tabular}{lllllllllll|ll}
\hline\hline
Object$^{\square}$ & $M_{\rm CSM}$ & $M_{\rm ej}$ & $n$ & $A_{V,{\rm host}}$ & $R_0$ & $\log\rho_{\rm CSM}$ & $s$  & $t_{\rm exp}^{\dag}$ & $\sigma$ & $v_{\rm ej}$ & $\log\dot{M}^{\dag\dag}$ & $R_{\rm outer, CSM}^{\dag\dag\dag}$\\
&  [$10^{-1}$\Modot{}] & [\Modot{}] & & [$10^{-2}$mag] & [$10^{14}$cm] & [g/cm$^{3}$] & & [days] & [mag] & [10$^{3}$ km/s] & [\Modot/yr] & [$10^{16}$cm]\\
\hline
10al & $1.77^{+0.15}_{-0.13}$ & $0.35^{+0.06}_{-0.03}$ & $7.34^{+0.43}_{-0.22}$ & $0.30^{+4.34}_{-0.29}$ & $0.61^{+0.39}_{-0.28}$ & $-11.44^{+0.23}_{-0.20}$ & $0.64^{+0.19}_{-0.16}$ & $-18.50^{+0.16}_{-0.17}$ & $0.17^{+0.02}_{-0.01}$ & $3.86^{+0.08}_{-0.09}$ & $+0.13^{+0.04}_{-0.03}$ & $0.04^{+0.00}_{-0.00}$ \\
OGLE-12 & $0.47^{+12.45}_{-0.44}$ & $14.70^{+7.93}_{-5.74}$ & $7.06^{+0.09}_{-0.04}$ & $0.42^{+19.15}_{-0.41}$ & $8.94^{+1.52}_{-1.30}$ & $-16.59^{+0.23}_{-0.21}$ & $1.47^{+0.04}_{-0.06}$ & $-12.78^{+0.25}_{-0.27}$ & $0.20^{+0.01}_{-0.01}$ & $3.54^{+0.40}_{-0.26}$ & $-2.66^{+0.47}_{-0.21}$ & $5.38^{+51.74}_{-4.27}$ \\
OGLE-14 & $1.29^{+35.97}_{-1.24}$ & $0.25^{+0.34}_{-0.12}$ & $9.88^{+1.39}_{-0.72}$ & $0.03^{+0.69}_{-0.03}$ & $33.77^{+1.57}_{-2.09}$ & $-16.88^{+0.14}_{-0.09}$ & $1.44^{+0.03}_{-0.04}$ & $-48.94^{+2.27}_{-3.20}$ & $0.14^{+0.02}_{-0.01}$ & $3.93^{+0.40}_{-0.47}$ & $-1.60^{+0.53}_{-0.24}$ & $4.65^{+34.39}_{-3.91}$ \\
14av & $1.99^{+52.12}_{-1.93}$ & $0.13^{+0.04}_{-0.02}$ & $7.02^{+0.03}_{-0.02}$ & $0.07^{+3.07}_{-0.07}$ & $24.98^{+0.76}_{-0.86}$ & $-16.94^{+0.08}_{-0.04}$ & $1.60^{+0.01}_{-0.02}$ & $-8.63^{+0.25}_{-0.31}$ & $0.12^{+0.01}_{-0.01}$ & $9.30^{+0.44}_{-0.46}$ & $-2.34^{+0.41}_{-0.22}$ & $12.71^{+120.44}_{-11.51}$ \\
14aki & $2.43^{+52.49}_{-2.35}$ & $1.43^{+5.61}_{-1.09}$ & $9.69^{+1.22}_{-0.97}$ & $0.09^{+5.06}_{-0.08}$ & $16.46^{+1.97}_{-1.71}$ & $-16.35^{+0.25}_{-0.26}$ & $1.39^{+0.06}_{-0.08}$ & $-9.14^{+0.31}_{-0.42}$ & $0.19^{+0.02}_{-0.01}$ & $6.63^{+1.19}_{-1.18}$ & $-1.68^{+0.51}_{-0.24}$ & $5.75^{+36.95}_{-5.06}$ \\
15U & $1.50^{+0.56}_{-0.26}$ & $1.42^{+0.49}_{-0.27}$ & $8.95^{+0.99}_{-0.49}$ & $0.01^{+0.02}_{-0.00}$ & $4.61^{+0.68}_{-0.62}$ & $-12.27^{+0.09}_{-0.09}$ & $1.28^{+0.11}_{-0.15}$ & $-10.31^{+0.41}_{-0.58}$ & $0.18^{+0.01}_{-0.01}$ & $9.24^{+0.57}_{-0.96}$ & $-0.13^{+0.11}_{-0.04}$ & $0.07^{+0.01}_{-0.01}$ \\
15ul & $0.35^{+0.19}_{-0.09}$ & $1.41^{+4.47}_{-0.85}$ & $8.65^{+1.97}_{-1.24}$ & $0.02^{+0.23}_{-0.01}$ & $0.70^{+1.37}_{-0.43}$ & $-11.20^{+0.74}_{-0.73}$ & $1.94^{+0.04}_{-0.07}$ & $-5.04^{+0.12}_{-0.17}$ & $0.42^{+0.05}_{-0.03}$ & $4.66^{+0.54}_{-0.49}$ & $-0.29^{+0.16}_{-0.10}$ & $0.03^{+0.01}_{-0.01}$ \\
15dpn & $0.04^{+0.07}_{-0.02}$ & $0.23^{+0.17}_{-0.09}$ & $8.12^{+0.97}_{-0.58}$ & $8.78^{+14.08}_{-6.49}$ & $8.64^{+0.73}_{-0.59}$ & $-15.52^{+0.17}_{-0.19}$ & $1.22^{+0.07}_{-0.11}$ & $-9.83^{+0.37}_{-0.68}$ & $0.15^{+0.01}_{-0.01}$ & $3.77^{+0.44}_{-0.35}$ & $-1.86^{+0.24}_{-0.14}$ & $0.25^{+0.18}_{-0.08}$ \\
18jmt & $6.86^{+90.00}_{-6.66}$ & $1.34^{+0.95}_{-0.58}$ & $7.13^{+0.11}_{-0.08}$ & $0.14^{+1.52}_{-0.13}$ & $22.50^{+2.55}_{-2.17}$ & $-16.54^{+0.23}_{-0.25}$ & $1.47^{+0.04}_{-0.05}$ & $-5.29^{+0.36}_{-0.38}$ & $0.17^{+0.01}_{-0.01}$ & $5.01^{+0.48}_{-0.41}$ & $-2.00^{+0.40}_{-0.24}$ & $13.46^{+62.65}_{-12.18}$ \\
19uo & $0.60^{+0.42}_{-0.27}$ & $0.18^{+0.11}_{-0.06}$ & $8.86^{+1.57}_{-1.11}$ & $0.43^{+5.51}_{-0.42}$ & $0.74^{+0.72}_{-0.39}$ & $-10.55^{+0.39}_{-0.48}$ & $1.01^{+0.65}_{-0.74}$ & $-11.04^{+0.61}_{-0.82}$ & $0.26^{+0.02}_{-0.02}$ & $3.73^{+0.41}_{-0.38}$ & $-0.03^{+0.09}_{-0.05}$ & $0.01^{+0.01}_{-0.01}$ \\
19deh & $0.67^{+28.91}_{-0.64}$ & $1.82^{+1.22}_{-0.53}$ & $7.06^{+0.19}_{-0.05}$ & $0.10^{+5.81}_{-0.09}$ & $22.39^{+1.00}_{-1.21}$ & $-16.88^{+0.13}_{-0.09}$ & $1.56^{+0.03}_{-0.03}$ & $-8.83^{+0.05}_{-0.06}$ & $0.12^{+0.01}_{-0.01}$ & $4.15^{+0.20}_{-0.19}$ & $-2.37^{+0.51}_{-0.21}$ & $5.45^{+68.67}_{-4.64}$ \\
19kbj & $0.14^{+45.47}_{-0.12}$ & $0.35^{+0.35}_{-0.15}$ & $7.16^{+0.34}_{-0.11}$ & $0.04^{+1.05}_{-0.04}$ & $16.52^{+6.49}_{-2.30}$ & $-16.06^{+0.22}_{-0.45}$ & $1.36^{+0.07}_{-0.03}$ & $-4.38^{+0.75}_{-4.06}$ & $0.17^{+0.01}_{-0.01}$ & $5.72^{+1.54}_{-0.81}$ & $-2.54^{+0.74}_{-0.21}$ & $0.67^{+35.02}_{-0.41}$ \\
20bqj & $0.22^{+8.92}_{-0.20}$ & $0.37^{+0.18}_{-0.10}$ & $7.02^{+0.02}_{-0.01}$ & $0.02^{+0.31}_{-0.02}$ & $44.52^{+0.26}_{-0.50}$ & $-16.54^{+0.03}_{-0.03}$ & $1.42^{+0.01}_{-0.01}$ & $-8.76^{+0.72}_{-0.85}$ & $0.16^{+0.01}_{-0.01}$ & $4.14^{+0.31}_{-0.39}$ & $-1.77^{+0.69}_{-0.25}$ & $0.92^{+6.78}_{-0.41}$ \\
20nxt & $0.72^{+0.39}_{-0.18}$ & $0.16^{+0.07}_{-0.04}$ & $8.65^{+1.44}_{-1.05}$ & $0.05^{+1.34}_{-0.04}$ & $0.46^{+0.55}_{-0.28}$ & $-10.35^{+0.19}_{-0.26}$ & $0.28^{+0.37}_{-0.20}$ & $-13.28^{+0.89}_{-0.88}$ & $0.40^{+0.02}_{-0.01}$ & $4.16^{+0.41}_{-0.40}$ & $+0.33^{+0.10}_{-0.05}$ & $0.01^{+0.00}_{-0.00}$ \\
20able & $2.92^{+1.44}_{-1.06}$ & $0.42^{+0.39}_{-0.24}$ & $9.30^{+1.43}_{-1.32}$ & $12.59^{+8.58}_{-5.76}$ & $1.85^{+1.61}_{-0.86}$ & $-10.59^{+0.38}_{-0.41}$ & $1.44^{+0.40}_{-0.56}$ & $-21.37^{+0.35}_{-0.44}$ & $0.44^{+0.02}_{-0.02}$ & $3.21^{+0.18}_{-0.13}$ & $+0.53^{+0.12}_{-0.08}$ & $0.03^{+0.01}_{-0.01}$ \\
21jpk & $0.02^{+0.01}_{-0.01}$ & $12.47^{+11.13}_{-8.51}$ & $7.30^{+0.34}_{-0.21}$ & $0.01^{+0.54}_{-0.01}$ & $2.08^{+1.50}_{-1.02}$ & $-16.20^{+0.92}_{-0.55}$ & $1.71^{+0.20}_{-0.20}$ & $-7.81^{+0.25}_{-0.28}$ & $0.28^{+0.03}_{-0.04}$ & $6.17^{+1.28}_{-1.30}$ & $-3.76^{+0.42}_{-0.21}$ & $2.82^{+4.44}_{-1.80}$ \\
22ihx & $0.35^{+7.47}_{-0.32}$ & $2.58^{+1.71}_{-2.05}$ & $7.11^{+0.16}_{-0.08}$ & $0.15^{+6.74}_{-0.15}$ & $21.13^{+1.84}_{-1.64}$ & $-16.80^{+0.18}_{-0.14}$ & $1.50^{+0.04}_{-0.04}$ & $-12.20^{+0.25}_{-0.25}$ & $0.37^{+0.03}_{-0.03}$ & $3.46^{+2.16}_{-0.38}$ & $-2.39^{+0.46}_{-0.21}$ & $2.98^{+19.85}_{-2.28}$ \\
22pda & $0.11^{+1.23}_{-0.09}$ & $14.69^{+7.49}_{-5.17}$ & $8.24^{+0.79}_{-0.61}$ & $0.61^{+5.61}_{-0.59}$ & $20.68^{+1.57}_{-1.44}$ & $-15.70^{+0.15}_{-0.15}$ & $1.11^{+0.07}_{-0.07}$ & $-21.25^{+0.16}_{-0.17}$ & $0.18^{+0.01}_{-0.01}$ & $3.68^{+0.56}_{-0.51}$ & $-1.57^{+0.55}_{-0.23}$ & $0.36^{+0.86}_{-0.13}$ \\
22ablq & $1.50^{+1.00}_{-0.51}$ & $0.82^{+0.70}_{-0.32}$ & $9.35^{+0.99}_{-1.29}$ & $0.00^{+0.01}_{-0.00}$ & $0.41^{+0.40}_{-0.32}$ & $-10.26^{+0.18}_{-0.23}$ & $1.13^{+0.53}_{-0.73}$ & $-8.66^{+0.44}_{-0.51}$ & $0.28^{+0.02}_{-0.02}$ & $5.38^{+0.80}_{-0.55}$ & $+0.93^{+0.09}_{-0.07}$ & $0.02^{+0.01}_{-0.00}$ \\
23emq & $2.11^{+53.85}_{-2.04}$ & $1.51^{+4.79}_{-1.14}$ & $9.00^{+0.79}_{-0.70}$ & $0.09^{+4.03}_{-0.09}$ & $22.47^{+1.01}_{-1.08}$ & $-16.93^{+0.09}_{-0.05}$ & $1.65^{+0.02}_{-0.03}$ & $-8.89^{+0.31}_{-0.39}$ & $0.29^{+0.03}_{-0.03}$ & $6.52^{+1.35}_{-1.06}$ & $-2.60^{+0.38}_{-0.21}$ & $16.69^{+170.16}_{-15.26}$ \\
23iuc & $1.96^{+53.25}_{-1.90}$ & $7.17^{+3.69}_{-2.27}$ & $7.10^{+0.16}_{-0.07}$ & $0.03^{+0.51}_{-0.02}$ & $15.24^{+1.10}_{-1.34}$ & $-16.79^{+0.20}_{-0.14}$ & $1.51^{+0.04}_{-0.04}$ & $-5.66^{+0.14}_{-0.18}$ & $0.17^{+0.02}_{-0.01}$ & $3.19^{+0.27}_{-0.14}$ & $-2.65^{+0.46}_{-0.23}$ & $13.16^{+120.16}_{-11.86}$ \\
23qre & $0.67^{+25.24}_{-0.63}$ & $0.25^{+0.17}_{-0.10}$ & $7.11^{+0.14}_{-0.08}$ & $0.05^{+1.08}_{-0.05}$ & $18.37^{+1.36}_{-1.46}$ & $-16.75^{+0.19}_{-0.15}$ & $1.55^{+0.03}_{-0.04}$ & $-6.49^{+0.25}_{-0.30}$ & $0.07^{+0.02}_{-0.01}$ & $5.75^{+0.59}_{-0.56}$ & $-2.37^{+0.48}_{-0.21}$ & $5.33^{+63.81}_{-4.52}$ \\
23rau & $6.94^{+88.89}_{-6.60}$ & $0.19^{+1.34}_{-0.07}$ & $7.76^{+0.48}_{-0.42}$ & $0.09^{+45.56}_{-0.09}$ & $17.72^{+1.85}_{-2.11}$ & $-16.64^{+0.46}_{-0.25}$ & $1.37^{+0.04}_{-0.12}$ & $-12.25^{+1.44}_{-2.09}$ & $0.15^{+0.02}_{-0.01}$ & $6.23^{+1.50}_{-1.54}$ & $-1.52^{+0.42}_{-0.22}$ & $13.06^{+65.41}_{-11.38}$ \\
23xgo & $11.91^{+100.13}_{-11.57}$ & $1.01^{+0.35}_{-0.24}$ & $7.02^{+0.03}_{-0.01}$ & $0.03^{+0.41}_{-0.03}$ & $24.78^{+0.65}_{-0.81}$ & $-16.95^{+0.07}_{-0.04}$ & $1.68^{+0.02}_{-0.02}$ & $-6.21^{+0.17}_{-0.22}$ & $0.22^{+0.02}_{-0.02}$ & $6.51^{+0.38}_{-0.38}$ & $-2.31^{+0.24}_{-0.20}$ & $61.38^{+266.17}_{-57.15}$ \\
\hline
\end{tabular}
}
\tablefoot{
\tablefoottext{\dag}{\footnotesize The lower boundary of the distribution of $t_{\rm exp}$ is the only value that changes between each SN modeling. The prior distributions of the \mosfit{} modeling are either Uniform ($\mathcal{U}$) or log-Uniform ($\log \mathcal{U}$), except for the ejecta velocity which is Gaussian ($\mathcal{G}$). The specific prior distribution per parameter of the CSI model are $\kappa$: $\mathcal{U}(0.1,0.4)$ [cm$^{2}$/g]; 
$M_{\rm CSM}$: $\log\mathcal{U}(10^{-3},30)$~[\Modot]; 
$M_{\rm ej}$: $\log\mathcal{U}(0.1,30)$~[\Modot]; 
$n$: $\mathcal{U}(7,12)$; 
$n_{\rm H,host}$: $\log\mathcal{U}(10^{16},6\times 10^{21})$~[cm$^{-2}$]; 
$R_{0}$: $\log\mathcal{U}(2\times10^{12}, 5\times 10^{15})$~[cm]; 
$\rho_{\rm CSM}$: $\log\mathcal{U}(10^{-17},10^{-10})$~[g/cm$^{3}$]; 
$s$: $\log\mathcal{U}(0,2)$;
$T_{\rm min}$: $\log \mathcal{U}(1000,20000)$~[K]; 
$t_{\rm exp}$ :$\mathcal{U}(-20,0)$~[days], 
and $v_{\rm ej}$:$\log\mathcal{G}(\mu=6000,\sigma=2000)$~[km/s].}
\\
\tablefoottext{\dag\dag}{\footnotesize Average mass-loss rates are estimated from Eq.~\ref{eq:mdot} assuming $v_w$ as the narrow component ($v_{\rm narrow}$) of the \ion{He}{i}~$\lambda 5876$~\AA{} line in Table~\ref{tab:spectroscopic}. See Sect.~\ref{sec:specana} for more details.}\\
\tablefoottext{\dag\dag\dag}{\footnotesize The outer radius of the CSM ($R_{\rm outer, CSM}$) is estimated from Eq.~\ref{eq:rcsm}.}\\
\tablefoottext{$\square$}{\footnotesize Abbreviations for SN~2010al (10al), OGLE-2012-SN-006 (OGLE-12), OGLE-2014-SN-131 (OGLE-14), SN~2014av (14av), iPTF14aki (14aki), SN~2015U (15U), iPTF15ul (15ul), PS15dpn (15dpn), SN~2018jmt (18jmt), SN~2019uo (19uo), SN~2019deh (19deh), SN~2019kbj (19kbj), SN~2020bqj (20bqj), SN~2020nxt (20nxt), SN~2020able (20able), SN~2021jpk (21jpk), SN~2022ihx (22ihx), SN~2022pda (22pda), SN~2022ablq (22ablq), SN~2023emq (23emq), SN~2023iuc (23iuc), SN~2023qre (23qre), SN~2023rau (23rau) and SN~2023xgo (23xgo). References for each SN are listed in Table~\ref{tab:photo_props}.}\\
}
\end{table*}

\begin{table*}
\centering
\caption{\label{tab:xray} Results from the X-ray analysis of the SNe in the X-RAY sample with {\it Swift} XRT observations.}
\begin{tabular}{llccc}
\hline\hline
Object & Phase$^{a}$ & $t_{X}^{b}$ & $L_{X}$ & $\dot{M}^{d}$\\
& (days) & (days) &  ($10^{41}$erg/s) & (M$_{\odot}$/yr)\\
\hline
2019lsm & $7-43$ & 17.0 & $<1.4$ & $<1.0$\\
2020able & $0-24$ & 16.2 & $<0.3$ & $<0.4$\\
2020nxt$^{\dag}$ & $13-44$ & 27.3 & $0.1 \pm 0.1$ & $0.4 \pm 0.1$\\
2021iyt & $6-17$ & 10.4 & $<4.9$ & $<1.5$\\
2021zfx & 10 & 10.4 & $<8.3$ & $<1.9$\\
2022ablq & $15-20$ & 17.5 & $0.5 \pm 0.2$ & $0.6 \pm 0.1$\\
2022ablq & $63-73$ & 67.9 & $0.6 \pm 0.2$ & $1.3 \pm 0.2$\\
2022ablq & $83-97$ & 91.1 & $<0.3$ & $<1.1$\\
2022ablq & $106-114$ & 110.8 & $0.2 \pm 0.1$ & $1.0 \pm 0.2$\\
2022eux & $2-35$ & 15.8 & $<0.6$ & $<0.6$\\
2022ihx & $6-22$ & 11.9 & $<0.3$ & $<0.4$\\
2022pda & $70-260$ & 142.3 & $<0.6$ & $<1.9$\\
2022qhy$^{\dag}$ & $2-8$ & 4.6 &  $0.04 \pm 0.02$ & $0.09 \pm 0.02$\\
2023emq & $7-10$ & 8.3 & $<1.2$ & $<0.7$\\
2023emq & 11 & 11.1 & $1.9 \pm 1.2$ & $0.9 \pm 0.3$\\
2023emq & 26 & 25.9 & $<1.6$ & $<1.3$\\
2023iuc & $1-53$ & 12.2 & $<4.3$ & $<1.5$\\
2023qre & $6-82$ & 26.0 & $<2.8$ & $<1.7$\\
2023rau & $17-26$ & 21.9 & $<5.7$ & $<2.3$\\
2023xgo & $6-64$ & 24.3 & $<0.1$ & $<0.2$\\
\hline
\end{tabular}
\tablefoot{
\tablefoottext{a}{\footnotesize Ranges of the dates of the XRT observations relative to the discovery date of each SN~(Tab.~\ref{tab:summary}).}\\
\tablefoottext{b}{\footnotesize Duration of the X-ray emission derived from the average date of the XRT observations with respect to the discovery date of each SN .}\\
\tablefoottext{c}{\footnotesize X-ray luminosity obtained from the XRT observations at 1 keV.}\\
\tablefoottext{d}{\footnotesize Mass-loss rates estimated through Eq.~\ref{eq:xray}.}\\
\tablefoottext{$\dag$}{\footnotesize X-ray emission from this source is likely from the host galaxy.}
}
\end{table*}
\begin{table*}
\centering
\caption{\label{tab:spectroscopic} Minimum/Maximum velocity of the narrow/broad components of \hei{}~$\lambda5876$\AA{} line of 59 SNe Ibn.$^{\square}$}
\begin{tabular}{lllc}
\hline\hline
Object & $v_{\rm narrow}$ & $v_{\rm broad}$ & References$^{\dag}$\\
& (km/s) & (km/s) & \\
\hline
OGLE-2014$^{\dag\dag}$ & $2902^{+119}_{-126}$ & $5286^{+300}_{-280}$ & 2\\
2015G & $\sim$1300 & $\sim$5500 & 1\\
2015U & $1041^{+31}_{-30}$ & $\sim$3450 & 1,2\\
iPTF15akq & $1168^{+213}_{-147}$ & $-$ & 2,3\\
PS15dpn & $\sim$3000 & $-$ & 4,5\\
iPTF15ul & $1234^{+121}_{-162}$ & $-$ & 2,3\\
15ed & $\sim$1200 & $\sim$7000 & 1\\
2018bcc & $1232^{+44}_{-56}$ & $-$ & 2,6\\
2018jmt & $625^{+100}_{-100}$ & $8551^{+1168}_{-1052}$ & 2,7\\
2019cj & $\sim$740 & $\sim$5600 & 7\\
2019deh & $1124^{+606}_{-458}$ & $-$ & 2,8\\
2019kbj & $426^{+68}_{-85}$ & $\sim$8000 & 2,9\\
2019lsm & $1799^{+407}_{-403}$ & $-$ & 2\\
2019uo & $683^{+72}_{-88}$ & $4998^{+502}_{-524}$ & 2,10\\
2019wep & $915^{+59}_{-65}$ & $-$ & 2,11\\
2020able & $980^{+149}_{-149}$ & $4226^{+800}_{-656}$ & 2\\
2020bqj & $230^{+130}_{-130}$ & $4119^{+400}_{-368}$ & 2,12\\
2020nxt & $1062^{+56}_{-56}$ & $5890^{+233}_{-221}$ & 2,13\\
2022pda & $\sim$1900 & $4466^{+244}_{-403}$ & 2,14\\
2022qhy & $2367^{+31}_{-66}$ & $5860^{+413}_{-506}$ & 2\\
2022ablq & $2998^{+453}_{-955}$ & $5173^{+542}_{-691}$ & 2,15\\
2023emq & $600^{+457}_{-62}$ & $3600^{+99}_{-96}$ & 2,16\\
\hline
\end{tabular}
\tablefoot{
\tablefoottext{$\square$}{\footnotesize Full table available in electronic form.}\\
\tablefoottext{\dag}{\footnotesize [1]~\citet{Pastorello_2014av},[2]~This work,[3]~\citet{Hosseinzadeh_2017},[4]~\citet{Smartt_15dpn},[5]~\citet{Wang_15dpn},[6]~\citet{2018bcc_Karamehmetoglu},[7]~\citet{Wang_2024_Ibn},[8]~\citet{Pellegrino_2022_19deh},[9]~\citet{BenAmi_2022},[10]~\citet{Gangopadhyay_2020},[11]~\citet{Gango_2022},[12]~\citet{Kool_2021_2020bqj},
[13]~\citet{Qinan_2020nxt},[14] Cai in prep,[15]~\citet{Pellegrino_2022ablq},[16]~\citet{Pursiainen_2023}.
}\\
\tablefoottext{\dag\dag}{\footnotesize Abbreviation for OGLE-2014-SN-131 (OGLE-14).}
}
\end{table*}
\longtab[6]{
\begin{landscape}
\begin{longtable}{lccccccccccc}
\caption{\label{tab:photo_props} Photometric properties of our F25 + Literature sample of 61 SNe Ibn$^{\dag\dag\dag}$.}\\
\hline\hline
Object$^{\dag}$ & Band & $z$ & ${E(B-V)_{\rm tot}}$  &  MJD$_{\rm max}$ & M$_{\rm peak}$ & $\gamma_{-20}$ & $\gamma_{-10}$ & $\gamma_{+10}$ & $\gamma_{+20}$ & $\gamma_{+30}$ & References$^{\dag\dag}$\\
\cline{7-11}
 & & & (mag) & (days) & (mag) & & & (mag/day)& \\
\hline
1999cq$^{\dag}$ & $R$ & 0.026 & 0.15 & $51348.0\pm 3.0$ & $-19.93\pm 0.40$ & -- &  ${-0.996}^{+0.053}_{-0.053}$ &  ${0.050}^{+0.010}_{-0.010}$ &  ${0.192}^{+0.018}_{-0.018}$ & -- & 1,2\\
2000er$^{\dag}$ & $R$ & 0.031 & 0.11 & $51868.5\pm 5.0$ & $-19.89\pm 0.46$ & -- & -- &  ${0.083}^{+0.002}_{-0.002}$ & -- & -- & 2\\
2002ao$^{\dag}$ & $R$ & 0.005 & 0.25 & $52282.5\pm 4.0$ & $-18.04\pm 0.35$ & -- & -- & -- & -- &  ${0.123}^{+0.003}_{-0.003}$ & 2\\
2005la$^{\dag}$ & $R$ & 0.018 & 0.01 & $53694.0\pm 1.5$ & $-17.31\pm 0.13$ & -- & -- &  ${0.022}^{+0.005}_{-0.005}$ &  ${0.048}^{+0.003}_{-0.003}$ &  ${0.108}^{+0.009}_{-0.009}$ & 3\\
2006jc$^{\dag}$ & $R$ & 0.006 & 0.05 & $54008.0\pm 8.0$ & $-19.32\pm 0.91$ & -- & -- &  ${0.081}^{+0.002}_{-0.002}$ &  ${0.095}^{+0.001}_{-0.001}$ &  ${0.113}^{+0.013}_{-0.013}$ & 4,5,6\\
2010al & $R$ & 0.017 & 0.06 & $55285.4\pm 0.1$ & $-18.89\pm 0.05$ &  ${-0.300}^{+0.012}_{-0.012}$ &  ${-0.033}^{+0.003}_{-0.003}$ &  ${0.055}^{+0.006}_{-0.006}$ &  ${0.112}^{+0.008}_{-0.008}$ & -- & 7\\
2011hw$^{\dag}$ & $R$ & 0.023 & 0.10 & $55874.0\pm 10.0$ & $-19.01\pm 0.33$ & -- & -- & -- &  ${0.019}^{+0.002}_{-0.002}$ &  ${-0.014}^{+0.003}_{-0.003}$ & 7,8\\
PTF11rfh$^{\dag}$ & $R$ & 0.060 & 0.00 & $55911.0\pm 5.0$ & $-20.49\pm 0.99$ & -- & -- & -- & -- & -- & 9\\
PS1-12sk & $r$ & 0.054 & 0.03 & $56005.6\pm 0.1$ & $-19.25\pm 0.02$ & -- &  ${-0.053}^{+0.006}_{-0.006}$ &  ${0.095}^{+0.002}_{-0.002}$ &  ${0.106}^{+0.010}_{-0.010}$ &  ${0.110}^{+0.058}_{-0.058}$ & 10\\
PTF12ldy$^{\dag}$ & $R$ & 0.106 & 0.05 & $56243.9\pm 0.2$ & $-19.18\pm 0.03$ & -- &  ${-0.218}^{+0.009}_{-0.009}$ &  ${0.095}^{+0.007}_{-0.007}$ &  ${0.139}^{+0.056}_{-0.056}$ & -- & 9\\
LSQ12btw$^{\dag}$ & $g$ & 0.058 & 0.02 & $56013.1\pm 1.0$ & $-19.40\pm 0.09$ & -- & -- &  ${0.071}^{+0.004}_{-0.004}$ &  ${0.088}^{+0.005}_{-0.005}$ &  ${0.103}^{+0.011}_{-0.011}$ & 11\\
OGLE-12$^{\square}$ & $I$ & 0.057 & 0.07 & $56218.2\pm 0.0$ & $-19.79\pm 0.01$ & -- &  ${-0.092}^{+0.002}_{-0.002}$ &  ${0.044}^{+0.001}_{-0.001}$ &  ${0.047}^{+0.002}_{-0.002}$ &  ${0.024}^{+0.002}_{-0.002}$ & 12\\
LSQ13ccw & $V$ & 0.060 & 0.04 & $56539.9\pm 0.2$ & $-18.36\pm 0.05$ & -- &  ${-0.320}^{+0.049}_{-0.048}$ &  ${0.189}^{+0.008}_{-0.008}$ &  ${0.114}^{+0.045}_{-0.045}$ &  ${0.162}^{+0.061}_{-0.061}$ & 11\\
iPTF13beo & $R$ & 0.091 & 0.04 & $56434.5\pm 0.1$ & $-18.54\pm 0.06$ & -- &  ${-0.307}^{+0.027}_{-0.026}$ &  ${-0.011}^{+0.008}_{-0.008}$ &  ${0.114}^{+0.005}_{-0.005}$ &  ${0.041}^{+0.026}_{-0.026}$ & 13\\
2014av & $g/r$ & 0.030 & 0.02 & $56770.5\pm 0.1$ & $-19.58\pm 0.04$ & -- &  ${-0.186}^{+0.012}_{-0.012}$ &  ${0.133}^{+0.007}_{-0.007}$ &  ${0.140}^{+0.010}_{-0.010}$ &  ${0.058}^{+0.013}_{-0.013}$ & 14\\
2014bk$^{\dag}$  & $R$ & 0.070 & 0.05 & $56808.0\pm 5.0$ & $-19.99\pm 0.65$ & -- & -- &  ${0.057}^{+0.028}_{-0.028}$ &  ${0.093}^{+0.015}_{-0.014}$ &  ${0.133}^{+0.076}_{-0.077}$ & 14\\
iPTF14aki & $R$ & 0.064 & 0.03 & $56765.4\pm 0.1$ & $-19.27\pm 0.02$ & -- &  ${-0.191}^{+0.006}_{-0.006}$ &  ${0.098}^{+0.002}_{-0.002}$ &  ${0.125}^{+0.005}_{-0.005}$ & -- & 9\\
14ms$^{\dag,\square}$  & $V$ & 0.054 & 0.01 & $57023.4\pm 0.0$ & $-20.58\pm 0.04$ & -- &  ${0.058}^{+0.002}_{-0.002}$ &  ${0.093}^{+0.002}_{-0.002}$ &  ${0.112}^{+0.005}_{-0.005}$ &  ${0.044}^{+0.008}_{-0.008}$ & 15,16\\
OGLE-14$^{\square}$ & $I$ & 0.085 & 0.16 & $56970.8\pm 0.4$ & $-18.43\pm 0.04$ &  ${-0.031}^{+0.007}_{-0.007}$ &  ${-0.017}^{+0.005}_{-0.005}$ &  ${-0.008}^{+0.005}_{-0.005}$ &  ${0.006}^{+0.004}_{-0.004}$ &  ${0.023}^{+0.004}_{-0.003}$ & 17\\
15ed$^{\square}$ & $r$ & 0.049 & 0.14 & $57085.9\pm 0.6$ & $-20.09\pm 0.22$ & -- & -- &  ${0.102}^{+0.013}_{-0.013}$ &  ${0.122}^{+0.011}_{-0.012}$ &  ${0.090}^{+0.044}_{-0.044}$ & 18\\
2015G$^{\dag}$  & $r$ & 0.005 & 0.38 & $57100.0\pm 5.0$ & $-17.89\pm 0.50$ & -- & -- & -- &  ${0.100}^{+0.002}_{-0.002}$ &  ${0.083}^{+0.006}_{-0.006}$ & 19\\
2015U & $r$ & 0.014 & 0.86 & $57071.9\pm 0.0$ & $-19.86\pm 0.02$ & -- &  ${-0.098}^{+0.003}_{-0.003}$ &  ${0.086}^{+0.002}_{-0.002}$ &  ${0.186}^{+0.006}_{-0.006}$ & -- & 20,21\\
iPTF15ul & $g$ & 0.066 & 0.41 & $57094.8\pm 0.0$ & $-21.09\pm 0.11$ & -- &  ${-0.283}^{+0.010}_{-0.010}$ &  ${0.182}^{+0.004}_{-0.004}$ & -- & -- & 9\\
PS15dpn & $r$ & 0.175 & 0.10 & $57392.9\pm 0.3$ & $-20.09\pm 0.03$ & -- &  ${-0.055}^{+0.009}_{-0.009}$ &  ${0.070}^{+0.005}_{-0.005}$ &  ${0.046}^{+0.010}_{-0.010}$ &  ${0.092}^{+0.011}_{-0.011}$ & 22,23\\
iPTF15akq$^{\dag}$  & $R$ & 0.109 & 0.01 & $57134.0\pm 2.5$ & $-18.62\pm 0.31$ & -- &  ${-0.189}^{+0.017}_{-0.017}$ & -- & -- & -- & 9\\
2018bcc & $r$ & 0.064 & 0.01 & $58231.2\pm 0.1$ & $-19.58\pm 0.01$ & -- &  ${-0.264}^{+0.005}_{-0.005}$ &  ${0.082}^{+0.009}_{-0.009}$ &  ${0.181}^{+0.010}_{-0.010}$ &  ${0.066}^{+0.063}_{-0.064}$ & 24\\
2018jmt & $r$ & 0.032 & 0.10 & $58465.1\pm 0.1$ & $-18.83\pm 0.02$ & -- &  ${-0.083}^{+0.005}_{-0.005}$ &  ${0.080}^{+0.002}_{-0.002}$ &  ${0.112}^{+0.002}_{-0.002}$ &  ${0.041}^{+0.006}_{-0.006}$ & 25,26,38\\
2019cj & $o$ & 0.044 & 0.02 & $58494.5\pm 0.4$ & $-18.64\pm 0.04$ & -- &  ${-0.078}^{+0.006}_{-0.006}$ &  ${0.059}^{+0.011}_{-0.011}$ &  ${0.181}^{+0.202}_{-0.203}$ &  ${0.003}^{+0.046}_{-0.046}$ & 25,26\\
2019deh & $r$ & 0.055 & 0.02 & $58588.5\pm 0.1$ & $-19.51\pm 0.02$ & -- &  ${-0.144}^{+0.004}_{-0.004}$ &  ${0.125}^{+0.004}_{-0.004}$ &  ${0.156}^{+0.011}_{-0.011}$ &  ${0.015}^{+0.060}_{-0.060}$ & 27\\
2019kbj & $r$ & 0.048 & 0.05 & $58669.6\pm 0.4$ & $-19.18\pm 0.01$ & -- & -- &  ${0.096}^{+0.002}_{-0.002}$ &  ${0.133}^{+0.007}_{-0.007}$ &  ${0.034}^{+0.083}_{-0.082}$ & 28\\
2019lsm & $o$ & 0.042 & 0.28 & $58694.8\pm 0.5$ & $-18.73\pm 0.47$ & -- &  ${-0.089}^{+0.013}_{-0.013}$ &  ${0.075}^{+0.025}_{-0.025}$ & -- & -- & 38\\
2019myn & $r$ & 0.100 & 0.04 & $58707.8\pm 0.1$ & $-19.36\pm 0.03$ & -- &  ${-0.121}^{+0.008}_{-0.008}$ &  ${0.137}^{+0.007}_{-0.007}$ &  ${0.111}^{+0.024}_{-0.025}$ & -- & 29\\
2019php & $r$ & 0.087 & 0.12 & $58731.1\pm 0.1$ & $-19.09\pm 0.04$ & -- &  ${-0.184}^{+0.018}_{-0.018}$ &  ${0.116}^{+0.008}_{-0.008}$ &  ${0.086}^{+0.042}_{-0.041}$ & -- & 29\\
2019qav & $r$ & 0.135 & 0.02 & $58741.9\pm 0.4$ & $-19.90\pm 0.09$ & -- &  ${-0.149}^{+0.013}_{-0.013}$ &  ${0.124}^{+0.009}_{-0.009}$ &  ${0.136}^{+0.036}_{-0.036}$ & -- & 29\\
2019rii & $r$ & 0.123 & 0.09 & $58757.1\pm 0.1$ & $-19.95\pm 0.03$ & -- &  ${-0.204}^{+0.016}_{-0.016}$ &  ${0.103}^{+0.005}_{-0.005}$ & -- & -- & 29\\
2019uo & $r$ & 0.020 & 0.08 & $58509.3\pm 0.1$ & $-18.41\pm 0.06$ & -- &  ${-0.082}^{+0.005}_{-0.005}$ &  ${0.169}^{+0.003}_{-0.002}$ &  ${0.030}^{+0.041}_{-0.041}$ & -- & 30\\
2019wep$^{\dag}$  & $r$ & 0.025 & 0.02 & $58828.5\pm 2.0$ & $-17.87\pm 0.03$ & -- & -- &  ${0.069}^{+0.002}_{-0.002}$ &  ${0.084}^{+0.002}_{-0.002}$ &  ${0.076}^{+0.008}_{-0.008}$ & 31\\
2020bqj$^{\dag}$  & $r/R$ & 0.066 & 0.02 & $58884.5\pm 0.0$ & $-19.07\pm 0.06$ & -- & -- & -- &  ${0.007}^{+0.002}_{-0.002}$ &  ${0.003}^{+0.001}_{-0.001}$ & 32\\
\hline
\endfirsthead\\
\caption{continued}\\
\hline
Object$^{\dag}$ & Band & $z$ & ${E(B-V)_{\rm tot}}$  &  MJD$_{\rm max}$ & M$_{\rm peak}$ & $\gamma_{-20}$ & $\gamma_{-10}$ & $\gamma_{+10}$ & $\gamma_{+20}$ & $\gamma_{+30}$ & References$^{\dag\dag}$\\
\cline{7-11}
 & & & (mag) & (days) & (mag) & & & (mag/day)& \\
\hline
2020nxt & $o$ & 0.022 & 0.07 & $59039.0\pm 0.0$ & $-19.13\pm 0.05$ & -- &  ${-0.214}^{+0.006}_{-0.006}$ &  ${0.171}^{+0.009}_{-0.009}$ &  ${0.157}^{+0.023}_{-0.023}$ &  ${0.024}^{+0.055}_{-0.055}$ & 33\\
2020taz & $o$ & 0.050 & 0.17 & $59112.5\pm 0.4$ & $-18.21\pm 0.14$ & -- &  ${-0.050}^{+0.009}_{-0.009}$ &  ${0.031}^{+0.008}_{-0.008}$ &  ${0.054}^{+0.012}_{-0.012}$ & -- & 38\\
2020able & $o$ & 0.025 & 0.26 & $59205.0\pm 0.1$ & $-19.17\pm 0.26$ &  ${-0.172}^{+0.007}_{-0.007}$ &  ${-0.019}^{+0.006}_{-0.006}$ &  ${0.031}^{+0.004}_{-0.004}$ &  ${0.066}^{+0.007}_{-0.007}$ &  ${0.056}^{+0.119}_{-0.120}$ & 38\\
2021foa$^{\dag\dag\dag}$ & $r$ & 0.009 & 0.20 & $59302.3\pm 0.1$ & $-18.34\pm 0.03$ &  ${-0.125}^{+0.020}_{-0.020}$ &  ${-0.037}^{+0.002}_{-0.002}$ &  ${0.042}^{+0.002}_{-0.002}$ &  ${0.032}^{+0.001}_{-0.001}$ &  ${0.096}^{+0.003}_{-0.003}$ & 34\\
2021iyt & $r$ & 0.071 & 0.19 & $59318.2\pm 0.1$ & $-19.24\pm 0.36$ & -- &  ${-0.257}^{+0.020}_{-0.019}$ &  ${0.195}^{+0.031}_{-0.031}$ &  ${0.061}^{+0.051}_{-0.051}$ & -- & 38\\
2021jpk & $r$ & 0.038 & 0.02 & $59324.2\pm 0.3$ & $-17.49\pm 0.06$ & -- &  ${-0.111}^{+0.016}_{-0.016}$ &  ${0.048}^{+0.023}_{-0.023}$ & -- & -- & 27\\
2021wvg & $o$ & 0.085 & 0.26 & $59454.3\pm 0.3$ & $-19.92\pm 0.28$ & -- &  ${-0.114}^{+0.030}_{-0.030}$ &  ${0.089}^{+0.013}_{-0.013}$ & -- & -- & 38\\
2021zfx & $o$ & 0.069 & 0.08 & $59478.4\pm 3.8$ & $-19.11\pm 0.11$ & -- & -- &  ${0.106}^{+0.013}_{-0.013}$ & -- & -- & 38\\
2022acm & $o$ & 0.056 & 0.19 & $59618.1\pm 0.2$ & $-20.62\pm 0.20$ & -- &  ${-0.035}^{+0.003}_{-0.003}$ &  ${0.025}^{+0.003}_{-0.003}$ &  ${0.062}^{+0.003}_{-0.003}$ &  ${0.081}^{+0.006}_{-0.006}$ & 38\\
2022eux & $o$ & 0.025 & 0.30 & $59659.7\pm 0.2$ & $-19.42\pm 0.12$ & -- &  ${-0.080}^{+0.006}_{-0.006}$ &  ${0.015}^{+0.035}_{-0.035}$ &  ${0.163}^{+0.043}_{-0.043}$ &  ${0.107}^{+0.014}_{-0.014}$ & 38\\
2022gzg & $r$ & 0.089 & 0.20 & $59682.9\pm 0.3$ & $-20.11\pm 0.42$ & -- &  ${-0.158}^{+0.028}_{-0.028}$ &  ${0.132}^{+0.056}_{-0.056}$ & -- & -- & 38\\
2022ihx & $r$ & 0.033 & 0.22 & $59701.6\pm 0.1$ & $-18.22\pm 0.24$ & -- &  ${-0.082}^{+0.009}_{-0.009}$ &  ${0.058}^{+0.009}_{-0.009}$ &  ${0.141}^{+0.025}_{-0.025}$ & -- & 38\\
2022pda & $o$ & 0.062 & 0.12 & $59861.1\pm 0.1$ & $-20.10\pm 0.22$ &  ${-0.160}^{+0.007}_{-0.007}$ &  ${-0.030}^{+0.003}_{-0.003}$ &  ${0.035}^{+0.003}_{-0.003}$ &  ${0.028}^{+0.004}_{-0.004}$ &  ${0.016}^{+0.005}_{-0.005}$ & 35,38\\
2022qhy$^{\star}$ & $V/r$ & 0.006 & 0.03 & $59794.0\pm -1.0$ & $-17.04\pm 0.04$ & -- & -- &  ${0.081}^{+0.009}_{-0.009}$ &  ${0.161}^{+0.012}_{-0.012}$ &  ${0.083}^{+0.048}_{-0.048}$ & 38\\
2022ablq & $g$ & 0.014 & 0.21 & $59914.1\pm 1.0$ & $-19.45\pm 0.02$ & -- &  ${-0.236}^{+0.017}_{-0.017}$ &  ${0.075}^{+0.001}_{-0.001}$ &  ${0.102}^{+0.002}_{-0.002}$ &  ${0.069}^{+0.005}_{-0.005}$ & 36,38\\
2023emq & $o$ & 0.033 & 0.11 & $60041.2\pm 0.1$ & $-18.63\pm 0.24$ & -- &  ${-0.213}^{+0.022}_{-0.022}$ &  ${0.173}^{+0.020}_{-0.020}$ & -- & -- & 37,38\\
2023iuc & $r$ & 0.076 & 0.08 & $60087.5\pm 0.2$ & $-19.76\pm 0.15$ & -- &  ${-0.048}^{+0.011}_{-0.011}$ &  ${0.077}^{+0.008}_{-0.008}$ &  ${0.098}^{+0.051}_{-0.050}$ &  ${0.050}^{+0.029}_{-0.029}$ & 38\\
2023qre & $r$ & 0.047 & 0.08 & $60185.3\pm 0.1$ & $-18.98\pm 0.04$ & -- &  ${-0.120}^{+0.015}_{-0.015}$ &  ${0.153}^{+0.008}_{-0.008}$ &  ${0.142}^{+0.033}_{-0.034}$ &  ${0.121}^{+0.050}_{-0.050}$ & 38\\
2023rau & $r$ & 0.067 & 0.21 & $60193.7\pm 0.3$ & $-19.14\pm 0.15$ & -- &  ${-0.071}^{+0.013}_{-0.013}$ &  ${0.067}^{+0.008}_{-0.008}$ &  ${0.032}^{+0.015}_{-0.015}$ &  ${0.063}^{+0.031}_{-0.031}$ & 38\\
2023ubp & $r$ & 0.056 & 0.05 & $60226.5\pm 0.1$ & $-19.46\pm 0.04$ & -- &  ${-0.180}^{+0.010}_{-0.010}$ &  ${0.123}^{+0.005}_{-0.005}$ &  ${0.058}^{+0.026}_{-0.026}$ & -- & 38\\
2023utc & $r$ & 0.013 & 0.01 & $60234.5\pm 0.1$ & $-16.35\pm 0.04$ & -- &  ${-0.112}^{+0.012}_{-0.012}$ &  ${0.095}^{+0.006}_{-0.006}$ &  ${0.086}^{+0.010}_{-0.010}$ &  ${0.111}^{+0.026}_{-0.027}$ & 38\\
2023vwh & $o/r$ & 0.060 & 0.08 & $60252.4\pm 0.1$ & $-19.89\pm 0.03$ & -- &  ${-0.138}^{+0.016}_{-0.016}$ &  ${0.087}^{+0.009}_{-0.009}$ &  ${0.067}^{+0.021}_{-0.022}$ & -- & 38\\
2023xgo & $c/o/r$ & 0.013 & 0.14 & $60262.4\pm 1.8$ & $-17.97\pm 0.23$ & -- &  ${-0.200}^{+0.014}_{-0.014}$ &  ${0.085}^{+0.036}_{-0.036}$ &  ${0.173}^{+0.033}_{-0.033}$ & -- & 38\\
2023abbd$^{\star}$ & $o$ & 0.042 & 0.10 & $60302.5\pm 0.0$ & $-18.78\pm 0.07$ & -- & -- &  ${0.172}^{+0.012}_{-0.012}$ &  ${0.028}^{+0.022}_{-0.021}$ & -- & 38\\
\hline
\end{longtable}
\tablefoot{
\tablefoottext{\dag}{\footnotesize The peak time and the absolute magnitude of the SNe marked with the {\dag} symbol were retrieved from the corresponding references. Nevertheless, the photometric slopes $\gamma_{-20},\gamma_{-10},\gamma_{+10},\gamma_{+20},\gamma_{+30}$ were always calculated in this work.}\\
\tablefoottext{\dag\dag}{\footnotesize [1]~\citet{Matheson_SN1999cq},
                                       [2]~\citet{Pastorello_00er_02ao_06jc},
                                       [3]~\citet{Pastorello_2008_2005la},
                                       [4]~\citet{Foley_2006jc},
                                       [5]~\citet{Pastorello_2007_2006jc},
                                       [6]~\citet{Anupama_2009_2006jc},
                                       [7]~\citet{Pastorello_2015_2011hw},
                                       [8]~\citet{Smith_2012_2011hw},
                                       [9]~\citet{Hosseinzadeh_2017},
                                       [10]~\citet{Sanders},
                                       [11]~\citet{Pastorello_2015_LSQ},
                                       [12]~\citet{Pastorello_OGLE12},
                                       [13]~\citet{Gorbikov_13beo},
                                       [14]~\citet{Pastorello_2014av},
                                       [15]~\citet{ASASN_14ms_Vallely},
                                       [16]~\citet{ASASSN14ms},
                                       [17]~\citet{Karamehmetoglu_2017},
                                       [18]~\citet{Pastorello_ASASSN15ed},
                                       [19]~\citet{Shivvers_2015G},
                                       [20]~\citet{Shivvers_2015U},
                                       [21]~\citet{Pasto_PSN},
                                       [22]~\citet{Smartt_15dpn},
                                       [23]~\citet{Wang_15dpn},
                                       [24]~\citet{2018bcc_Karamehmetoglu},
                                       [25]~\citet{Wang_2024_Ibn},
                                       [26]~\citet{2018jmt_photometry}
                                       [27]~\citet{Pellegrino_2022_19deh},
                                       [28]~\citet{BenAmi_2022},
                                       [29]~\citet{Ho_2021},
                                       [30]~\citet{Gangopadhyay_2020},
                                       [31]~\citet{Gango_2022},
                                       [32]~\citet{Kool_2021_2020bqj},
                                       [33]~\citet{Qinan_2020nxt},
                                       [34]~\citet{Farias_2024},
                                       [35]~Cai in prep.,
                                       [36]~\citet{Pellegrino_2022ablq},
                                       [37]~\citet{Pursiainen_2023},
                                       [38]~This work.
}\\
\tablefoottext{$\square$}{\footnotesize Abbreviations for OGLE-2012-SN-006 (OGLE-12), ASASSN-14ms (14ms), OGLE-2014-SN-131 (OGLE-14) and ASASSN-15ed (15ed).}\\
\tablefoottext{$\star$}{\footnotesize The peak time and corresponding magnitude of SN~2022qhy and SN~2023abbd correspond to the first observation in $V$ and $o$ bands of those SNe, respectively.}\\
\tablefoottext{$\dag\dag\dag$}{\footnotesize SN~2021foa is a transitional type IIn/Ibn transient. We do not include this source in the sample of 61 SNe Ibn.}
}
\end{landscape}
}
\begin{sidewaystable*}
\centering
\caption{\label{tab:nimosfit}Median and 1$\sigma$ confidence intervals of  10 (+ 3 derived) parameters of the RD+CSI model implemented in {\tt MOSFiT} for the 24 SNe Ibn in the \mosfit{} sample.}
\resizebox{\textwidth}{!}{%
\begin{tabular}{lllllllllll|lll}
\hline\hline
Object$^{\square}$ & $M_{\rm CSM}$ & $M_{\rm ej}$ & $n$ & $A_{V,{\rm host}}$ & $R_0$ & $\log\rho_{\rm CSM}$ & $s$  & $t_{\rm exp}^{\dag}$ & $\sigma$ & $v_{\rm ej}$ & $M_{\rm Ni}^{\dag\dag}$ & $\log\dot{M}^{\dag\dag\dag}$ & $R_{\rm outer,CSM}^{\dag\dag\dag\dag}$\\
&  [$10^{-1}$\Modot{}] & [\Modot{}] & & [$10^{-2}$mag] & [$10^{14}$cm] & [g/cm$^{3}$] & & [days] & [mag] & [10$^{3}$ km/s] & [$10^{-2}$\Modot] & [\Modot/yr] & [$10^{16}$cm]\\
\hline
10al & $1.46^{+0.16}_{-0.11}$ & $0.47^{+0.07}_{-0.06}$ & $8.48^{+0.58}_{-0.57}$ & $0.39^{+7.64}_{-0.39}$ & $0.07^{+0.16}_{-0.04}$ & $-11.31^{+0.32}_{-0.31}$ & $0.48^{+0.10}_{-0.10}$ & $-19.20^{+0.28}_{-0.25}$ & $0.16^{+0.01}_{-0.01}$ & $3.73^{+0.10}_{-0.09}$ & $0.26^{+0.05}_{-0.04}$ & $+0.02^{+0.05}_{-0.03}$ & $0.04^{+0.00}_{-0.00}$\\
OGLE-12 & $48.26^{+37.28}_{-22.25}$ & $0.19^{+0.08}_{-0.06}$ & $10.12^{+0.44}_{-0.49}$ & $133.77^{+35.99}_{-111.39}$ & $1.89^{+3.81}_{-1.35}$ & $-15.37^{+0.62}_{-0.50}$ & $1.09^{+0.17}_{-0.10}$ & $-13.43^{+0.39}_{-0.34}$ & $0.20^{+0.01}_{-0.02}$ & $8.36^{+1.28}_{-0.66}$ & $3.84^{+2.56}_{-2.76}$ & $-1.08^{+0.30}_{-0.25}$ & $17.93^{+33.54}_{-9.77}$\\
OGLE-14 & $1.52^{+48.33}_{-1.47}$ & $1.39^{+0.28}_{-0.28}$ & $9.64^{+1.54}_{-1.69}$ & $0.13^{+2.89}_{-0.13}$ & $0.21^{+2.41}_{-0.17}$ & $-16.26^{+0.80}_{-0.52}$ & $1.10^{+0.10}_{-0.08}$ & $-47.86^{+0.79}_{-0.94}$ & $0.25^{+0.04}_{-0.03}$ & $10.76^{+1.14}_{-1.21}$ & $0.58^{+0.07}_{-0.06}$ & $-2.26^{+0.89}_{-0.27}$ & $26.39^{+201.97}_{-23.20}$\\
14av & $0.18^{+0.04}_{-0.03}$ & $0.13^{+0.04}_{-0.02}$ & $11.12^{+0.59}_{-0.78}$ & $0.45^{+3.44}_{-0.44}$ & $0.29^{+0.11}_{-0.08}$ & $-10.16^{+0.10}_{-0.14}$ & $0.97^{+0.52}_{-0.58}$ & $-8.19^{+0.19}_{-0.20}$ & $0.14^{+0.01}_{-0.01}$ & $6.22^{+0.28}_{-0.29}$ & $0.55^{+0.02}_{-0.02}$ & $-0.08^{+0.09}_{-0.06}$ & $0.01^{+0.00}_{-0.00}$\\
14aki & $1.65^{+0.61}_{-0.43}$ & $0.25^{+0.23}_{-0.11}$ & $9.45^{+1.43}_{-1.13}$ & $66.26^{+10.59}_{-11.22}$ & $1.69^{+1.52}_{-0.81}$ & $-11.82^{+0.38}_{-0.31}$ & $1.01^{+0.34}_{-0.38}$ & $-9.45^{+0.64}_{-0.88}$ & $0.19^{+0.02}_{-0.02}$ & $4.97^{+0.83}_{-0.84}$ & $8.00^{+4.12}_{-3.31}$ & $+0.23^{+0.15}_{-0.09}$ & $0.05^{+0.01}_{-0.01}$\\
15U & $3.10^{+0.98}_{-0.86}$ & $2.81^{+2.23}_{-1.07}$ & $8.33^{+0.74}_{-0.69}$ & $0.00^{+0.02}_{-0.00}$ & $5.01^{+2.38}_{-1.88}$ & $-11.85^{+0.25}_{-0.22}$ & $1.39^{+0.27}_{-0.31}$ & $-11.80^{+0.72}_{-0.86}$ & $0.11^{+0.01}_{-0.01}$ & $7.58^{+0.58}_{-0.50}$ & $0.03^{+0.20}_{-0.03}$ & $+0.19^{+0.12}_{-0.10}$ & $0.06^{+0.02}_{-0.01}$\\
15ul & $1.95^{+2.26}_{-0.94}$ & $1.08^{+1.65}_{-0.62}$ & $8.92^{+1.82}_{-1.28}$ & $0.13^{+0.91}_{-0.13}$ & $6.62^{+6.64}_{-3.19}$ & $-10.75^{+0.50}_{-0.70}$ & $1.39^{+0.44}_{-0.54}$ & $-4.97^{+0.08}_{-0.10}$ & $0.35^{+0.03}_{-0.03}$ & $4.48^{+0.71}_{-0.51}$ & $2.62^{+5.46}_{-2.40}$ & $+0.04^{+0.13}_{-0.08}$ & $0.07^{+0.07}_{-0.03}$\\
15dpn & $2.88^{+65.12}_{-2.78}$ & $0.28^{+0.24}_{-0.14}$ & $8.48^{+0.47}_{-0.58}$ & $71.25^{+11.35}_{-9.41}$ & $18.91^{+2.22}_{-2.67}$ & $-16.60^{+0.38}_{-0.26}$ & $1.41^{+0.05}_{-0.07}$ & $-9.71^{+0.47}_{-0.46}$ & $0.11^{+0.01}_{-0.01}$ & $8.76^{+0.82}_{-1.28}$ & $0.11^{+0.16}_{-0.06}$ & $-1.47^{+0.55}_{-0.25}$ & $7.62^{+44.23}_{-6.59}$\\
18jmt & $3.69^{+1.07}_{-0.80}$ & $4.94^{+5.32}_{-1.97}$ & $7.44^{+0.37}_{-0.32}$ & $95.42^{+9.35}_{-8.88}$ & $1.38^{+0.78}_{-0.59}$ & $-11.32^{+0.39}_{-0.33}$ & $1.64^{+0.08}_{-0.07}$ & $-5.67^{+0.53}_{-0.59}$ & $0.20^{+0.01}_{-0.01}$ & $3.55^{+0.54}_{-0.29}$ & $1.33^{+0.80}_{-0.52}$ & $+0.07^{+0.11}_{-0.07}$ & $0.06^{+0.01}_{-0.01}$\\
19uo & $1.14^{+0.80}_{-0.45}$ & $0.16^{+0.08}_{-0.05}$ & $8.61^{+1.31}_{-1.02}$ & $0.02^{+0.58}_{-0.02}$ & $2.62^{+1.94}_{-1.09}$ & $-11.14^{+0.50}_{-0.34}$ & $1.45^{+0.32}_{-0.40}$ & $-10.49^{+0.38}_{-0.53}$ & $0.18^{+0.02}_{-0.01}$ & $3.49^{+0.26}_{-0.27}$ & $0.26^{+0.04}_{-0.03}$ & $-0.10^{+0.13}_{-0.09}$ & $0.03^{+0.02}_{-0.01}$\\
19deh & $0.37^{+0.04}_{-0.03}$ & $0.30^{+0.08}_{-0.09}$ & $10.44^{+1.01}_{-1.50}$ & $25.31^{+5.43}_{-6.04}$ & $0.10^{+0.04}_{-0.03}$ & $-10.21^{+0.13}_{-0.17}$ & $0.82^{+0.18}_{-0.17}$ & $-9.16^{+0.07}_{-0.07}$ & $0.12^{+0.01}_{-0.01}$ & $3.13^{+0.07}_{-0.07}$ & $0.46^{+0.06}_{-0.06}$ & $+0.03^{+0.03}_{-0.02}$ & $0.01^{+0.00}_{-0.00}$\\
19kbj & $0.40^{+0.14}_{-0.10}$ & $0.17^{+0.04}_{-0.03}$ & $7.19^{+0.20}_{-0.13}$ & $16.06^{+4.93}_{-3.67}$ & $0.62^{+0.34}_{-0.18}$ & $-10.64^{+0.20}_{-0.15}$ & $0.51^{+0.34}_{-0.30}$ & $-4.59^{+0.32}_{-0.31}$ & $0.15^{+0.01}_{-0.01}$ & $3.14^{+0.14}_{-0.10}$ & $0.71^{+0.10}_{-0.08}$ & $-0.31^{+0.06}_{-0.05}$ & $0.01^{+0.00}_{-0.00}$\\
20bqj & $4.96^{+0.79}_{-0.77}$ & $22.66^{+3.64}_{-3.00}$ & $10.15^{+0.64}_{-0.67}$ & $152.93^{+8.19}_{-5.07}$ & $0.40^{+0.33}_{-0.16}$ & $-10.42^{+0.28}_{-0.28}$ & $0.91^{+0.10}_{-0.12}$ & $-8.58^{+1.39}_{-1.79}$ & $0.29^{+0.01}_{-0.01}$ & $4.30^{+1.14}_{-0.45}$ & $0.64^{+0.94}_{-0.41}$ & $+1.07^{+0.05}_{-0.04}$ & $0.03^{+0.01}_{-0.00}$\\
20nxt & $0.68^{+0.55}_{-0.25}$ & $0.60^{+0.58}_{-0.46}$ & $9.00^{+1.13}_{-1.18}$ & $56.10^{+34.50}_{-26.38}$ & $0.41^{+1.53}_{-0.29}$ & $-10.43^{+0.27}_{-0.35}$ & $0.75^{+1.01}_{-0.53}$ & $-10.12^{+1.56}_{-1.87}$ & $0.47^{+0.01}_{-0.02}$ & $5.46^{+1.21}_{-1.34}$ & $0.82^{+0.83}_{-0.44}$ & $+0.27^{+0.05}_{-0.05}$ & $0.01^{+0.01}_{-0.00}$\\
20able & $2.74^{+1.36}_{-1.08}$ & $0.33^{+0.31}_{-0.14}$ & $8.52^{+1.50}_{-0.98}$ & $0.19^{+4.09}_{-0.18}$ & $2.28^{+1.78}_{-1.08}$ & $-10.68^{+0.43}_{-0.41}$ & $1.13^{+0.53}_{-0.64}$ & $-21.47^{+0.36}_{-0.36}$ & $0.40^{+0.02}_{-0.02}$ & $3.17^{+0.20}_{-0.12}$ & $0.11^{+0.51}_{-0.11}$ & $+0.43^{+0.10}_{-0.07}$ & $0.03^{+0.01}_{-0.01}$\\
21jpk & $0.08^{+9.93}_{-0.06}$ & $10.19^{+10.20}_{-8.61}$ & $8.15^{+1.84}_{-0.54}$ & $53.62^{+22.03}_{-50.78}$ & $20.63^{+2.75}_{-7.10}$ & $-16.62^{+0.51}_{-0.23}$ & $1.40^{+0.06}_{-0.16}$ & $-10.45^{+0.47}_{-0.99}$ & $0.18^{+0.03}_{-0.02}$ & $4.47^{+0.59}_{-0.52}$ & $0.03^{+0.04}_{-0.02}$ & $-2.67^{+1.11}_{-0.18}$ & $0.99^{+8.26}_{-0.52}$\\
22ihx & $2.45^{+1.20}_{-0.77}$ & $2.92^{+7.00}_{-2.43}$ & $8.72^{+1.18}_{-1.11}$ & $113.54^{+14.28}_{-14.85}$ & $0.63^{+0.93}_{-0.30}$ & $-10.69^{+0.43}_{-0.62}$ & $1.80^{+0.10}_{-0.36}$ & $-12.42^{+0.23}_{-0.25}$ & $0.34^{+0.03}_{-0.03}$ & $3.56^{+0.58}_{-0.43}$ & $0.40^{+2.62}_{-0.38}$ & $+0.28^{+0.14}_{-0.07}$ & $0.04^{+0.01}_{-0.01}$\\
22pda & $1.79^{+32.45}_{-1.68}$ & $0.19^{+0.04}_{-0.04}$ & $10.83^{+0.85}_{-1.29}$ & $0.91^{+10.67}_{-0.90}$ & $24.03^{+6.99}_{-2.85}$ & $-15.94^{+0.40}_{-0.73}$ & $1.71^{+0.18}_{-0.23}$ & $-21.14^{+0.26}_{-0.30}$ & $0.18^{+0.01}_{-0.01}$ & $3.84^{+1.01}_{-0.55}$ & $4.26^{+0.60}_{-0.37}$ & $-1.36^{+0.44}_{-0.23}$ & $3.38^{+27.17}_{-2.93}$\\
22ablq & $4.57^{+1.79}_{-2.10}$ & $0.91^{+0.54}_{-0.50}$ & $9.23^{+1.23}_{-1.11}$ & $0.00^{+0.03}_{-0.00}$ & $2.39^{+1.00}_{-0.70}$ & $-11.22^{+0.31}_{-0.17}$ & $1.74^{+0.19}_{-0.39}$ & $-8.85^{+0.61}_{-0.65}$ & $0.22^{+0.03}_{-0.02}$ & $4.52^{+0.80}_{-0.52}$ & $2.65^{+0.50}_{-0.45}$ & $+0.95^{+0.09}_{-0.11}$ & $0.05^{+0.01}_{-0.01}$\\
23emq & $0.21^{+0.14}_{-0.09}$ & $0.25^{+0.08}_{-0.07}$ & $9.26^{+1.53}_{-1.32}$ & $3.88^{+6.13}_{-2.98}$ & $0.16^{+0.24}_{-0.10}$ & $-10.52^{+0.37}_{-0.43}$ & $0.35^{+0.27}_{-0.20}$ & $-9.67^{+0.53}_{-0.53}$ & $0.26^{+0.02}_{-0.02}$ & $3.35^{+0.44}_{-0.25}$ & $0.09^{+0.02}_{-0.01}$ & $-0.31^{+0.08}_{-0.07}$ & $0.01^{+0.00}_{-0.00}$\\
23iuc & $2.20^{+4.70}_{-1.23}$ & $0.30^{+0.24}_{-0.14}$ & $8.60^{+0.95}_{-0.87}$ & $95.06^{+17.13}_{-17.12}$ & $3.88^{+8.66}_{-2.38}$ & $-10.86^{+0.55}_{-0.47}$ & $1.32^{+0.37}_{-0.39}$ & $-5.75^{+0.24}_{-0.38}$ & $0.50^{+0.02}_{-0.01}$ & $4.52^{+0.84}_{-0.81}$ & $0.15^{+2.17}_{-0.15}$ & $-0.10^{+0.06}_{-0.07}$ & $0.04^{+0.08}_{-0.02}$\\
23qre & $1.48^{+44.21}_{-1.43}$ & $0.20^{+0.13}_{-0.07}$ & $7.16^{+0.20}_{-0.11}$ & $0.11^{+1.25}_{-0.10}$ & $17.82^{+1.47}_{-1.74}$ & $-16.72^{+0.25}_{-0.19}$ & $1.56^{+0.04}_{-0.05}$ & $-6.37^{+0.24}_{-0.27}$ & $0.10^{+0.02}_{-0.02}$ & $6.05^{+0.59}_{-0.57}$ & $0.00^{+0.01}_{-0.00}$ & $-2.26^{+0.46}_{-0.22}$ & $9.19^{+92.13}_{-8.24}$\\
23rau & $2.02^{+50.60}_{-1.96}$ & $0.14^{+0.06}_{-0.03}$ & $9.67^{+1.52}_{-1.70}$ & $0.08^{+2.93}_{-0.08}$ & $0.20^{+1.62}_{-0.16}$ & $-15.92^{+0.95}_{-0.72}$ & $1.35^{+0.21}_{-0.23}$ & $-12.43^{+0.81}_{-0.95}$ & $0.14^{+0.02}_{-0.01}$ & $7.21^{+1.70}_{-1.34}$ & $0.72^{+0.06}_{-0.05}$ & $-2.78^{+0.99}_{-0.28}$ & $64.23^{+951.12}_{-58.63}$\\
23xgo & $0.36^{+0.34}_{-0.21}$ & $0.31^{+0.16}_{-0.11}$ & $8.90^{+1.38}_{-1.17}$ & $0.45^{+28.17}_{-0.44}$ & $1.59^{+1.77}_{-1.11}$ & $-11.14^{+0.86}_{-0.55}$ & $0.78^{+0.75}_{-0.47}$ & $-7.31^{+0.59}_{-0.61}$ & $0.37^{+0.03}_{-0.02}$ & $4.41^{+0.27}_{-0.29}$ & $0.49^{+0.11}_{-0.07}$ & $-0.32^{+0.09}_{-0.08}$ & $0.02^{+0.02}_{-0.02}$\\
\hline
\end{tabular}
}
\tablefoot{
\tablefoottext{\dag}{\footnotesize Similar to Table~\ref{tab:mosfit}, the lower boundary of the distribution of $t_{\rm exp}$ is the only value that changes between each SN modeling. The prior distributions of the shared parameters between the CSI and RD+CSI models remain unchanged. The RD+CSI model contains two additional parameters: the $\gamma$-ray opacity ($\kappa_{\gamma}$) and the fraction of the nickel mass in the SN ejecta ($f_{\rm Ni}$). The prior distributions of each parameter are $\kappa_{\gamma}:\log{\mathcal{U}(0.1,10000)}$~[cm$^{2}$/g], and $f_{\rm Ni}:\log \mathcal{U}(10^{-5},1)$.  
\\
\tablefoottext{\dag\dag}{\footnotesize Nickel mass ($M_{\rm Ni}$) is estimated as $M_{\rm Ni}=f_{\rm Ni}\times M_{\rm ej}$, where $f_{\rm Ni}$ is a free parameter in the RD+CSI model corresponding to the fraction of nickel mass in the SN ejecta.}\\
\tablefoottext{\dag\dag\dag}{\footnotesize Average mass-loss rates are estimated from Eq.~\ref{eq:mdot} assuming $v_w$ as the narrow component ($v_{\rm narrow}$) in Table~\ref{tab:spectroscopic}}.\\
\tablefoottext{\dag\dag\dag\dag}{\footnotesize Outer radius of the CSM is estimated from Eq.~\ref{eq:rcsm}.}\\
\tablefoottext{$\square$}{\footnotesize Abbreviations for each SN are listed in Table~\ref{tab:mosfit}.}\\
}
}
\end{sidewaystable*}

\end{appendix}

\label{LastPage} 

\end{document}